\documentclass[11pt,a4paper]{article}
\usepackage[margin=23mm]{geometry}
\usepackage[T1]{fontenc}
\usepackage{lmodern,amsmath,amssymb,mathtools,slashed,booktabs,array,float,xcolor,tikz}
\usetikzlibrary{decorations.pathmorphing,decorations.markings,arrows.meta}
\usepackage[colorlinks=true,linkcolor=blue!45!black,citecolor=blue!45!black,urlcolor=blue!45!black]{hyperref}
\numberwithin{equation}{section}
\allowdisplaybreaks[2]
\newcommand{\dd}{\mathrm d}
\newcommand{\tr}{\operatorname{tr}}
\newcommand{\bx}{\boldsymbol x}
\newcommand{\bp}{\boldsymbol p}
\newcommand{\bq}{\boldsymbol q}
\newcommand{\bk}{\boldsymbol k}
\newcommand{\bell}{\boldsymbol\ell}
\newcommand{\be}{\boldsymbol\epsilon}
\newcommand{\Born}{\mathrm{Born}}
\newcommand{\Reh}{\operatorname{Re}_{\epsilon}}

\newcommand{\Rij}[2]{\mathcal R_{#1\times #2}}
\tikzset{
 fermion/.style={thick,postaction={decorate},decoration={markings,mark=at position .74 with {\arrow{Latex[length=1.7mm]}}}},
 antifermion/.style={thick,postaction={decorate},decoration={markings,mark=at position .74 with {\arrowreversed{Latex[length=1.7mm]}}}},
 photon/.style={thick,decorate,decoration={snake,amplitude=1.2pt,segment length=5pt}},
 gluon/.style={thick,decorate,decoration={coil,aspect=.7,amplitude=2.0pt,segment length=5pt}},
 shockvertex/.style={circle,fill=black,inner sep=1.35pt},
 energyinsertion/.style={rectangle,fill=black,draw=black,minimum width=4.5pt,minimum height=4.5pt,inner sep=0pt}
}

\newcommand{\gluonshockvertex}[1]{%
 \node[shockvertex] at (2.65,#1) {};}
\newcommand{\virtualselfenergy}[3]{%
 \pgfmathsetmacro{\loopangle}{#3*atan(1/5.1)}
 \pgfmathsetmacro{\loopradius}{(#2-#1)*sqrt(1+1/(5.1*5.1))/2}
 \begin{scope}[shift={(#1,{#3*#1/5.1})},rotate=\loopangle]
  \draw[gluon] (0,0) arc[start angle=180,end angle={180-#3*180},radius=\loopradius];
 \end{scope}
 \pgfmathparse{(#1<2.65) && (#2>2.65)}
 \ifnum\pgfmathresult=1\relax
  \pgfmathsetmacro{\loopcrossy}{#3*((#1+#2)/(2*5.1)+sqrt(\loopradius^2-(2.65-(#1+#2)/2)^2))}
  \gluonshockvertex{\loopcrossy}
 \fi}
\newcommand{\virtualtopology}[1]{%
\begin{center}\begin{tikzpicture}[x=.78cm,y=.78cm]
 \draw[photon] (-1.1,0)--(0,0);
 \node[left] at (-1.1,0) {$\gamma_T^*(l)$};
 \draw[fermion] (0,0)--(5.1,1);
 \draw[antifermion] (0,0)--(5.1,-1);
 \node[right] at (5.1,1) {$q(p)$};
 \node[energyinsertion] at (5.1,-1) {};
 \node[right] at (5.1,-1) {$\bar q(q)$};
 \draw[line width=2pt,black!55] (2.65,-2.65)--(2.65,2.65);
 \node[above] at (2.65,2.65) {shockwave};
 #1
 \node[shockvertex] at (2.65,{2.65/5.1}) {};
 \node[shockvertex] at (2.65,{-2.65/5.1}) {};
\end{tikzpicture}\end{center}}
\newcommand{\realamplitudediagram}[2]{%
\begin{center}\begin{tikzpicture}[x=.91cm,y=.64cm,font=\small]
 \draw[photon] (-1.15,0)--(0,0);
 \node[left] at (-1.15,0) {$\gamma_T^*(l)$};
 \draw[line width=2pt,black!55] (3,-2.75)--(3,2.75);
 \node[above] at (3,2.75) {shockwave};
 #2
 \node[anchor=north east] at (-.55,-2.25) {$i\mathcal M_{#1}^{a}$};
\end{tikzpicture}\end{center}}

\title{Energy-weighted inclusive DIS at NLO in the CGC}
\newcommand{\paperauthor}{Jichen Pan}
\author{\paperauthor}
\newcommand{\affiliation}[1]{\par{\small\itshape #1\par}}
\date{}

\begin{document}

\hypersetup{
 pdftitle={Energy-weighted inclusive DIS at NLO in the CGC},
 pdfauthor={\paperauthor},
 pageanchor=false
}
\pagenumbering{gobble}
\begin{titlepage}
\vspace*{\fill}
\begin{center}
{\LARGE Energy-weighted inclusive DIS\\[2mm]at NLO in the CGC\par}
\vspace{8mm}
{\large \paperauthor\par}
\vspace{5mm}
\affiliation{State Key Laboratory of Nuclear Physics and Technology, Institute of Quantum Matter, South China Normal University, Guangzhou 510006, China\vspace{0.2cm}}
\affiliation{Guangdong Basic Research Center of Excellence for Structure and Fundamental Interactions of Matter, Guangdong Provincial Key Laboratory of Nuclear Science, Guangzhou 510006, China\vspace{0.2cm}}
\vspace{2mm}
{\small E-mail: \href{mailto:panjichen@pku.edu.cn}{\nolinkurl{panjichen@pku.edu.cn}}\par}
\end{center}
\vspace{12mm}
\begin{abstract}
We present the first unified NLO calculation of the full angular
distribution of one-point energy flow in unpolarized DIS within the CGC
framework, combining all quark, antiquark and gluon energy insertions for
both transverse and longitudinal virtual photons. Ultraviolet, ordinary-soft and final-state
collinear singularities cancel between the real and virtual terms, and
the remaining rapidity divergence is absorbed into the JIMWLK evolution
of the target. The full-angle weighted moment reproduces the inclusive
DIS cross section and independently tests the surviving finite impact
factors against the known NLO result. At small $x$, the angular
distribution is sensitive to the transverse dynamics of the dense gluon
field, and the result provides the starting point for extracting the NLO
nucleon energy correlator through a small-angle expansion and matching
to the inclusive DIS coefficient functions.

\end{abstract}
\vfill
\end{titlepage}
\hypersetup{pageanchor=true}
\pagenumbering{roman}
\tableofcontents
\clearpage
\pagenumbering{arabic}

\section{Introduction}
\label{sec:introduction}

Deep-inelastic scattering has established the partonic structure of the
nucleon~\cite{Bloom1969,Breidenbach1969}. Resolving how quarks and gluons are organized within it, and how
their motion is correlated with nucleon spin, remains a major challenge.
Three-dimensional imaging lies at the heart of the Electron--Ion Collider
(EIC) physics program~\cite{Accardi2016,EICYellow2022}. Its reach at small
Bjorken $x$ opens a further line of inquiry: the nonlinear dynamics of dense
gluon fields. Complementing this program, the Electron--Ion
Collider in China (EicC) focuses on the sea-quark region at lower
center-of-mass energies, including the transition to valence
dynamics~\cite{Anderle2021,XiaoZhaoZhou2026}.

This program motivates observables that retain three-dimensional information
without giving up the inclusiveness that makes ordinary deep-inelastic
scattering (DIS) theoretically and experimentally robust. The nucleon energy
correlator (NEEC) is such an object: it inserts an energy-flow operator into a
forward nucleon matrix element and resolves the energy deposited at an angle
$\theta$ from the target beam \cite{LiuZhu2023}. Experimentally, the
corresponding one-point observable is obtained by supplementing inclusive DIS
with a forward calorimetric measurement.

After averaging the azimuth about the photon--target axis for an unpolarized
target, the physical lepton--target cross section in the one-photon exchange
approximation is
\begin{equation}
 \frac{\dd\Sigma_E^{\ell A}}
 {\dd x_B\,\dd Q^2\,\dd\theta^2}
 =\frac{\alpha_{\rm em}}{\pi x_BQ^2}
 \left[\left(1-y+\frac{y^2}{2}\right)\Sigma_E^T(\theta)
       +(1-y)\Sigma_E^L(\theta)\right],\qquad
 y=\frac{Q^2}{x_Bs}.
 \label{eq:unpolarized-energy-flow}
\end{equation}
Here $s$ is the lepton--target center-of-mass energy squared; lepton and target
mass corrections are neglected. The photon--target cross section
$\Sigma_E^T$ is averaged over the two physical transverse states, while
$\Sigma_E^L$ contains the single longitudinal state. Each includes its Born
and NLO terms and its electromagnetic coupling, in the normalization of
Ref.~\cite{Beuf2017}. A distribution normalized to the inclusive DIS rate
requires a consistent NLO expansion of both numerator and denominator.

At small $x$, the angular distribution of the energy flow is sensitive to
the transverse dynamics of the dense gluon field. The Color Glass Condensate
(CGC) framework describes this field through Wilson-line correlators and
their nonlinear small-$x$ evolution
\cite{Gelis2010,Balitsky1996,Kovchegov1999,Iancu2001}. Leading-order CGC
studies have shown that, in the forward regime $Q\theta\sim Q_s\ll Q$,
the saturation scale $Q_s$ is imprinted in the NEEC angular distribution
\cite{Liu2023}. Its angular shape can therefore constrain the saturation
scale generated by nonlinear target dynamics. Turning this sensitivity into
precision predictions for future measurements requires the NEEC in the CGC
beyond leading order, including the effects of real and virtual radiation.

Extracting that NEEC requires a distinction between the nucleon correlator
and the measured cross section. In the forward small-angle limit, the
energy-weighted inclusive DIS cross section factorizes at leading power
into inclusive partonic DIS coefficients convoluted with NEECs
\cite{LiuZhu2023,Liu2023}. Within this factorization, logarithmic
resummation has been developed for NEEC measurements~\cite{Cao2023}.
Energy sum rules further connect these correlators to fracture
functions~\cite{ChenMaTong2024}.

With leading-order CGC predictions now available for related one-point
energy observables with both photon polarizations~\cite{Kang2026}, the
next step is the complete NLO energy-weighted cross section. Extracting
the NLO CGC NEEC then requires a small-angle expansion of that cross
section and matching to the factorization formula. The present work
provides the full angular result, establishing the starting point for
this extraction.

Precision calculations at small $x$ have developed along several connected
directions. For inclusive DIS, the photon impact factor was obtained using
the high-energy operator expansion~\cite{BalitskyChirilli2011,BalitskyChirilli2013}
and light-front perturbation theory~\cite{Beuf2012,Beuf2016,Beuf2017,Hanninen2018}.
The NLO dipole cross section has been implemented numerically~\cite{Ducloue2017}
and combined with improved BK evolution in fits to HERA data~\cite{BeufHERA2020}.
Rapidity-regularization schemes have also been investigated directly in
the NLO longitudinal structure function~\cite{AltinolukRapidity2026}.

For identified hadrons, NLO dihadron production has been calculated for
longitudinal and transverse virtual photons~\cite{Bergabo2022,Bergabo2023}.
In the back-to-back regime, the treatment of TMD fragmentation was
developed in Ref.~\cite{CaucalDihadron2024}. NLO finite cross sections are
also available for diffractive single- and double-hadron
production~\cite{FucillaSingle2024,FucillaDihadron2023}.

Integrating over one of the final-state partons gives access to
single-inclusive hadron production. The quark and antiquark fragmentation
channels were calculated for longitudinal photons~\cite{BergaboSingle2023},
followed by the gluon fragmentation contribution for the same
polarization~\cite{BergaboGluon2024}. For transverse photons, the
single-quark channel is obtained by integrating over the unobserved
antiquark and convoluting the produced-quark spectrum with a fragmentation
function~\cite{Altinoluk2025}. At large virtuality, Sudakov logarithms in
single-hadron production have also been extracted~\cite{AltinolukSudakov2024}.
Recent SIDIS phenomenology has used the LO CGC cross
section~\cite{CasugaSIDIS2026}, while one-loop fracture-function
coefficients address hadron production in the target-fragmentation
region~\cite{ChenGluon2024}.

On the jet side, the inclusive dijet impact factor was derived in the
CGC~\cite{CaucalDijet2021}, and a light-cone-wavefunction treatment
supplied the real three-parton contribution~\cite{IancuMulian2023}.
Subsequent work in the back-to-back limit established the NLO
Weizs\"acker--Williams gluon distribution and Sudakov
factorization~\cite{CaucalSudakov2022,CaucalWW2023,CaucalComplete2024}.
Single-inclusive jet production has been calculated for both photon
polarizations, with fermion and gluon jets treated in the small-radius
approximation~\cite{CaucalSingleJet2024}. The finite terms depend on the
jet clustering prescription and radius. Further studies examined the
jet definition required by TMD factorization at large
virtuality~\cite{CaucalJetTMD2025} and longitudinal-photon jet production
in the target-fragmentation region, including its relation to the extended
quark fracture function~\cite{CaucalTarget2026,CaucalFracture2026}.

With the calorimetric measurement in
Eq.~\eqref{eq:unpolarized-energy-flow}, the final-state radiation is
organized by an additive weight: each particle contributes its energy at
its own angle. The real $q\bar qg$ state therefore carries three energy
insertions; the virtual $q\bar q$ contribution carries the two fermion
insertions. These insertions are applied before integrating the unobserved
momenta. In a collinear splitting, the daughter energy weights add to the
parent weight, completing the final-state collinear cancellation when all
parton species are combined.

In this work, we present the first unified NLO calculation of the
angular energy-flow cross section in unpolarized DIS in the CGC
framework, combining all quark, antiquark and gluon channels for
transverse and longitudinal virtual photons. For massless quarks,
we retain finite-$N_c$ target correlators and the full polar-angle
dependence at leading eikonal order. The measured real and
virtual terms are combined with a consistent dimensional prescription. The
UV, ordinary soft and final-state collinear singularities cancel, while the
remaining rapidity singularity is absorbed into JIMWLK evolution. The
surviving impact factors in weighted angular moments are matched
analytically to the inclusive NLO result of Ref.~\cite{Beuf2017}, including
their finite coefficients and normalization. The transverse and
longitudinal contributions enter the
physical cross section in Eq.~\eqref{eq:unpolarized-energy-flow}.

The paper is organized as follows. The fixed-order transverse calculation
and its singularity structure are
presented in Secs.~\ref{sec:fixed-order} and
\ref{sec:singularity-structure}. Section~\ref{sec:finite-validation}
checks the finite impact factors against inclusive DIS, and
Sec.~\ref{sec:longitudinal} completes the longitudinal contribution.
Section~\ref{sec:outlook} summarizes the results and discusses further
steps toward NEEC predictions and experimental comparisons.

\section{Fixed-order calculation}
\label{sec:fixed-order}

The fixed-order calculation is organized by the particle selected by the
one-point energy-flow operator.  We first keep the antiquark momentum fixed and
integrate over the remaining final state.  The quark-tagged and gluon-tagged
terms are then constructed directly from the same unintegrated amplitudes, with
their own detector insertions applied before either unobserved transverse
momentum is integrated.  Thus the real $q\bar qg$ contribution contains all
three alternative one-particle tags.

\subsection{Conventions, kinematics, and the energy-flow measurement}
\label{sec:fixed-order-conventions}

We use the reflected Breit frame in which the virtual photon moves in the plus
direction and the target moves in the minus direction.  The target field is
localized near $x^+=0$.  Light-cone components and the transverse scalar
product are defined by
\begin{equation}
 v^\pm=\frac{v^0\pm v^3}{\sqrt2},\qquad
 v\cdot w=v^+w^-+v^-w^+-\boldsymbol v\cdot\boldsymbol w,
 \qquad v^2=2v^+v^--\boldsymbol v^2 .
 \label{eq:lightcone-conventions}
\end{equation}
The photon momentum is
\begin{equation}
 l=\left(l^+,-\frac{Q^2}{2l^+},\boldsymbol0\right),
 \qquad l^+=\frac{Q}{\sqrt2}.
 \label{eq:photon-momentum}
\end{equation}
For the real correction, the on-shell quark, antiquark, and gluon momenta obey
\begin{align}
 p^+&=z_1l^+,&p^-&=\frac{\bp^2}{2z_1l^+},\nonumber\\
 q^+&=z_2l^+,&q^-&=\frac{\bq^2}{2z_2l^+},\nonumber\\
 k^+&=zl^+,&k^-&=\frac{\bk^2}{2zl^+},
 &z_1+z_2+z&=1 .
 \label{eq:threefractions}
\end{align}
The virtual final state contains no gluon and obeys $z_1+z_2=1$.  In every
virtual graph below, $z$ denotes the plus-momentum fraction carried by the loop
gluon; it is integrated over the range stated for that graph and must not be set
to zero.  In a real graph, $z$ instead denotes the plus-momentum fraction of the
on-shell final-state gluon.  The two longitudinal sectors of the exchange graphs
with a negative signed loop momentum are stated separately.

The calculation is performed in $A^+=0$ gauge with
$n^\mu=(0,1,\boldsymbol0)$ and $n\cdot r=r^+$.  For massless quarks, the vacuum
propagators and perturbative vertices are
\begin{align}
 S_F(r)&=\frac{i\slashed r}{r^2+i0},\nonumber\\
 G_{ab}^{\mu\nu}(r)&=\frac{i\delta_{ab}}{r^2+i0}
 \left[-g^{\mu\nu}
 +\frac{r^\mu n^\nu+r^\nu n^\mu}{r^+}\right]
 \equiv\frac{i\delta_{ab}}{r^2+i0}d^{\mu\nu}(r),\nonumber\\
 V_{qg}^{a\mu}&=ig\gamma^\mu t^a,\qquad
 V_{\gamma q\bar q}^{\rho}=iee_f\gamma^\rho,\qquad
 \tr(t^at^b)=\frac12\delta^{ab},\qquad
 C_F=\frac{N_c^2-1}{2N_c}.
 \label{eq:vacuum-rules}
\end{align}
This is the projectile light-cone gauge for the plus-moving photon.
In the eikonal CGC description, the minus-moving target is represented by
a color current $J_a^-=\rho_a(x^+,\bx)$ independent of $x^-$.
The choice $A^+=0$ makes this static current compatible with covariant
current conservation and permits a background of the form
$A_{\rm cl}^\mu=\delta^{\mu-}\alpha(x^+,\bx)$ localized near $x^+=0$.
Multiple interactions of a fast projectile parton with this field are
ordered along $x^+$ and resum into Wilson lines of $A^-$, while the
radiated gluons have transverse physical polarizations with
$n\cdot\epsilon=\epsilon^+=0$~\cite{Gelis2010,Kolbe2021}.

The target interaction is encoded in fundamental and adjoint Wilson lines,
\begin{align}
 V(\bx)&=\mathcal P\exp\left[ig\int_{-\infty}^{+\infty}
 \dd x^+ A_a^-(x^+,\bx)t^a\right],\nonumber\\
 U^{ab}(\bx)&=2\tr[t^aV(\bx)t^bV^\dagger(\bx)].
 \label{eq:wilson-lines}
\end{align}
For outgoing momentum $r$ and incoming momentum $r'$, the full shockwave
insertions are
\begin{align}
 \mathcal T_F(r,r')={}&2\pi\delta(r^+-r'^+)\gamma^+
 \int\dd^2\bx\,
 e^{-i(\boldsymbol r-\boldsymbol r')\cdot\bx}
 \left[\Theta(r^+)V(\bx)-\Theta(-r^+)V^\dagger(\bx)\right],\nonumber\\
 \mathcal T_{A,\mu\nu}^{ab}(r,r')={}&-2\pi\delta(r^+-r'^+)
 (2r^+)g_{\mu\nu}
 \int\dd^2\bx\,
 e^{+i(\boldsymbol r-\boldsymbol r')\cdot\bx}
 \left[\Theta(r^+)U^{ab}(\bx)
 -\Theta(-r^+)U^{\dagger ab}(\bx)\right].
 \label{eq:shockrules}
\end{align}
These vertices contain the identity contribution.  We therefore subtract the
globally no-target amplitude once.  With $V_i\equiv V(\bx_i)$, the target
operators used below are
\begin{equation}
 S_{ij}=\frac1{N_c}\tr(V_iV_j^\dagger),\qquad
 S_{ijkl}=\frac1{N_c}\tr(V_iV_j^\dagger V_kV_l^\dagger),
 \qquad \bx_{ij}=\bx_i-\bx_j .
 \label{eq:targetops}
\end{equation}
Products of traces remain inside a single average
$\langle O\rangle_Y=\int D\rho\,W_Y[\rho]O[\rho]$.

The same prescription for the axial pole is used before the longitudinal
sectors are separated.  Loop transverse momenta are continued to
$2-2\epsilon$ dimensions in the four-dimensional helicity (FDH) scheme, with spin dimension $D_s=4$
and momentum dimension $D=4-2\epsilon$. Measured external momenta and the
two physical photon polarizations remain four dimensional. Hatted loop
components are retained until the dimensionally regulated integration is
complete; their finite contribution to the virtual correction numbered
$A_{14}$ is included below.  A single dimensional regulator $\epsilon$ is used.
The labels $1/\epsilon_{\rm UV}$ and $1/\epsilon_{\rm coll}$ identify the
momentum region that produces a pole. In particular, a scaleless transverse integral is resolved as
\begin{equation}
 \mu^{2\epsilon}\int\frac{\dd^{2-2\epsilon}\bell}{(2\pi)^{2-2\epsilon}}
 \frac1{\ell^2}
 =\frac1{4\pi}\left(\frac1{\epsilon_{\rm UV}}
 -\frac1{\epsilon_{\rm coll}}\right)
 \quad\text{at pole level}.
 \label{eq:scaleless-uv-coll}
\end{equation}
Every virtual-gluon longitudinal integration and every real-gluon phase-space
integration is regulated analytically by inserting
\begin{equation}
 \frac{\dd z}{z}\longrightarrow
 \frac{\dd z}{z}\left(\frac{\nu^+}{zl^+}\right)^{\eta_Y},
 \qquad \operatorname{Re}\eta_Y<0
 \label{eq:analytic-rapidity-regulator}
\end{equation}
for potential rapidity divergences.
The same factor applies to both longitudinal measures $\dd z$ and $\dd z/z$.  Here $zl^+=k^+$ is the emitted gluon plus momentum and
 $\nu^+$ is the rapidity-renormalization scale.  The regulator $\eta_Y$ is kept
 independent of the transverse dimensional regulator $\epsilon$.  The
 $\eta_Y$ expansion is performed before the $\epsilon$ expansion, while every
 $1/(\epsilon\eta_Y)$ term is retained until the real and virtual graph sums
 have been formed.  For any
kernel $F(z)$ regular at $z=0$, analytic continuation from
$\operatorname{Re}\eta_Y<0$ gives
\begin{align}
 &\int_0^{z_{\max}}\frac{\dd z}{z}
 \left(\frac{\nu^+}{zl^+}\right)^{\eta_Y}F(z)
 \nonumber\\[-2pt]
 &\quad=-\frac{F(0)}{\eta_Y}
 +F(0)\ln\frac{z_{\max}l^+}{\nu^+}
 +\int_0^{z_{\max}}\frac{\dd z}{z}\,[F(z)-F(0)]
 +\mathcal O(\eta_Y).
 \label{eq:analytic-rapidity-expansion}
\end{align}
Rapidity minimal subtraction removes only the $1/\eta_Y$ pole.  The logarithm
of $\nu^+$ and the last integral in
Eq.~\eqref{eq:analytic-rapidity-expansion} are finite contributions to the NLO
impact factor.

For a measured massless particle $a$ with momentum fraction $z_a$ and
transverse momentum $\boldsymbol p_a$, its energy and target-referenced polar
angle are
\begin{equation}
 \frac{E_a}{E_A}=x_B\left(z_a+
 \frac{\boldsymbol p_a^2}{Q^2z_a}\right),\qquad
 \vartheta_a=2\arctan\frac{Qz_a}{|\boldsymbol p_a|},
 \qquad E_A=\frac{Q}{2x_B}.
 \label{eq:particle-energy-angle}
\end{equation}
The one-particle detector insertion is
\begin{equation}
 \mathcal M_a(\theta;z_a,\boldsymbol p_a)=
 x_B\left(z_a+\frac{\boldsymbol p_a^2}{Q^2z_a}\right)
 \delta(\theta^2-\vartheta_a^2).
 \label{eq:oneparticlemeasurement}
\end{equation}
We use the original linear energy-flow measurement, with all final-state
particles included. The longitudinal fractions lie on the two- or three-body
simplex, and transverse momenta are integrated over the full plane in the
leading eikonal approximation. The real and virtual insertions are
\begin{align}
 \mathcal M_E^{(3)}&=
 \mathcal M_q(\theta;z_1,\bp)
 +\mathcal M_{\bar q}(\theta;z_2,\bq)
 +\mathcal M_g(\theta;z,\bk),
 \label{eq:fullrealmeasurement}\\
 \mathcal M_E^{(2)}&=
 \mathcal M_q(\theta;z_1,\bp)
 +\mathcal M_{\bar q}(\theta;z_2,\bq).
 \label{eq:fullvirtualmeasurement}
\end{align}
Each detector insertion is applied before an unobserved momentum is integrated
or a singular limit is taken.  The sum in
Eqs.~\eqref{eq:fullrealmeasurement}--\eqref{eq:fullvirtualmeasurement} is a sum
over the alternative one-particle energy tags.

When a final-state quark or antiquark becomes collinear with the gluon,
the two daughter energy weights add to the parent weight at the same
angle. The gluon tag therefore supplies the contribution needed for
the real--virtual collinear cancellation in
Sec.~\ref{sec:final-state-collinear}.

\subsection{Born amplitude and normalization}
\label{sec:born}

The Born process fixes the normalization shared by all real and virtual
corrections.  Applying the photon vertex and the two fermion shockwave
insertions gives\footnote{Transverse Fourier transforms use
$\int_{\bk}\equiv\int\dd^2\bk/(2\pi)^2$ and
$f(\bx)=\int_{\bk}e^{i\bk\cdot\bx}f(\bk)$.
The inverse transform has phase $e^{-i\bk\cdot\bx}$.}
\begin{align}
 i\mathcal A_0={}&iee_f
 \int\dd^2\bx_1\dd^2\bx_2\,
 e^{-i\bp\cdot\bx_1-i\bq\cdot\bx_2}
 (V_1V_2^\dagger-1)
 \int_{k_1^-,\bk_1}
 \frac{e^{i\bk_1\cdot\bx_{12}}\mathcal N_0}
 {D(k_1)D(k_1-l)},\nonumber\\
 \mathcal N_0={}&
 \bar u(p)\gamma^+\slashed k_1\slashed\epsilon_T
 (\slashed k_1-\slashed l)\gamma^+v(q),
 \qquad k_1^+=z_1l^+,
 \label{eq:born-amplitude}
\end{align}
where $D(r)=r^2+i0$ and
$\int_{k_1^-,\bk_1}\equiv
\int\dd k_1^-/(2\pi)\int\dd^2\bk_1/(2\pi)^2$.
The lower $k_1^2$ pole, or equivalently the upper
$(k_1-l)^2$ pole, yields
\begin{equation}
 \int\frac{\dd k_1^-}{2\pi}
 \frac1{D(k_1)D(k_1-l)}
 =\frac{i}{2l^+(\bk_1^2+Q_\gamma^2)},
 \qquad Q_\gamma^2=z_1z_2Q^2 .
 \label{eq:born-energy-integral}
\end{equation}
For photon helicity $+$, the two nonvanishing fermion-spin numerators
are\footnote{The transverse polarization vectors are
$\be_\pm=(1,\pm i)/\sqrt2$, with $\epsilon_T^+=\epsilon_T^-=0$.
We write $\be\equiv\be_+$, $\be^*\equiv\be_-$,
$a_\epsilon=\boldsymbol a\cdot\be$ and
$a_{\epsilon^*}=\boldsymbol a\cdot\be^*$, so that
$a_\epsilon a_{\epsilon^*}=\boldsymbol a^2/2$.
$\Reh$ conjugates only the helicity vectors; $\operatorname{Re}$ also
conjugates phases and target operators.}
\begin{equation}
 \mathcal N_{0,+}=-8(l^+)^2z_1\sqrt{z_1z_2}\,k_{1\epsilon},
 \qquad
 \mathcal N_{0,-}=+8(l^+)^2z_2\sqrt{z_1z_2}\,k_{1\epsilon}.
 \label{eq:born-spin}
\end{equation}
The opposite photon helicity follows by reversing the fermion helicities and
conjugating the transverse polarization.

The remaining transverse transform,
\begin{equation}
 \int_{\bk_1}
 \frac{k_{1\epsilon}e^{i\bk_1\cdot\bx_{12}}}
 {\bk_1^2+Q_\gamma^2}
 =\frac{iQ_\gamma}{2\pi}
 \frac{x_{12,\epsilon}}{|\bx_{12}|}
 K_1(Q_\gamma|\bx_{12}|),
 \label{eq:born-transverse-integral}
\end{equation}
gives the two transverse-photon wave functions
\begin{align}
 \psi_{0,+}(\bx_{12})&=
 \frac{4iee_fl^+Q_\gamma}{2\pi}
 z_1\sqrt{z_1z_2}\,
 \frac{x_{12,\epsilon}}{|\bx_{12}|}
 K_1(Q_\gamma|\bx_{12}|),\nonumber\\
 \psi_{0,-}(\bx_{12})&=
 -\frac{4iee_fl^+Q_\gamma}{2\pi}
 z_2\sqrt{z_1z_2}\,
 \frac{x_{12,\epsilon}}{|\bx_{12}|}
 K_1(Q_\gamma|\bx_{12}|).
 \label{eq:born-wave-functions}
\end{align}
The target-dependent Born amplitude is the photon wave function multiplied by
$V_1V_2^\dagger-1$.  The subtraction makes the amplitude vanish when all
Wilson lines are set to the identity.

The incident flux is $2l^+$, and the two-body phase space is normalized as
\begin{equation}
 \frac{\dd\Phi^{(2)}}{2l^+}
 =\frac{\dd^2\bp\,\dd^2\bq\,\dd y_1\dd y_2}
 {8(l^+)^2(2\pi)^5}\,
 \delta(1-z_1-z_2).
 \label{eq:born-phase-space}
\end{equation}
For a one-sided virtual graph $i$, including the average over the two
transverse photon polarizations, we define
\begin{equation}
 \mathcal D_i=
 \frac1{8(l^+)^2(2\pi)^5}\frac12
 \sum_{\lambda,h,\bar h}
 \left\langle
 \tr\!\left[(i\mathcal A_{i,\lambda h\bar h})
 (i\mathcal A_{0,\lambda h\bar h})^\dagger\right]
 \right\rangle_Y .
 \label{eq:interference-normalization}
\end{equation}
The conjugate-side correction is included only after forming
$2\operatorname{Re}\mathcal D_i$.  When the antiquark is measured, the
unobserved-quark phase space defines
\begin{equation}
 \mathcal V_i(z_2,\bq)=
 \int_0^1\frac{\dd z_1}{z_1}\int\dd^2\bp\,
 \delta(1-z_1-z_2)\mathcal D_i .
 \label{eq:spectator-definition}
\end{equation}
Graph-dependent momentum phases must be retained until the $\bp$ integral is
performed; the Born graph alone sets $\bx_{1'}=\bx_1$ directly.

With $z_1=1-z_2$, the Born photon overlap after the spectator integration is
\begin{align}
 \mathcal W_T(\bx_1,\bx_2,\bx_{2'};z_1,z_2,\bq)
 ={}&\frac{e^2e_f^2Q^2N_c}{(2\pi)^5}
 z_1z_2^2(z_1^2+z_2^2)
 e^{i\bq\cdot\bx_{2'2}}\nonumber\\
 &\times
 \frac{\bx_{12}\cdot\bx_{12'}}
 {|\bx_{12}|\,|\bx_{12'}|}
 K_1(Q_\gamma|\bx_{12}|)
 K_1(Q_\gamma|\bx_{12'}|).
 \label{eq:born-overlap}
\end{align}
The antiquark-tagged Born kernel then reads
\begin{equation}
 \mathcal V_{\Born}(z_2,\bq)=
 \int\dd^2\bx_1\dd^2\bx_2\dd^2\bx_{2'}\,
 \mathcal W_T
 \left\langle
 S_{2'2}-S_{12}-S_{2'1}+1
 \right\rangle_Y .
 \label{eq:born-result}
\end{equation}
The measured Born contribution is
\begin{equation}
 \Sigma^T_{\bar q,\Born}(\theta)=
 \int\frac{\dd z_2}{z_2}\int\dd^2\bq\,
 \mathcal M_{\bar q}(\theta;z_2,\bq)
 \mathcal V_{\Born}(z_2,\bq).
 \label{eq:measured-born}
\end{equation}
The angular delta function first fixes the transverse momentum:
\begin{equation}
 |\bq|_*=Qz_2\cot\frac\theta2,\qquad
 \left|\frac{\partial\vartheta_{\bar q}}
 {\partial|\bq|}\right|_*
 =\frac{\sin\theta}{|\bq|_*}.
 \label{eq:born-angular-root}
\end{equation}
At this root the energy fraction is
\begin{align}
 \left.\frac{E_{\bar q}}{E_A}\right|_*
 &=x_Bz_2\left(1+\cot^2\frac\theta2\right)
 =x_Bz_2\csc^2\frac\theta2 .
 \label{eq:born-measurement-support}
\end{align}
The longitudinal simplex supplies the support
\begin{equation}
 0<z_2<1 .
 \label{eq:born-z2-support}
\end{equation}
The detector delta function fixes the radial momentum at each longitudinal fraction.
Since
\begin{align}
 &\int_0^\infty |\bq|\dd|\bq|\,
 \delta(\theta^2-\vartheta_{\bar q}^2)F(|\bq|)
 =\frac{|\bq|_*^2}{2\theta\sin\theta}F(|\bq|_*),\nonumber\\
 &\left.\frac{E_{\bar q}}{E_A}\right|_*
 =x_Bz_2\csc^2\frac\theta2,
 \label{eq:born-angular-jacobian}
\end{align}
the radial momentum integral can be carried out with exact angular kinematics.  Setting
\begin{equation}
 z_2=v\sin^2\frac\theta2,\qquad
 \bq_*=\frac{Qv}{2}\sin\theta\,\boldsymbol e_\phi,\qquad
 \boldsymbol e_\phi=(\cos\phi,\sin\phi),
 \label{eq:born-angular-variables}
\end{equation}
gives
\begin{equation}
 \boxed{\displaystyle
 \Sigma^T_{\bar q,\Born}(\theta)=
 \frac{x_BQ^2\sin\theta}{8\theta}
 \int_0^{\csc^2(\theta/2)}\dd v\,v^2\int_0^{2\pi}\dd\phi\,
 \mathcal V_{\Born}\left(
 v\sin^2\frac\theta2,
 \frac{Qv}{2}\sin\theta\,\boldsymbol e_\phi
 \right).}
 \label{eq:measured-born-final}
\end{equation}
All dependence on the target remains in the transverse-coordinate convolution
inside $\mathcal V_{\Born}$.  Equation~\eqref{eq:measured-born-final} completes
the radial integral fixed by the detector and maps the remaining measured
phase space to $0<v<\csc^2(\theta/2)$ and $0<\phi<2\pi$.

The quark-tagged term follows from the same amplitude by interchanging the
fermion labels, momenta, and Wilson-line orientations before integrating the
unobserved antiquark.  Denoting the resulting fixed-quark kernel by
$\mathcal V_{q,\Born}$ and the kernel in Eq.~\eqref{eq:born-result} by
$\mathcal V_{\bar q,\Born}$, the complete Born energy flow is
\begin{align}
 \Sigma^T_{E,\Born}(\theta)
 ={}&\frac{x_BQ^2\sin\theta}{8\theta}
 \int_0^{\csc^2(\theta/2)}\dd v\,v^2\int_0^{2\pi}\dd\phi\,
 \Bigg[
 \mathcal V_{q,\Born}\left(
 v\sin^2\frac\theta2,
 \frac{Qv}{2}\sin\theta\,\boldsymbol e_\phi\right)
 \nonumber\\[-2pt]
 &\hspace{43mm}+
 \mathcal V_{\bar q,\Born}\left(
 v\sin^2\frac\theta2,
 \frac{Qv}{2}\sin\theta\,\boldsymbol e_\phi\right)
 \Bigg].
 \label{eq:measured-born-energy-sum}
\end{align}
For a charge-conjugation-even target average, the two terms are equal after
relabeling the integration variables.

\subsection{Antiquark-tagged channel}
\label{sec:antiquark-channel}

We first keep the antiquark momentum $q$ fixed and integrate over the quark
momentum only after forming the relevant interference.  There are ten
independent one-loop topologies:
\begin{center}
\begin{tabular}{@{}ccl@{}}
\toprule
Graphs & Number & Topology \\
\midrule
$A_5$--$A_8$ & $4$ & gluon propagation or exchange across the shockwave \\
$A_9,A_{10}$ & $2$ & external self-energies after the shockwave \\
$A_{11},A_{12}$ & $2$ & self-energies before the shockwave \\
$A_{13}$ & $1$ & quark--antiquark exchange after the shockwave \\
$A_{14}$ & $1$ & quark--antiquark exchange before the shockwave \\
\bottomrule
\end{tabular}
\end{center}
Graphs $A_{13}$ and $A_{14}$ each split into two longitudinal momentum
sectors.  The ten topologies therefore produce twelve one-sided virtual
contributions.  The one-loop amplitude is written as
\begin{equation}
 \mathcal A_V=\sum_{i=5}^{14}\mathcal A_i ,
 \label{eq:virtual-amplitude-sum}
\end{equation}
with both sectors included in $\mathcal A_{13}$ and $\mathcal A_{14}$.  The
graph numbering follows Refs.~\cite{Bergabo2023,Altinoluk2025}.  Each complex
conjugate enters the cross section through $2\operatorname{Re}$.

\subsubsection{Virtual graphs and their measured contributions}
\label{sec:antiquark-virtual-amplitudes}

Graph $A_5$ illustrates the complete derivation.  The remaining virtual
graphs will be presented as a diagram, a final amplitude, and their
contribution to the measured cross section, using the same momentum routing and
Wilson-line convention.

\paragraph{Example: gluon exchange across the shockwave on the quark line.}
The quark emits a virtual gluon before the target and absorbs it after the
target.  Both the daughter quark and the gluon cross the shockwave.
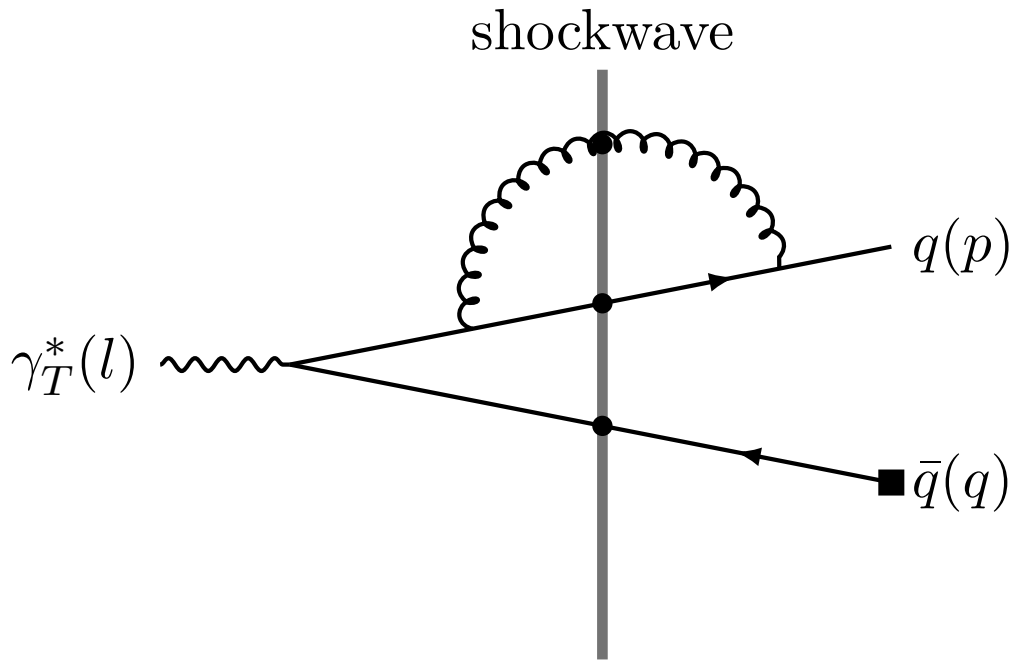
\begin{figure}[t]
\centering
\begin{tikzpicture}[x=.78cm,y=.78cm]
 \draw[thick,decorate,decoration={snake,amplitude=1.2pt,segment length=5pt}]
 (-1.1,0)--(0,0);
 \node[left] at (-1.1,0) {$\gamma_T^*(l)$};
 \draw[fermion] (0,0)--(5.1,1);
 \draw[antifermion] (0,0)--(5.1,-1);
 \node[right] at (5.1,1) {$q(p)$};
 \node[energyinsertion] at (5.1,-1) {};
 \node[right] at (5.1,-1) {$\bar q(q)$};
 \draw[line width=2pt,black!55] (2.65,-2.5)--(2.65,2.5);
 \node[above] at (2.65,2.5) {shockwave};
 \virtualselfenergy{1.55}{4.15}{1}
 \node[shockvertex] at (2.65,{2.65/5.1}) {};
 \node[shockvertex] at (2.65,{-2.65/5.1}) {};
\end{tikzpicture}
\caption{Virtual graph $A_5$.  The gluon and daughter quark both cross the
target shockwave. The black square marks the antiquark energy-flow
insertion; black dots mark shockwave interactions.}
\label{fig:virtual-a5}
\end{figure}

Choose the plus-momentum routing
\begin{equation}
 k_1^+=z_1l^+,\qquad
 k_2^+=k_3^+=(z_1-z)l^+,\qquad
 0<z<z_1 .
 \label{eq:g5-plus-routing}
\end{equation}
The two gluon arrows carry momenta $k_2-k_1$ and $k_3-p$, both
with plus component $-zl^+$.  The negative-plus adjoint Wilson line and the
factor $2(k_3^+-p^+)=-2zl^+$ from the adjoint shockwave insertion combine into
the positive factor $2zl^+$ below.  Direct multiplication of the photon
vertex, the two QCD vertices, four quark propagators, two gluon propagators,
and three shockwave insertions gives
\begin{align}
 i\mathcal A_5={}&iee_fg^2
 \int_0^{z_1}\left(\frac{\nu^+}{zl^+}\right)^{\eta_Y}\frac{l^+\dd z}{2\pi}
 \int\dd^2\bx_1\dd^2\bx_2\dd^2\bx_3
 \int_{k_1^-,\bk_1}\int_{k_2^-,\bk_2}
 \int_{k_3^-,\bk_3}\nonumber\\[-2pt]
 &\times
 \frac{2zl^+\,e^{i\Phi_5}\mathcal N_5
 [V_3t^aV_3^\dagger V_1t^aV_2^\dagger-C_F\boldsymbol1]}
 {D(k_1)D(k_1-l)D(k_2)D(k_2-k_1)D(k_3)D(k_3-p)},\nonumber\\
 \Phi_5={}&
 \bk_1\cdot\bx_{32}-\bk_2\cdot\bx_{31}
 +\bk_3\cdot\bx_{31}-\bp\cdot\bx_3-\bq\cdot\bx_2,\nonumber\\
 \mathcal N_5={}&
 \bar u(p)\gamma^\mu\slashed k_3\gamma^+
 \slashed k_2\gamma^\nu\slashed k_1\slashed\epsilon_T
 (\slashed k_1-\slashed l)\gamma^+v(q)\,
 d_{\nu\rho}(k_2-k_1)d_\mu{}^\rho(k_3-p).
 \label{eq:g5-insertion}
\end{align}
The subtraction proportional to $C_F\boldsymbol1$ removes the globally
no-target amplitude in the same full-vertex convention used at Born level.

The three minus-momentum integrations can be done sequentially.  Closing the
$k_3^-$ and $k_2^-$ contours below and the $k_1^-$ contour above selects
\begin{align}
 k_3^-&=\frac{\bk_3^2}{2(z_1-z)l^+}-i0,\qquad
 k_2^-=\frac{\bk_2^2}{2(z_1-z)l^+}-i0,\nonumber\\
 k_1^-&=l^--\frac{\bk_1^2}{2z_2l^+}+i0 .
 \label{eq:g5-energy-poles}
\end{align}
The product of the three residues is
\begin{equation}
 \frac{-i}{2(z_1-z)l^+}
 \frac{-i}{2(z_1-z)l^+}
 \frac{+i}{-2z_2l^+}
 \left.
 \frac1{D(k_3-p)D(k_1)D(k_2-k_1)}
 \right|_{\rm poles}.
 \label{eq:g5-scalar-residues}
\end{equation}
Introduce the relative momenta of the quark--gluon pair on the two sides of
the target,
\begin{equation}
 \bell_{\rm out}=\bk_3-\frac{z_1-z}{z_1}\bp,\qquad
 \bell_{\rm in}=\bk_2-\frac{z_1-z}{z_1}\bk_1 .
 \label{eq:g5-relative-momenta}
\end{equation}
At the selected poles, the remaining inverse propagators become
\begin{align}
 D(k_3-p)&=-\frac{z_1}{z_1-z}\ell_{\rm out}^2,\nonumber\\
 D(k_1)&=-\frac{\bk_1^2+Q_\gamma^2}{z_2},\nonumber\\
 D(k_2-k_1)&=-\frac{z_1}{z_1-z}
 \left[
 \ell_{\rm in}^2+
 \frac{z(z_1-z)}{z_1^2z_2}
 (\bk_1^2+Q_\gamma^2)
 \right].
 \label{eq:g5-pole-denominators}
\end{align}
Equations~\eqref{eq:g5-scalar-residues} and
\eqref{eq:g5-pole-denominators} give
\begin{equation}
 \frac{-i}{8z_1^2(l^+)^3}
 \frac1{
 \ell_{\rm out}^2(\bk_1^2+Q_\gamma^2)
 [\ell_{\rm in}^2+
 z(z_1-z)(\bk_1^2+Q_\gamma^2)/(z_1^2z_2)]}.
 \label{eq:g5-residue}
\end{equation}

The two light-cone-gauge numerators may be contracted before evaluating the
spinor chain:
\begin{align}
 \gamma^\mu X\gamma^\nu
 d_{\nu\rho}(r)d_\mu{}^\rho(r')
 ={}&\gamma^\mu X\gamma_\mu
 -\frac{\slashed rX\gamma^+}{r^+}
 -\frac{\gamma^+X\slashed r'}{r'^+}
 +\frac{r\cdot r'}{r^+r'^+}\gamma^+X\gamma^+ .
 \label{eq:double-gluon-contraction}
\end{align}
Evaluation at the poles gives the two nonzero spin components
\begin{align}
 \mathcal N_{5,+}={}&
 \frac{32(l^+)^2z_1\sqrt{z_1z_2}}{z^2}
 \left[
 z_1^2\ell_{{\rm out},\epsilon}
 \ell_{{\rm in},\epsilon^*}
 +(z_1-z)^2\ell_{{\rm out},\epsilon^*}
 \ell_{{\rm in},\epsilon}
 \right]k_{1\epsilon},\nonumber\\
 \mathcal N_{5,-}={}&
 -\frac{32(l^+)^2z_2\sqrt{z_1z_2}}{z^2}
 \left[
 z_1^2\ell_{{\rm out},\epsilon^*}
 \ell_{{\rm in},\epsilon}
 +(z_1-z)^2\ell_{{\rm out},\epsilon}
 \ell_{{\rm in},\epsilon^*}
 \right]k_{1\epsilon}\nonumber\\
 &+\frac{16(l^+)^2\sqrt{z_2/z_1}(z_1-z)^2}{z}
 (\bk_1^2+Q_\gamma^2)\ell_{{\rm out},\epsilon}.
 \label{eq:g5-spin}
\end{align}
The last line cancels the photon denominator in
Eq.~\eqref{eq:g5-residue}.  Retaining this contact term is necessary for the
finite part of the amplitude.  Combining the vertex factors and the three
residues gives
\begin{align}
 i\mathcal A_{5,h}={}&
 \frac{ee_fg^2}{4z_1^2l^+}
 \int_0^{z_1}\left(\frac{\nu^+}{zl^+}\right)^{\eta_Y}\frac{z\dd z}{2\pi}
 \int\dd^2\bx_1\dd^2\bx_2\dd^2\bx_3\,
 [V_3t^aV_3^\dagger V_1t^aV_2^\dagger-C_F\boldsymbol1]
 \nonumber\\
 &\times
 \int_{\bk_1,\bell_{\rm in},\bell_{\rm out}}
 \frac{e^{i\Phi_5}\mathcal N_{5,h}}
 {\ell_{\rm out}^2(\bk_1^2+Q_\gamma^2)
 [\ell_{\rm in}^2+
 z(z_1-z)(\bk_1^2+Q_\gamma^2)/(z_1^2z_2)]}.
 \label{eq:g5-post-contour-amplitude}
\end{align}

The transverse integrations are most transparent after shifting to the
relative momenta in Eq.~\eqref{eq:g5-relative-momenta}.  Define the parent
position, the photon separation, and the target color matrix
\begin{equation}
 \boldsymbol B_5=\bx_1+\frac{z}{z_1}\bx_{31},\qquad
 \boldsymbol R_5=\bx_{12}+\frac{z}{z_1}\bx_{31},\qquad
 \mathcal U_5=V_3t^aV_3^\dagger V_1t^aV_2^\dagger
 -C_F\boldsymbol1 .
 \label{eq:g5-coordinate-definitions}
\end{equation}
The shifted phase is
\begin{equation}
 \Phi_5=
 \bk_1\cdot\boldsymbol R_5
 -\bell_{\rm in}\cdot\bx_{31}
 +\bell_{\rm out}\cdot\bx_{31}
 -\bp\cdot\boldsymbol B_5-\bq\cdot\bx_2 .
 \label{eq:g5-shifted-phase}
\end{equation}

The outgoing relative momentum is integrated first:
\begin{equation}
 \int_{\bell_{\rm out}}
 \frac{\ell_{{\rm out},i}
 e^{i\bell_{\rm out}\cdot\bx_{31}}}{\ell_{\rm out}^2}
 =\frac{ix_{31,i}}{2\pi x_{31}^2}.
 \label{eq:g5-outgoing-fourier}
\end{equation}
Substitution into Eq.~\eqref{eq:g5-post-contour-amplitude} gives a double
transverse integral for each spin component,
\begin{align}
 i\mathcal A_{5,+}
 ={}&\frac{8iee_fg^2l^+\sqrt{z_1z_2}}
 {z_1(2\pi)^2}
 \int_0^{z_1}\left(\frac{\nu^+}{zl^+}\right)^{\eta_Y}\frac{\dd z}{z}
 \int\dd^2\bx_1\dd^2\bx_2\dd^2\bx_3\,
 \mathcal U_5e^{-i\bp\cdot\boldsymbol B_5-i\bq\cdot\bx_2}
 \frac1{x_{31}^2}
 \nonumber\\
 &\times\int_{\bk_1,\bell_{\rm in}}
 \frac{e^{i\bk_1\cdot\boldsymbol R_5
 -i\bell_{\rm in}\cdot\bx_{31}}k_{1\epsilon}}
 {(\bk_1^2+Q_\gamma^2)
 [\ell_{\rm in}^2+m_5^2(\bk_1)]}
 \left[
 z_1^2x_{31,\epsilon}\ell_{{\rm in},\epsilon^*}
 +(z_1-z)^2x_{31,\epsilon^*}\ell_{{\rm in},\epsilon}
 \right],
 \label{eq:g5-after-lout-plus}\\[3pt]
 i\mathcal A_{5,-}
 ={}&-\frac{8iee_fg^2l^+z_2\sqrt{z_1z_2}}
 {z_1^2(2\pi)^2}
 \int_0^{z_1}\left(\frac{\nu^+}{zl^+}\right)^{\eta_Y}\frac{\dd z}{z}
 \int\dd^2\bx_1\dd^2\bx_2\dd^2\bx_3\,
 \mathcal U_5e^{-i\bp\cdot\boldsymbol B_5-i\bq\cdot\bx_2}
 \frac1{x_{31}^2}
 \nonumber\\
 &\times\int_{\bk_1,\bell_{\rm in}}
 \frac{e^{i\bk_1\cdot\boldsymbol R_5
 -i\bell_{\rm in}\cdot\bx_{31}}k_{1\epsilon}}
 {(\bk_1^2+Q_\gamma^2)
 [\ell_{\rm in}^2+m_5^2(\bk_1)]}
 \left[
 z_1^2x_{31,\epsilon^*}\ell_{{\rm in},\epsilon}
 +(z_1-z)^2x_{31,\epsilon}\ell_{{\rm in},\epsilon^*}
 \right]
 \nonumber\\
 &+\frac{4iee_fg^2l^+\sqrt{z_2/z_1}}
 {z_1^2(2\pi)^2}
 \int_0^{z_1}\left(\frac{\nu^+}{zl^+}\right)^{\eta_Y}\dd z\,(z_1-z)^2
 \int\dd^2\bx_1\dd^2\bx_2\dd^2\bx_3\,
 \mathcal U_5e^{-i\bp\cdot\boldsymbol B_5-i\bq\cdot\bx_2}
 \frac{x_{31,\epsilon}}{x_{31}^2}
 \nonumber\\
 &\times\int_{\bk_1,\bell_{\rm in}}
 \frac{e^{i\bk_1\cdot\boldsymbol R_5
 -i\bell_{\rm in}\cdot\bx_{31}}}
 {\ell_{\rm in}^2+m_5^2(\bk_1)},
 \label{eq:g5-after-lout-minus}
\end{align}
where the remaining transverse mass is
\begin{equation}
 m_5^2(\bk_1)=
 \frac{z(z_1-z)}{z_1^2z_2}
 (\bk_1^2+Q_\gamma^2).
 \label{eq:g5-transverse-mass}
\end{equation}

The incoming relative momentum is integrated next.  The two required
transforms are
\begin{align}
 \int_{\bell_{\rm in}}
 \frac{\ell_{{\rm in},i}
 e^{-i\bell_{\rm in}\cdot\bx_{31}}}
 {\ell_{\rm in}^2+m_5^2}
 &=-\frac{im_5}{2\pi}
 \frac{x_{31,i}}{x_{31}}K_1(m_5x_{31}),\nonumber\\
 \int_{\bell_{\rm in}}
 \frac{e^{-i\bell_{\rm in}\cdot\bx_{31}}}
 {\ell_{\rm in}^2+m_5^2}
 &=\frac1{2\pi}K_0(m_5x_{31}).
 \label{eq:g5-incoming-fourier}
\end{align}
The amplitudes now contain one transverse loop integral:
\begin{align}
 i\mathcal A_{5,+}
 ={}&\frac{4ee_fg^2l^+\sqrt{z_1z_2}}
 {z_1(2\pi)^3}
 \int_0^{z_1}\left(\frac{\nu^+}{zl^+}\right)^{\eta_Y}\frac{\dd z}{z}
 [z_1^2+(z_1-z)^2]
 \int\dd^2\bx_1\dd^2\bx_2\dd^2\bx_3\,
 \mathcal U_5e^{-i\bp\cdot\boldsymbol B_5-i\bq\cdot\bx_2}
 \nonumber\\
 &\times\int_{\bk_1}
 e^{i\bk_1\cdot\boldsymbol R_5}
 \frac{k_{1\epsilon}m_5K_1(m_5x_{31})}
 {x_{31}(\bk_1^2+Q_\gamma^2)},
 \label{eq:g5-after-lin-plus}\\[3pt]
 i\mathcal A_{5,-}
 ={}&-\frac{4ee_fg^2l^+z_2\sqrt{z_1z_2}}
 {z_1^2(2\pi)^3}
 \int_0^{z_1}\left(\frac{\nu^+}{zl^+}\right)^{\eta_Y}\frac{\dd z}{z}
 [z_1^2+(z_1-z)^2]
 \int\dd^2\bx_1\dd^2\bx_2\dd^2\bx_3\,
 \mathcal U_5e^{-i\bp\cdot\boldsymbol B_5-i\bq\cdot\bx_2}
 \nonumber\\
 &\times\int_{\bk_1}
 e^{i\bk_1\cdot\boldsymbol R_5}
 \frac{k_{1\epsilon}m_5K_1(m_5x_{31})}
 {x_{31}(\bk_1^2+Q_\gamma^2)}
 \nonumber\\
 &+\frac{4iee_fg^2l^+\sqrt{z_2/z_1}}
 {z_1^2(2\pi)^3}
 \int_0^{z_1}\left(\frac{\nu^+}{zl^+}\right)^{\eta_Y}\dd z\,(z_1-z)^2
 \int\dd^2\bx_1\dd^2\bx_2\dd^2\bx_3\,
 \mathcal U_5e^{-i\bp\cdot\boldsymbol B_5-i\bq\cdot\bx_2}
 \frac{x_{31,\epsilon}}{x_{31}^2}
 \nonumber\\
 &\times\int_{\bk_1}
 e^{i\bk_1\cdot\boldsymbol R_5}K_0(m_5x_{31}).
 \label{eq:g5-after-lin-minus}
\end{align}

The last $\bk_1$ integral can be evaluated by introducing Schwinger
parameters for the two coupled denominators,
\begin{align}
 &\frac1{(\bk_1^2+Q_\gamma^2)
 [\ell_{\rm in}^2+
 z(z_1-z)(\bk_1^2+Q_\gamma^2)/(z_1^2z_2)]}
 \nonumber\\
 &\quad=
 \int_0^\infty\dd t\,\dd s\,
 \exp\left\{
 -t(\bk_1^2+Q_\gamma^2)
 -s\left[
 \ell_{\rm in}^2+
 \frac{z(z_1-z)}{z_1^2z_2}
 (\bk_1^2+Q_\gamma^2)
 \right]\right\}.
 \label{eq:g5-schwinger}
\end{align}
With $T=t+sz(z_1-z)/(z_1^2z_2)$, the remaining parameter integrals are
\begin{align}
 \int_0^{Tz_1^2z_2/[z(z_1-z)]}\frac{\dd s}{s^2}
 e^{-x_{31}^2/(4s)}
 &=
 \frac4{x_{31}^2}
 e^{-z(z_1-z)x_{31}^2/(4Tz_1^2z_2)},\nonumber\\
 \int_0^\infty\frac{\dd T}{T^2}
 e^{-Q_\gamma^2T-X_5^2/(4z_1z_2T)}
 &=
 \frac{4Qz_1z_2}{X_5}K_1(QX_5),
 \label{eq:g5-parameter-integrals}
\end{align}
where
\begin{equation}
 X_5^2=
 z_2(z_1-z)x_{12}^2+
 z(z_1-z)x_{13}^2+
 z_2z x_{23}^2 .
 \label{eq:g5-transverse-size}
\end{equation}
They imply the two final Fourier transforms
\begin{align}
 \int_{\bk_1}
 e^{i\bk_1\cdot\boldsymbol R_5}
 \frac{k_{1i}m_5K_1(m_5x_{31})}
 {x_{31}(\bk_1^2+Q_\gamma^2)}
 &=
 \frac{iQz_1z_2}{2\pi}
 \frac{K_1(QX_5)}{X_5}
 \frac{R_{5,i}}{x_{31}^2},\nonumber\\
 \int_{\bk_1}
 e^{i\bk_1\cdot\boldsymbol R_5}K_0(m_5x_{31})
 &=
 \frac{Qz_1z_2}{2\pi}\frac{K_1(QX_5)}{X_5}.
 \label{eq:g5-final-fourier}
\end{align}
Substitution in Eqs.~\eqref{eq:g5-after-lin-plus} and
\eqref{eq:g5-after-lin-minus} gives the coordinate-space amplitudes with no
remaining loop momentum:
\begin{align}
 i\mathcal A_{5,+}={}&
 \frac{4iee_fg^2l^+Qz_2\sqrt{z_1z_2}}{(2\pi)^4}
 \int_0^{z_1}\left(\frac{\nu^+}{zl^+}\right)^{\eta_Y}\frac{\dd z}{z}
 \int\dd^2\bx_1\dd^2\bx_2\dd^2\bx_3\,
 \mathcal U_5e^{-i\bp\cdot\boldsymbol B_5-i\bq\cdot\bx_2}
 \frac{K_1(QX_5)}{X_5x_{31}^2}
 \nonumber\\
 &\times[z_1^2+(z_1-z)^2]R_{5,\epsilon},
 \label{eq:g5-final-amplitude-plus}\\[3pt]
 i\mathcal A_{5,-}={}&
 -\frac{4iee_fg^2l^+Qz_2^2\sqrt{z_1z_2}}
 {z_1(2\pi)^4}
 \int_0^{z_1}\left(\frac{\nu^+}{zl^+}\right)^{\eta_Y}\frac{\dd z}{z}
 \int\dd^2\bx_1\dd^2\bx_2\dd^2\bx_3\,
 \mathcal U_5e^{-i\bp\cdot\boldsymbol B_5-i\bq\cdot\bx_2}
 \frac{K_1(QX_5)}{X_5x_{31}^2}
 \nonumber\\
 &\times\left\{
 [z_1^2+(z_1-z)^2]R_{5,\epsilon}
 -\frac{z(z_1-z)^2}{z_1z_2}x_{31,\epsilon}
 \right\}.
 \label{eq:g5-final-amplitude-minus}
\end{align}
The external phase fixes the parent-quark position to
$\boldsymbol B_5=\bx_1+(z/z_1)\bx_{31}$.  The unobserved-quark integral must
therefore not be replaced by $\bx_{1'}=\bx_1$ before the Born interference is
formed.

\paragraph{Contribution of $A_5$ to the measured cross section.}
The color matrix in the amplitude reduces at finite $N_c$ to
\begin{equation}
 V_3t^aV_3^\dagger V_1t^aV_2^\dagger
 =\frac{N_c}{2}S_{13}V_3V_2^\dagger
 -\frac{V_1V_2^\dagger}{2N_c}.
 \label{eq:g5-amplitude-color}
\end{equation}
Interference with the Born color matrix
$V_{2'}V_{1'}^\dagger-\boldsymbol1$ gives the normalized target operator
\begin{equation}
 \mathcal C_5=
 \left\langle
 \frac{N_c}{2}S_{13}(S_{322'1'}-S_{32})
 +C_F(1-S_{2'1'})
 -\frac{S_{122'1'}-S_{12}}{2N_c}
 \right\rangle_Y .
 \label{eq:g5-color}
\end{equation}
The external phase retained above combines with the Born conjugate amplitude to give
\begin{equation}
 \int\dd^2\bp\,
 e^{i\bp\cdot[\bx_{1'}-\bx_1-(z/z_1)\bx_{31}]}
 =(2\pi)^2
 \delta^{(2)}\left(
 \bx_{1'}-\bx_1-\frac{z}{z_1}\bx_{31}\right).
 \label{eq:g5-spectator-delta}
\end{equation}
The longitudinal spectator integral then sets $z_1=1-z_2$ and supplies the
factor $1/z_1$.  After the photon-helicity average and fermion-spin sum, the
one-sided graph kernel is
\begin{align}
 \mathcal V_5(z_2,\bq)={}&
 \frac{e^2e_f^2g^2Q^2N_c}{(2\pi)^8}
 \frac{z_2^{5/2}}{\sqrt{z_1}}
 \int\dd^2\bx_1\dd^2\bx_2\dd^2\bx_3\dd^2\bx_{2'}\,
 e^{i\bq\cdot\bx_{2'2}}
 \int_0^{z_1}\left(\frac{\nu^+}{zl^+}\right)^{\eta_Y}\frac{\dd z}{z}\,
 \left.\mathcal C_5\right|_{
 \bx_{1'}=\bx_1+(z/z_1)\bx_{31}}
 \nonumber\\
 &\times
 \frac{K_1(QX_5)K_1(Q_\gamma|\bx_{1'2'}|)}
 {X_5x_{31}^2|\bx_{1'2'}|}\,
 \bx_{1'2'}\cdot
 \Bigg\{
 (z_1^2+z_2^2)
 [z_1^2+(z_1-z)^2]
 \left[\bx_{12}+\frac{z}{z_1}\bx_{31}\right]
 \nonumber\\
 &\hspace{42mm}
 -\frac{z_2z(z_1-z)^2}{z_1}\bx_{31}
 \Bigg\},
 \qquad z_1=1-z_2 .
 \label{eq:g5-result}
\end{align}
The coordinate replacement following from
Eq.~\eqref{eq:g5-spectator-delta} applies to every occurrence of
$\bx_{1'}$, including the conjugate photon wave function and the target
operator.

The antiquark measurement is inserted only after the fixed-$(z_2,\bq)$ kernel
has been obtained:
\begin{equation}
 \Sigma^T_{\bar q,5}(\theta)=
 2\operatorname{Re}
 \int\frac{\dd z_2}{z_2}\int\dd^2\bq\,
 \mathcal M_{\bar q}(\theta;z_2,\bq)\,
 \mathcal V_5(z_2,\bq).
 \label{eq:g5-measured}
\end{equation}
Using the measurement support and Jacobian derived in
Eqs.~\eqref{eq:born-angular-root}--\eqref{eq:born-angular-jacobian}, the exact
angular integral becomes
\begin{equation}
 \boxed{\displaystyle
 \Sigma^T_{\bar q,5}(\theta)=
 2\operatorname{Re}
 \frac{x_BQ^2\sin\theta}{8\theta}
 \int_0^{\csc^2(\theta/2)}\dd v\,v^2\int_0^{2\pi}\dd\phi\,
 \mathcal V_5\left(
 v\sin^2\frac\theta2,
 \frac{Qv}{2}\sin\theta\,\boldsymbol e_\phi
 \right).}
 \label{eq:g5-measured-final}
\end{equation}
The $z$ integral inside $\mathcal V_5$ remains independent of the detector
integration; after the substitution in Eq.~\eqref{eq:g5-measured-final}, its
upper endpoint is $z_1=1-v\sin^2(\theta/2)$.

\paragraph{Graph $A_6$: antiquark emission across the shockwave.}
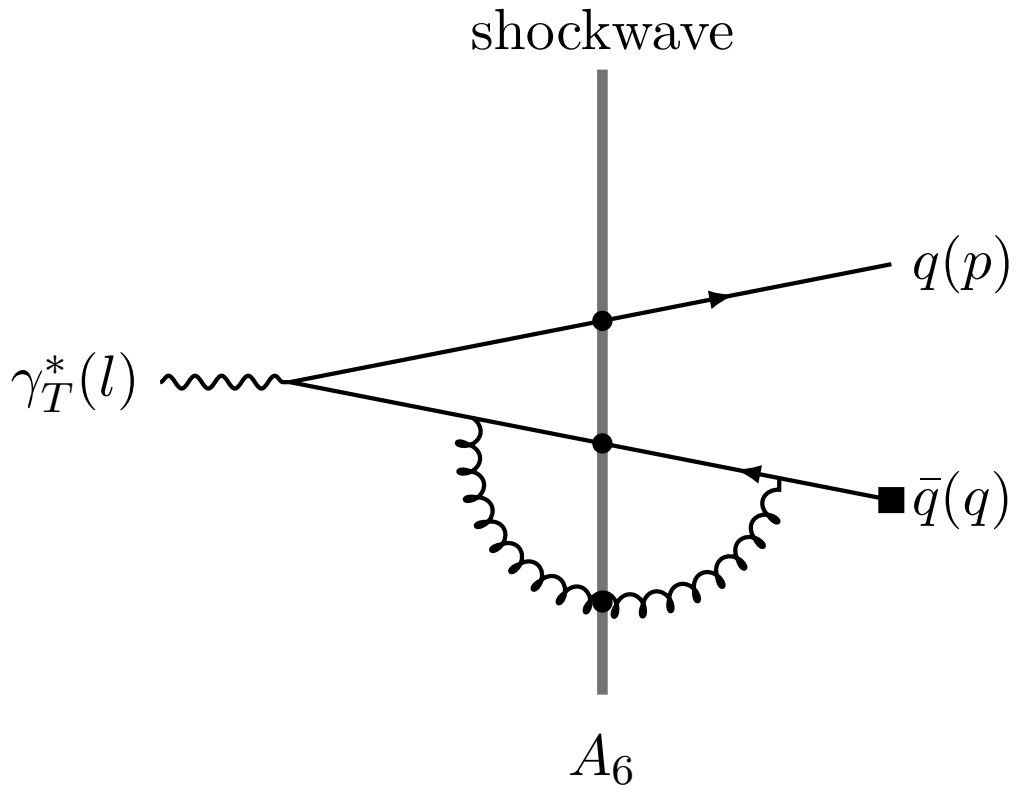
\virtualtopology{\virtualselfenergy{1.55}{4.15}{-1} \node at (2.65,-3.2) {$A_6$};}
The definitions that keep the amplitude compact are
\begin{align}
 a_6&=z_2-z,&
 \boldsymbol B_6&=\bx_2+\frac{z}{z_2}\bx_{32},&
 \boldsymbol R_6&=\bx_{21}+\frac{z}{z_2}\bx_{32},\nonumber\\
 X_6^2&=z_1a_6x_{12}^2+za_6x_{23}^2+z_1zx_{13}^2,&
 \mathcal U_6&=V_1t^aV_2^\dagger V_3t^aV_3^\dagger-C_F\boldsymbol1 .
 \label{eq:g6-coordinate-definitions}
\end{align}
The corresponding finite-$N_c$ target operator is
\begin{equation}
 \mathcal C_6=\left\langle
 \frac{N_c}{2}S_{32}(S_{2'3}-S_{13})
 +C_F(1-S_{2'1})-\frac{S_{2'2}-S_{12}}{2N_c}
 \right\rangle_Y .
 \label{eq:g6-color}
\end{equation}
After the three transverse integrations, the two nonzero spin amplitudes are
\begin{align}
 i\mathcal A_{6,+}={}&-
 \frac{4iee_fg^2l^+Qz_1^2\sqrt{z_1z_2}}{z_2(2\pi)^4}
 \nonumber\\[-2pt]
 &\times
 \int_0^{z_2}\left(\frac{\nu^+}{zl^+}\right)^{\eta_Y}\frac{\dd z}{z}
 \int\dd^2\bx_1\dd^2\bx_2\dd^2\bx_3\,
 \mathcal U_6e^{-i\bp\cdot\bx_1-i\bq\cdot\boldsymbol B_6}
 \frac{K_1(QX_6)}{X_6x_{32}^2}
 \nonumber\\
 &\times\left\{[z_2^2+a_6^2]R_{6,\epsilon}
 -\frac{za_6^2}{z_1z_2}x_{32,\epsilon}\right\},
 \label{eq:g6-final-amplitude-plus}\\[3pt]
 i\mathcal A_{6,-}={}&
 \frac{4iee_fg^2l^+Qz_1\sqrt{z_1z_2}}{(2\pi)^4}
 \nonumber\\[-2pt]
 &\times
 \int_0^{z_2}\left(\frac{\nu^+}{zl^+}\right)^{\eta_Y}\frac{\dd z}{z}
 \int\dd^2\bx_1\dd^2\bx_2\dd^2\bx_3\,
 \mathcal U_6e^{-i\bp\cdot\bx_1-i\bq\cdot\boldsymbol B_6}
 \frac{K_1(QX_6)}{X_6x_{32}^2}
 [z_2^2+a_6^2]R_{6,\epsilon}.
 \label{eq:g6-final-amplitude-minus}
\end{align}
Interference with the Born amplitude gives
\begin{align}
 \mathcal V_6(z_2,\bq)={}&-
 \frac{e^2e_f^2g^2Q^2N_c}{(2\pi)^8}z_1^{3/2}\sqrt{z_2}
 \int\dd^2\bx_1\dd^2\bx_2\dd^2\bx_3\dd^2\bx_{2'}
 \int_0^{z_2}\left(\frac{\nu^+}{zl^+}\right)^{\eta_Y}\frac{\dd z}{z}\,
 e^{i\bq\cdot[\bx_{2'2}-(z/z_2)\bx_{32}]}\mathcal C_6
 \nonumber\\
 &\times\frac{K_1(QX_6)K_1(Q_\gamma|\bx_{1 2'}|)}
 {X_6x_{32}^2|\bx_{1 2'}|}\,
 \bx_{1 2'}\cdot\left\{
 (z_1^2+z_2^2)[z_2^2+a_6^2]\boldsymbol R_6
 -\frac{z_1za_6^2}{z_2}\bx_{32}\right\}.
 \label{eq:g6-result}
\end{align}

\paragraph{Graph $A_7$: exchange across the shockwave, ending on the quark.}
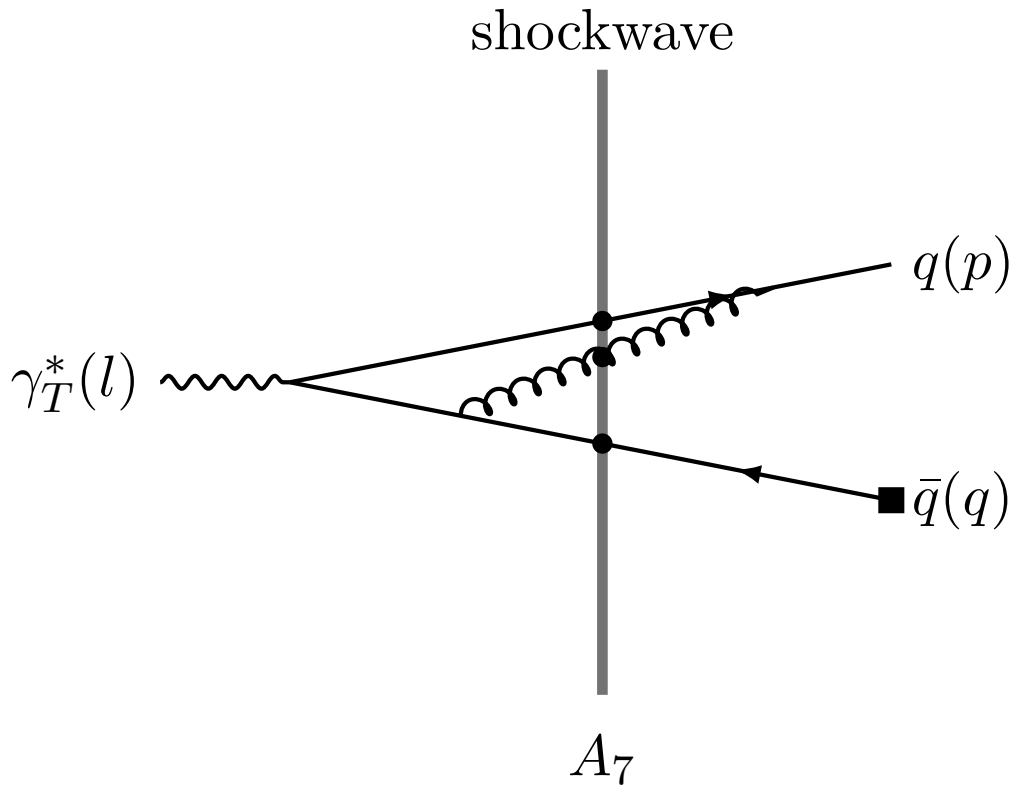
\virtualtopology{\draw[gluon] (1.45,{-1*1.45/5.1}) .. controls (1.85,-.10) and
(3.25,.45) .. (4.1,{1*4.1/5.1}); \gluonshockvertex{.213894}
\node at (2.65,-3.2) {$A_7$};}
For this graph we use
\begin{align}
 a_7&=z_1-z,& b_7&=z_2+z,&
 \boldsymbol B_7&=\bx_1+\frac{z}{z_1}\bx_{31},\nonumber\\
 \boldsymbol R_7&=\bx_{31}-\frac{z_2}{b_7}\bx_{32},&
 X_7^2&=z_2a_7x_{12}^2+za_7x_{13}^2+z_2zx_{23}^2,&
 \mathcal U_7&=V_3t^aV_3^\dagger V_1t^aV_2^\dagger-C_F\boldsymbol1 .
 \label{eq:g7-coordinate-definitions}
\end{align}
The color trace is
\begin{equation}
 \mathcal C_7=\left\langle
 \frac{N_c}{2}S_{13}(S_{322'1'}-S_{32})
 +C_F(1-S_{2'1'})-\frac{S_{122'1'}-S_{12}}{2N_c}
 \right\rangle_Y .
 \label{eq:g7-color}
\end{equation}
It is useful to collect the polarization tensors into
\begin{align}
 \mathcal T_{7,+}={}&R_{7,\epsilon}
 \left[z_1z_2x_{31,\epsilon}x_{32,\epsilon^*}
 +a_7b_7x_{31,\epsilon^*}x_{32,\epsilon}\right]
 -\frac{zz_1z_2}{2a_7b_7}x_{31,\epsilon}x_{32}^2,
 \nonumber\\
 \mathcal T_{7,-}={}&R_{7,\epsilon}
 \left[z_1z_2x_{31,\epsilon^*}x_{32,\epsilon}
 +a_7b_7x_{31,\epsilon}x_{32,\epsilon^*}\right].
 \label{eq:g7-spin-tensors}
\end{align}
The coordinate-space amplitudes are then
\begin{align}
 i\mathcal A_{7,+}={}&
 \frac{8iee_fg^2l^+Qa_7^2\sqrt{z_1z_2}}{z_1(2\pi)^4}
 \nonumber\\[-2pt]
 &\times
 \int_0^{z_1}\left(\frac{\nu^+}{zl^+}\right)^{\eta_Y}\frac{\dd z}{z}
 \int\dd^2\bx_1\dd^2\bx_2\dd^2\bx_3\,
 \mathcal U_7e^{-i\bp\cdot\boldsymbol B_7-i\bq\cdot\bx_2}
 \frac{K_1(QX_7)}{X_7x_{31}^2x_{32}^2}\mathcal T_{7,+},
 \label{eq:g7-final-amplitude-plus}\\
 i\mathcal A_{7,-}={}&-
 \frac{8iee_fg^2l^+Qa_7b_7\sqrt{z_1z_2}}{z_1(2\pi)^4}
 \nonumber\\[-2pt]
 &\times
 \int_0^{z_1}\left(\frac{\nu^+}{zl^+}\right)^{\eta_Y}\frac{\dd z}{z}
 \int\dd^2\bx_1\dd^2\bx_2\dd^2\bx_3\,
 \mathcal U_7e^{-i\bp\cdot\boldsymbol B_7-i\bq\cdot\bx_2}
 \frac{K_1(QX_7)}{X_7x_{31}^2x_{32}^2}\mathcal T_{7,-}.
 \label{eq:g7-final-amplitude-minus}
\end{align}
The spectator constraint is
$\bx_{1'}=\bx_1+(z/z_1)\bx_{31}$.  With this replacement made in every
primed factor, the one-sided kernel is
\begin{align}
 \mathcal V_7(z_2,\bq)={}&
 \frac{e^2e_f^2g^2Q^2N_c}{(2\pi)^8}\sqrt{z_1}z_2^{3/2}
 \int\dd^2\bx_1\dd^2\bx_2\dd^2\bx_3\dd^2\bx_{2'}
 e^{i\bq\cdot\bx_{2'2}}
 \int_0^{z_1}\left(\frac{\nu^+}{zl^+}\right)^{\eta_Y}\frac{\dd z\,a_7}{z}\,
 \left.\mathcal C_7\right|_{\bx_{1'}=\boldsymbol B_7}
 \nonumber\\
 &\times\frac{K_1(QX_7)K_1(Q_\gamma|\bx_{1'2'}|)}
 {X_7x_{31}^2|\bx_{1'2'}|}
 \Biggl\{\frac{4\Reh}{x_{32}^2}
 \Biggl[x_{1'2',\epsilon^*}R_{7,\epsilon}
 \Bigl(\frac{z_2a_7}{z_1}[z_1^2+b_7^2]
 x_{31,\epsilon}x_{32,\epsilon^*}
 \nonumber\\
 &\hspace{55mm}
 +b_7[z_2^2+a_7^2]x_{32,\epsilon}x_{31,\epsilon^*}\Bigr)\Biggr]
 -\frac{z_1z_2z}{b_7}\bx_{31}\cdot\bx_{1'2'}\Biggr\}.
 \label{eq:g7-result}
\end{align}

\paragraph{Graph $A_8$: exchange across the shockwave, ending on the antiquark.}
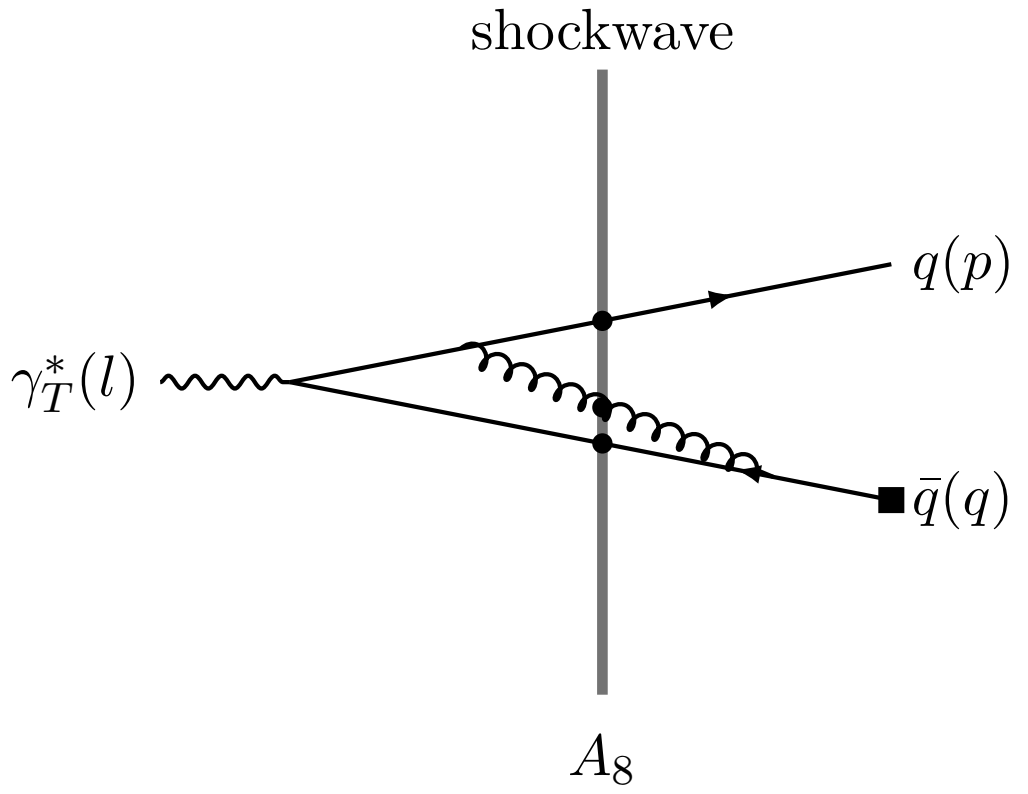
\virtualtopology{\draw[gluon] (1.45,{1*1.45/5.1}) .. controls (1.85,.10) and
(3.25,-.45) .. (4.1,{-1*4.1/5.1}); \gluonshockvertex{-.213894}
\node at (2.65,-3.2) {$A_8$};}
We define
\begin{align}
 a_8&=z_2-z,& b_8&=z_1+z,&
 \boldsymbol B_8&=\bx_1+\frac{z}{z_2}\bx_{32},\nonumber\\
 \boldsymbol R_8&=\bx_{32}-\frac{z_1}{b_8}\bx_{31},&
 X_8^2&=z_1a_8x_{21}^2+za_8x_{23}^2+z_1zx_{13}^2,&
 \mathcal U_8&=V_1t^aV_2^\dagger V_3t^aV_3^\dagger-C_F\boldsymbol1 .
 \label{eq:g8-coordinate-definitions}
\end{align}
Its target trace is identical to that of $A_6$,
\begin{equation}
 \mathcal C_8=\left\langle
 \frac{N_c}{2}S_{32}(S_{2'3}-S_{13})
 +C_F(1-S_{2'1})-\frac{S_{2'2}-S_{12}}{2N_c}
 \right\rangle_Y .
 \label{eq:g8-color}
\end{equation}
The two polarization tensors are
\begin{align}
 \mathcal T_{8,+}={}&R_{8,\epsilon}
 \left[a_8b_8x_{32,\epsilon}x_{31,\epsilon^*}
 +z_1z_2x_{32,\epsilon^*}x_{31,\epsilon}\right],
 \nonumber\\
 \mathcal T_{8,-}={}&R_{8,\epsilon}
 \left[a_8b_8x_{32,\epsilon^*}x_{31,\epsilon}
 +z_1z_2x_{32,\epsilon}x_{31,\epsilon^*}\right]
 -\frac{zz_1z_2}{2a_8b_8}x_{32,\epsilon}x_{31}^2.
 \label{eq:g8-spin-tensors}
\end{align}
The final amplitudes read
\begin{align}
 i\mathcal A_{8,+}={}&-
 \frac{8iee_fg^2l^+Qa_8b_8\sqrt{z_1z_2}}{z_2(2\pi)^4}
 \nonumber\\[-2pt]
 &\times
 \int_0^{z_2}\left(\frac{\nu^+}{zl^+}\right)^{\eta_Y}\frac{\dd z}{z}
 \int\dd^2\bx_1\dd^2\bx_2\dd^2\bx_3\,
 \mathcal U_8e^{-i\bp\cdot\bx_2-i\bq\cdot\boldsymbol B_8}
 \frac{K_1(QX_8)}{X_8x_{32}^2x_{31}^2}\mathcal T_{8,+},
 \label{eq:g8-final-amplitude-plus}\\
 i\mathcal A_{8,-}={}&
 \frac{8iee_fg^2l^+Qa_8^2\sqrt{z_1z_2}}{z_2(2\pi)^4}
 \nonumber\\[-2pt]
 &\times
 \int_0^{z_2}\left(\frac{\nu^+}{zl^+}\right)^{\eta_Y}\frac{\dd z}{z}
 \int\dd^2\bx_1\dd^2\bx_2\dd^2\bx_3\,
 \mathcal U_8e^{-i\bp\cdot\bx_2-i\bq\cdot\boldsymbol B_8}
 \frac{K_1(QX_8)}{X_8x_{32}^2x_{31}^2}\mathcal T_{8,-}.
 \label{eq:g8-final-amplitude-minus}
\end{align}
The unobserved-quark integral sets $\bx_{1'}=\bx_1$.  The shifted antiquark
phase remains in the fixed-momentum kernel,
\begin{align}
 \mathcal V_8(z_2,\bq)={}&-
 \frac{e^2e_f^2g^2Q^2N_c}{(2\pi)^8}\sqrt{z_1}z_2^{3/2}
 \int\dd^2\bx_1\dd^2\bx_2\dd^2\bx_3\dd^2\bx_{2'}
 \int_0^{z_2}\left(\frac{\nu^+}{zl^+}\right)^{\eta_Y}\frac{\dd z\,a_8}{z}\,
 e^{i\bq\cdot[\bx_{2'2}-(z/z_2)\bx_{32}]}\mathcal C_8
 \nonumber\\
 &\times\frac{K_1(QX_8)K_1(Q_\gamma|\bx_{1 2'}|)}
 {X_8x_{32}^2|\bx_{1 2'}|}
 \Biggl\{\frac{4\Reh}{x_{31}^2}
 \Biggl[x_{1 2',\epsilon^*}R_{8,\epsilon}
 \Bigl(\frac{z_1a_8}{z_2}[z_2^2+b_8^2]
 x_{32,\epsilon}x_{31,\epsilon^*}
 \nonumber\\
 &\hspace{55mm}
 +b_8[z_1^2+a_8^2]x_{31,\epsilon}x_{32,\epsilon^*}\Bigr)\Biggr]
 -\frac{z_1z_2z}{b_8}\bx_{32}\cdot\bx_{1 2'}\Biggr\}.
 \label{eq:g8-result}
\end{align}

\paragraph{Graphs $A_9$ and $A_{10}$: external self-energies after the shockwave.}
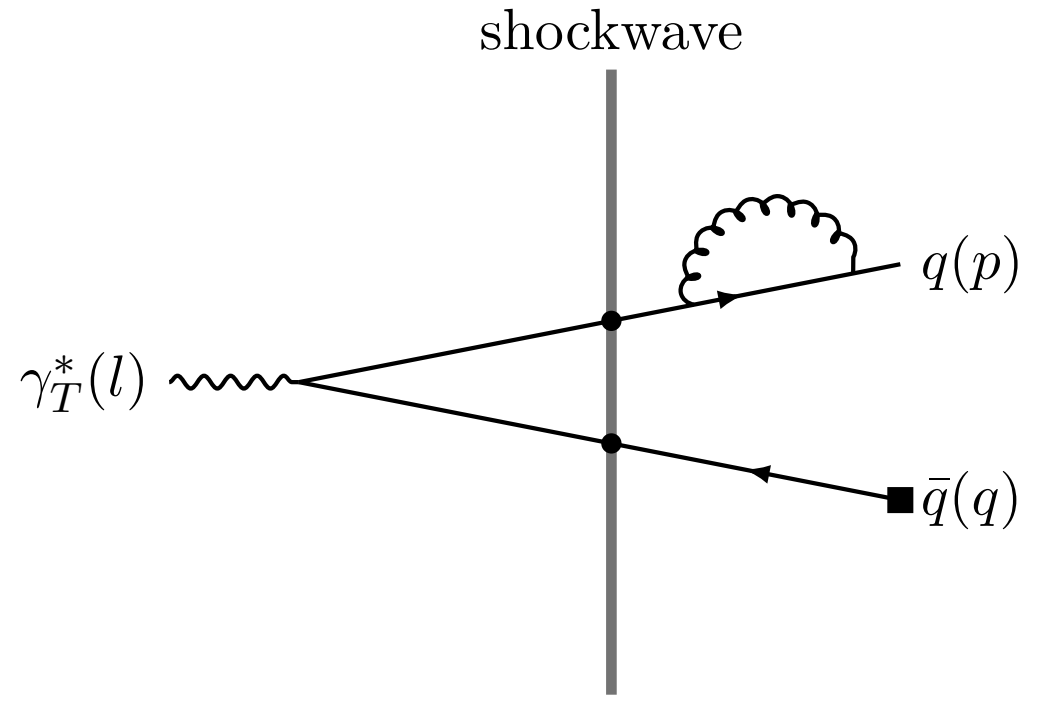
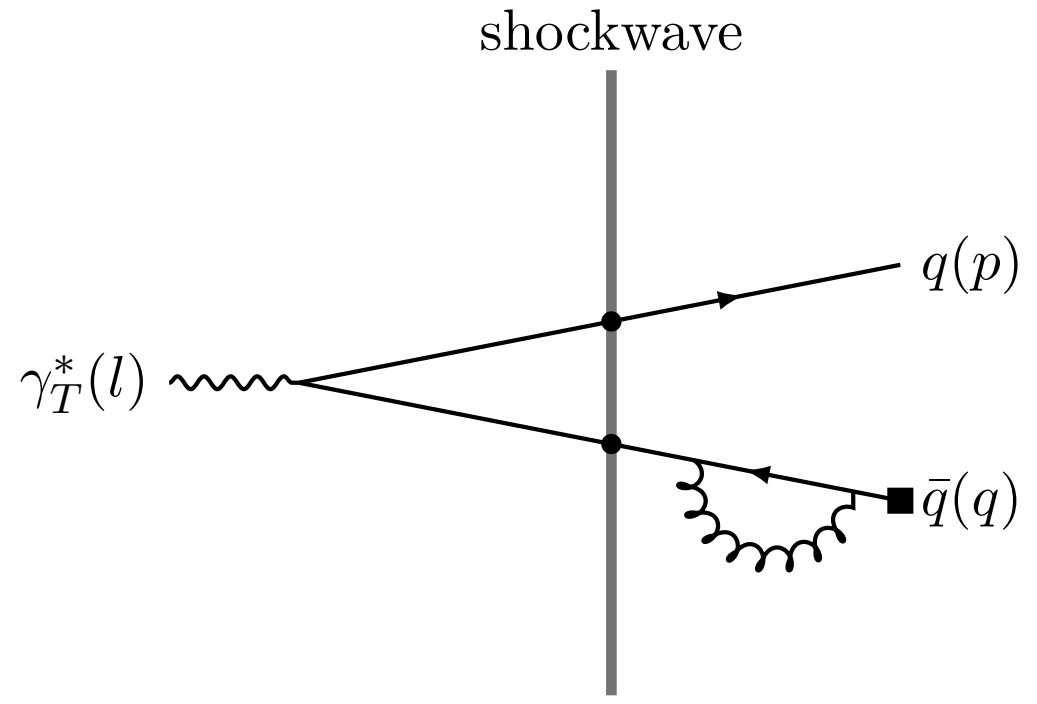
\virtualtopology{\virtualselfenergy{3.35}{4.7}{1} \node at (2.65,-3.2) {$A_9$};}
\virtualtopology{\virtualselfenergy{3.35}{4.7}{-1} \node at (2.65,-3.2) {$A_{10}$};}
Both graphs carry the Born color matrix and the Born target operator,
\begin{align}
 \mathcal U_0&=V_1V_2^\dagger-\boldsymbol1,
 &\mathcal C_0&=\left\langle S_{2'2}-S_{12}-S_{2'1}+1\right\rangle_Y.
 \label{eq:born-color-matrices}
\end{align}
With
\begin{align}
 \bell_9&=\bk-\frac{z}{z_1}\bp,&
 \mathcal P_{qg}&=\frac{g^2C_F}{2\pi}
 \int_0^{z_1}\left(\frac{\nu^+}{zl^+}\right)^{\eta_Y}\frac{\dd z}{z}
 \frac{z_1^2+(z_1-z)^2}{z_1^2}
 \mu^{2\epsilon}\int\frac{\dd^{2-2\epsilon}\bell_9}{(2\pi)^{2-2\epsilon}}
 \frac1{\ell_9^2},\nonumber\\
 \bell_{10}&=\bk-\frac{z}{z_2}\bq,&
 \mathcal P_{\bar qg}&=\frac{g^2C_F}{2\pi}
 \int_0^{z_2}\left(\frac{\nu^+}{zl^+}\right)^{\eta_Y}\frac{\dd z}{z}
 \frac{z_2^2+(z_2-z)^2}{z_2^2}
 \mu^{2\epsilon}\int\frac{\dd^{2-2\epsilon}\bell_{10}}{(2\pi)^{2-2\epsilon}}
 \frac1{\ell_{10}^2},
 \label{eq:g9-g10-definitions}
\end{align}
LSZ normalization gives the amplitudes
\begin{align}
 i\mathcal A_{9,+}&=-\frac12\mathcal P_{qg}\,i\mathcal A_{0,+},&
 i\mathcal A_{9,-}&=-\frac12\mathcal P_{qg}\,i\mathcal A_{0,-},
 \label{eq:g9-final-amplitudes}\\
 i\mathcal A_{10,+}&=-\frac12\mathcal P_{\bar qg}\,i\mathcal A_{0,+},&
 i\mathcal A_{10,-}&=-\frac12\mathcal P_{\bar qg}\,i\mathcal A_{0,-}.
 \label{eq:g10-final-amplitudes}
\end{align}
Consequently,
\begin{align}
 \mathcal V_9={}&-\frac{g^2C_F}{4\pi}\mathcal V_\Born
 \int_0^{z_1}\left(\frac{\nu^+}{zl^+}\right)^{\eta_Y}\frac{\dd z}{z}
 \frac{z_1^2+(z_1-z)^2}{z_1^2}
 \mu^{2\epsilon}\int\frac{\dd^{2-2\epsilon}\bell}{(2\pi)^{2-2\epsilon}}
 \frac1{\ell^2},
 \label{eq:g9-result}\\
 \mathcal V_{10}={}&-\frac{g^2C_F}{4\pi}\mathcal V_\Born
 \int_0^{z_2}\left(\frac{\nu^+}{zl^+}\right)^{\eta_Y}\frac{\dd z}{z}
 \frac{z_2^2+(z_2-z)^2}{z_2^2}
 \mu^{2\epsilon}\int\frac{\dd^{2-2\epsilon}\bell}{(2\pi)^{2-2\epsilon}}
 \frac1{\ell^2}.
 \label{eq:g10-result}
\end{align}

\paragraph{Graphs $A_{11}$ and $A_{12}$: self-energies before the shockwave.}
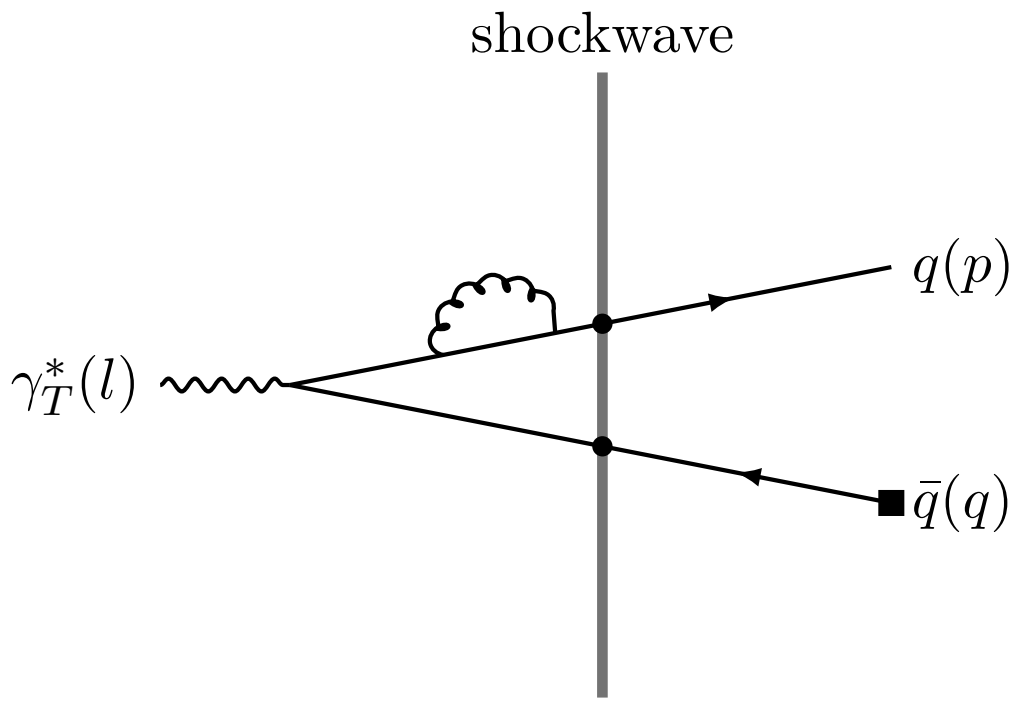
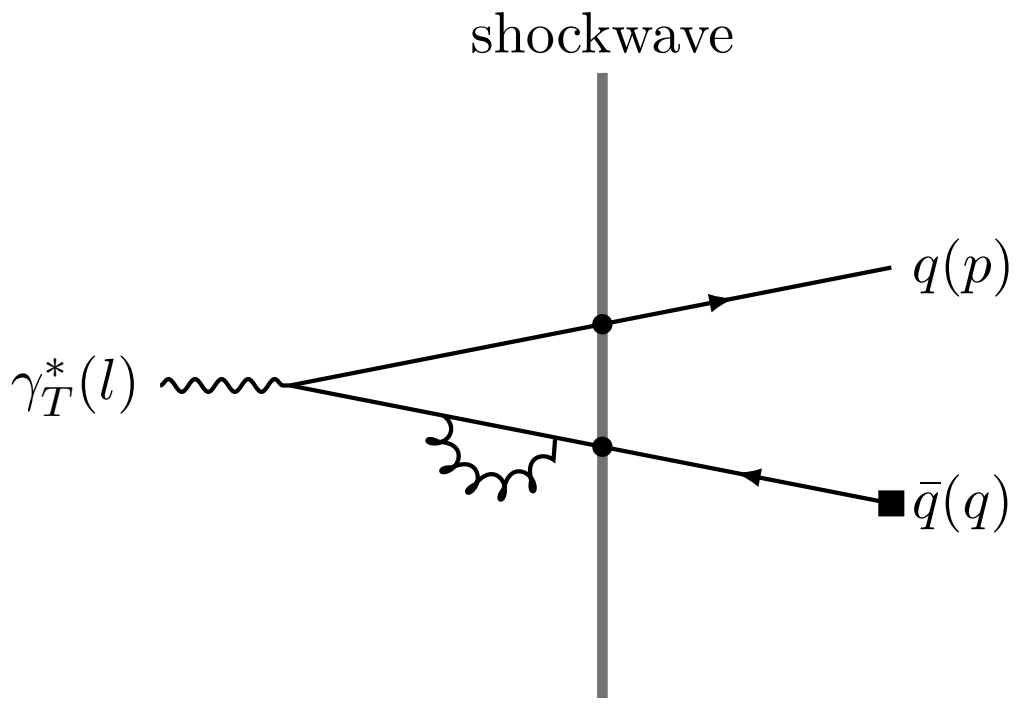
\virtualtopology{\virtualselfenergy{1.3}{2.25}{1} \node at (2.65,-3.2) {$A_{11}$};}
\virtualtopology{\virtualselfenergy{1.3}{2.25}{-1} \node at (2.65,-3.2) {$A_{12}$};}
These graphs again carry $\mathcal U_0$ and $\mathcal C_0$.  Define the
regulated vector transform
\begin{align}
 \mathfrak I_{{\rm SE},\epsilon}(\boldsymbol r;a)
 \equiv{}&\mu^{2\epsilon}
 \int_{\bk}\int\frac{\dd^{2-2\epsilon}\bell}{(2\pi)^{2-2\epsilon}}
 \frac{k_\epsilon e^{i\bk\cdot\boldsymbol r}}
 {(\bk^2+Q_\gamma^2)[\ell^2+a(\bk^2+Q_\gamma^2)]}
 \nonumber\\
 ={}&\frac{iQ_\gamma\mu^{2\epsilon}\Gamma(\epsilon)}
 {2\pi(4\pi)^{1-\epsilon}\Gamma(1+\epsilon)}
 a^{-\epsilon}\frac{r_\epsilon}{r}
 \left(\frac{r}{2Q_\gamma}\right)^\epsilon
 K_{1-\epsilon}(Q_\gamma r).
 \label{eq:self-energy-vector-transform}
\end{align}
For $A_{11}$ set
$a_{11}=z(z_1-z)/(z_1^2z_2)$ and $\boldsymbol r_{11}=\bx_{12}$.
The two amplitudes are
\begin{align}
 i\mathcal A_{11,+}={}&-
 \frac{4ee_fg^2C_Fl^+\sqrt{z_1z_2}}{z_1}
 \nonumber\\[-2pt]
 &\times
 \int_0^{z_1}\left(\frac{\nu^+}{zl^+}\right)^{\eta_Y}\frac{\dd z}{2\pi z}
 [z_1^2+(z_1-z)^2]
 \int\dd^2\bx_1\dd^2\bx_2\,
 \mathcal U_0e^{-i\bp\cdot\bx_1-i\bq\cdot\bx_2}
 \mathfrak I_{{\rm SE},\epsilon}(\bx_{12};a_{11}),
 \label{eq:g11-final-amplitude-plus}\\
 i\mathcal A_{11,-}={}&
 \frac{4ee_fg^2C_Fl^+z_2\sqrt{z_1z_2}}{z_1^2}
 \nonumber\\[-2pt]
 &\times
 \int_0^{z_1}\left(\frac{\nu^+}{zl^+}\right)^{\eta_Y}\frac{\dd z}{2\pi z}
 [z_1^2+(z_1-z)^2]
 \int\dd^2\bx_1\dd^2\bx_2\,
 \mathcal U_0e^{-i\bp\cdot\bx_1-i\bq\cdot\bx_2}
 \mathfrak I_{{\rm SE},\epsilon}(\bx_{12};a_{11}).
 \label{eq:g11-final-amplitude-minus}
\end{align}
Their interference gives
\begin{align}
 \mathcal V_{11}={}&-
 \frac{e^2e_f^2g^2Q^2N_cC_F}{(2\pi)^6}
 \frac{z_2^2}{z_1}(z_1^2+z_2^2)
 \frac{\mu^{2\epsilon}\Gamma(\epsilon)}
 {(4\pi)^{1-\epsilon}\Gamma(1+\epsilon)}
 \int\dd^2\bx_1\dd^2\bx_2\dd^2\bx_{2'}
 e^{i\bq\cdot\bx_{2'2}}\mathcal C_0
 \nonumber\\
 &\times\frac{\bx_{12}\cdot\bx_{1 2'}}{|\bx_{12}|\,|\bx_{1 2'}|}
 K_1(Q_\gamma|\bx_{1 2'}|)
 \left(\frac{|\bx_{12}|}{2Q_\gamma}\right)^\epsilon
 K_{1-\epsilon}(Q_\gamma|\bx_{12}|)
 \nonumber\\[-2pt]
 &\times
 \int_0^{z_1}\left(\frac{\nu^+}{zl^+}\right)^{\eta_Y}\frac{\dd z}{z}
 [z_1^2+(z_1-z)^2]
 \left[\frac{z(z_1-z)}{z_1^2z_2}\right]^{-\epsilon}.
 \label{eq:g11-result}
\end{align}
For $A_{12}$ set
$a_{12}=z(z_2-z)/(z_2^2z_1)$ and $\boldsymbol r_{12}=\bx_{21}$.
The amplitudes are
\begin{align}
 i\mathcal A_{12,+}={}&
 \frac{4ee_fg^2C_Fl^+z_1\sqrt{z_1z_2}}{z_2^2}
 \nonumber\\[-2pt]
 &\times
 \int_0^{z_2}\left(\frac{\nu^+}{zl^+}\right)^{\eta_Y}\frac{\dd z}{2\pi z}
 [z_2^2+(z_2-z)^2]
 \int\dd^2\bx_1\dd^2\bx_2\,
 \mathcal U_0e^{-i\bp\cdot\bx_1-i\bq\cdot\bx_2}
 \mathfrak I_{{\rm SE},\epsilon}(\bx_{21};a_{12}),
 \label{eq:g12-final-amplitude-plus}\\
 i\mathcal A_{12,-}={}&-
 \frac{4ee_fg^2C_Fl^+\sqrt{z_1z_2}}{z_2}
 \nonumber\\[-2pt]
 &\times
 \int_0^{z_2}\left(\frac{\nu^+}{zl^+}\right)^{\eta_Y}\frac{\dd z}{2\pi z}
 [z_2^2+(z_2-z)^2]
 \int\dd^2\bx_1\dd^2\bx_2\,
 \mathcal U_0e^{-i\bp\cdot\bx_1-i\bq\cdot\bx_2}
 \mathfrak I_{{\rm SE},\epsilon}(\bx_{21};a_{12}).
 \label{eq:g12-final-amplitude-minus}
\end{align}
The one-sided kernel is
\begin{align}
 \mathcal V_{12}={}&-
 \frac{e^2e_f^2g^2Q^2N_cC_F}{(2\pi)^6}
 z_1(z_1^2+z_2^2)
 \frac{\mu^{2\epsilon}\Gamma(\epsilon)}
 {(4\pi)^{1-\epsilon}\Gamma(1+\epsilon)}
 \int\dd^2\bx_1\dd^2\bx_2\dd^2\bx_{2'}
 e^{i\bq\cdot\bx_{2'2}}\mathcal C_0
 \nonumber\\
 &\times\frac{\bx_{12}\cdot\bx_{1 2'}}{|\bx_{12}|\,|\bx_{1 2'}|}
 K_1(Q_\gamma|\bx_{1 2'}|)
 \left(\frac{|\bx_{12}|}{2Q_\gamma}\right)^\epsilon
 K_{1-\epsilon}(Q_\gamma|\bx_{12}|)
 \nonumber\\[-2pt]
 &\times
 \int_0^{z_2}\left(\frac{\nu^+}{zl^+}\right)^{\eta_Y}\frac{\dd z}{z}
 [z_2^2+(z_2-z)^2]
 \left[\frac{z(z_2-z)}{z_2^2z_1}\right]^{-\epsilon}.
 \label{eq:g12-result}
\end{align}

\paragraph{Graph $A_{13}$: quark--antiquark exchange after the shockwave.}
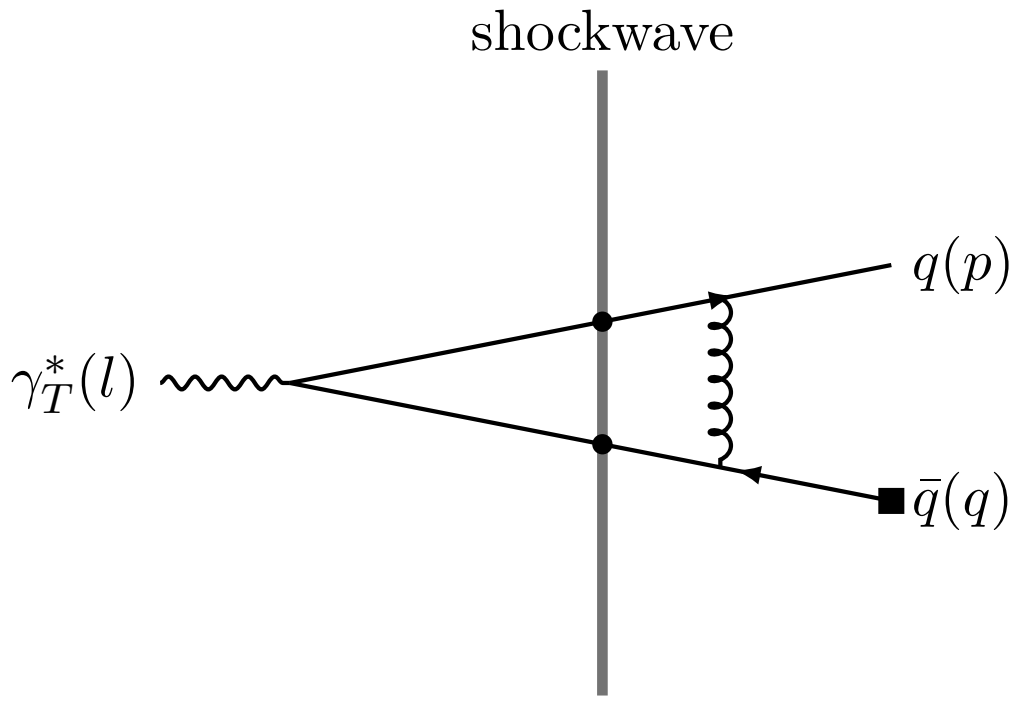
\virtualtopology{\draw[gluon] (3.65,{3.65/5.1})--(3.65,{-3.65/5.1});
\node at (2.65,-3.2) {$A_{13}$};}
The amplitude-side color matrix and the target operator are
\begin{align}
 \mathcal U_{13}&=t^aV_1V_2^\dagger t^a-C_F\boldsymbol1,
 &t^aV_1V_2^\dagger t^a
 &=\frac{N_c}{2}S_{12}\boldsymbol1-\frac{V_1V_2^\dagger}{2N_c},
 \nonumber\\
 \mathcal C_{13}&=\left\langle
 \frac{N_c}{2}S_{12}S_{2'1'}+C_F(1-S_{12}-S_{2'1'})
 -\frac{S_{122'1'}}{2N_c}\right\rangle_Y .
 \label{eq:g13-color}
\end{align}
Both signs of the exchanged gluon plus momentum occur.  We keep the two
longitudinal sectors separate.

For sector $13(1)$, with $0<z<z_2$, define
\begin{align}
 \alpha_{13,1}&=z_1+z,& \beta_{13,1}&=z_2-z,&
 \Lambda_{13,1}&=|z_2\bk-z\bq|^2,\nonumber\\
 \Delta_{13,1}&=
 \frac{|z_1\bk-z\bp|^2}{z_1\alpha_{13,1}}
 -\frac{|z_2\bk-z\bq|^2}{z_2\beta_{13,1}} .
 \label{eq:g131-definitions}
\end{align}
The noncontact helicity polynomial is
\begin{align}
 n_{13,1}^{\epsilon\epsilon^*}={}&
 z_1z\left(\frac{p_\epsilon}{z_1}-\frac{q_\epsilon}{z_2}\right)
 \left(\frac{p_{\epsilon^*}}{z_1}-\frac{k_{\epsilon^*}}z\right)
 +z^2\left(\frac{k_\epsilon}z-\frac{q_\epsilon}{z_2}\right)
 \left(\frac{p_{\epsilon^*}}{z_1}-\frac{k_{\epsilon^*}}z\right)
 \nonumber\\
 &-z_2z\left(\frac{k_\epsilon}z-\frac{q_\epsilon}{z_2}\right)
 \left(\frac{p_{\epsilon^*}}{z_1}-\frac{q_{\epsilon^*}}{z_2}\right)
 -\frac12\left(\frac{z_1}{z_2}\bq^2+
 \frac{z_2}{z_1}\bp^2-2\bp\cdot\bq\right),
 \label{eq:g131-spin-polynomial}
\end{align}
and $n_{13,1}^{\epsilon^*\epsilon}$ is obtained by interchanging the two
transverse polarizations.  We include the contact term in
\begin{equation}
 \widehat n_{13,1}^{\lambda\bar\lambda}
 =n_{13,1}^{\lambda\bar\lambda}
 +\frac{\alpha_{13,1}\beta_{13,1}}{2z^2}\Delta_{13,1}.
 \label{eq:g131-complete-spin-polynomial}
\end{equation}
Here $(\lambda,\bar\lambda)$ denotes either
$(\epsilon,\epsilon^*)$ or $(\epsilon^*,\epsilon)$.
The common coordinate and exchange-momentum kernel is
\begin{align}
 \mathcal J_{13,1}[\mathcal F]\equiv{}&
 \int\dd^2\bx_1\dd^2\bx_2\,
 \mathcal U_{13}e^{-i\bp\cdot\bx_1-i\bq\cdot\bx_2}
 \frac{x_{12,\epsilon}}{|\bx_{12}|}
 \nonumber\\
 &\times K_1\!\left(
 Q\sqrt{\alpha_{13,1}\beta_{13,1}}|\bx_{12}|\right)
 \nonumber\\
 &\times
 \int_{\bk}e^{-i\bk\cdot\bx_{12}}
 \frac{\mathcal F}{\Lambda_{13,1}(\Delta_{13,1}+i0)}.
 \label{eq:g131-amplitude-kernel}
\end{align}
After the photon transverse transform, the amplitudes are
\begin{align}
 i\mathcal A_{13(1),+}={}&
 \frac{8iee_fg^2l^+Qz_2\sqrt{z_1z_2}}{(2\pi)^2}
 \nonumber\\
 &\times
 \int_0^{z_2}\left(\frac{\nu^+}{zl^+}\right)^{\eta_Y}\dd z\,z
 \sqrt{\alpha_{13,1}\beta_{13,1}}
 \nonumber\\
 &\times
 \mathcal J_{13,1}
 \!\left[\widehat n_{13,1}^{\epsilon\epsilon^*}\right],
 \label{eq:g131-final-amplitude-plus}\\
 i\mathcal A_{13(1),-}={}&-
 \frac{8iee_fg^2l^+Qz_2\sqrt{z_1z_2}}{(2\pi)^2}
 \nonumber\\
 &\times
 \int_0^{z_2}\left(\frac{\nu^+}{zl^+}\right)^{\eta_Y}\dd z\,z
 \frac{\beta_{13,1}}{\alpha_{13,1}}
 \sqrt{\alpha_{13,1}\beta_{13,1}}\,
 \nonumber\\
 &\times
 \mathcal J_{13,1}
 \!\left[\widehat n_{13,1}^{\epsilon^*\epsilon}\right].
 \label{eq:g131-final-amplitude-minus}
\end{align}

For sector $13(2)$, with $0<z<z_1$, put
\begin{align}
 \alpha_{13,2}&=z_1-z,& \beta_{13,2}&=z_2+z,&
 \Lambda_{13,2}&=|z_1\bk-z\bp|^2,\nonumber\\
 \Delta_{13,2}&=
 \frac{|z_1\bk-z\bp|^2}{z_1\alpha_{13,2}}
 -\frac{|z_2\bk-z\bq|^2}{z_2\beta_{13,2}} .
 \label{eq:g132-definitions}
\end{align}
Here
\begin{align}
 n_{13,2}^{\epsilon^*\epsilon}={}&
 z_1z\left(\frac{p_{\epsilon^*}}{z_1}-\frac{q_{\epsilon^*}}{z_2}\right)
 \left(\frac{p_\epsilon}{z_1}-\frac{k_\epsilon}z\right)
 -z^2\left(\frac{k_{\epsilon^*}}z-\frac{q_{\epsilon^*}}{z_2}\right)
 \left(\frac{p_\epsilon}{z_1}-\frac{k_\epsilon}z\right)
 \nonumber\\
 &-z_2z\left(\frac{k_{\epsilon^*}}z-\frac{q_{\epsilon^*}}{z_2}\right)
 \left(\frac{p_\epsilon}{z_1}-\frac{q_\epsilon}{z_2}\right)
 +\frac12\left(\frac{z_1}{z_2}\bq^2+
 \frac{z_2}{z_1}\bp^2-2\bp\cdot\bq\right),
 \label{eq:g132-spin-polynomial}
\end{align}
with $n_{13,2}^{\epsilon\epsilon^*}$ again obtained by polarization exchange.
Define
\begin{equation}
 \widehat n_{13,2}^{\lambda\bar\lambda}
 =n_{13,2}^{\lambda\bar\lambda}
 +\frac{\alpha_{13,2}\beta_{13,2}}{2z^2}\Delta_{13,2}.
 \label{eq:g132-complete-spin-polynomial}
\end{equation}
Again $(\lambda,\bar\lambda)$ denotes either ordering of
$(\epsilon,\epsilon^*)$.
The second common kernel is
\begin{align}
 \mathcal J_{13,2}[\mathcal F]\equiv{}&
 \int\dd^2\bx_1\dd^2\bx_2\,
 \mathcal U_{13}e^{-i\bp\cdot\bx_1-i\bq\cdot\bx_2}
 \frac{x_{12,\epsilon}}{|\bx_{12}|}
 \nonumber\\
 &\times K_1\!\left(
 Q\sqrt{\alpha_{13,2}\beta_{13,2}}|\bx_{12}|\right)
 \nonumber\\
 &\times
 \int_{\bk}e^{+i\bk\cdot\bx_{12}}
 \frac{\mathcal F}{\Lambda_{13,2}(\Delta_{13,2}-i0)}.
 \label{eq:g132-amplitude-kernel}
\end{align}
The second-sector amplitudes become
\begin{align}
 i\mathcal A_{13(2),+}={}&
 \frac{8iee_fg^2l^+Qz_1\sqrt{z_1z_2}}{(2\pi)^2}
 \nonumber\\
 &\times
 \int_0^{z_1}\left(\frac{\nu^+}{zl^+}\right)^{\eta_Y}\dd z\,z
 \frac{\alpha_{13,2}}{\beta_{13,2}}
 \sqrt{\alpha_{13,2}\beta_{13,2}}\,
 \nonumber\\
 &\times
 \mathcal J_{13,2}
 \!\left[\widehat n_{13,2}^{\epsilon\epsilon^*}\right],
 \label{eq:g132-final-amplitude-plus}\\
 i\mathcal A_{13(2),-}={}&-
 \frac{8iee_fg^2l^+Qz_1\sqrt{z_1z_2}}{(2\pi)^2}
 \nonumber\\
 &\times
 \int_0^{z_1}\left(\frac{\nu^+}{zl^+}\right)^{\eta_Y}\dd z\,z
 \sqrt{\alpha_{13,2}\beta_{13,2}}
 \nonumber\\
 &\times
 \mathcal J_{13,2}
 \!\left[\widehat n_{13,2}^{\epsilon^*\epsilon}\right].
 \label{eq:g132-final-amplitude-minus}
\end{align}

To state the cross-section kernels, define
\begin{align}
 \boldsymbol R&=\bx_{1'}-\bx_1,&
 m_{13,1}^2&=-\frac{z_1\alpha_{13,1}}{z^2}
 \left[\frac{|z_2\bk-z\bq|^2}{z_2\beta_{13,1}}-i0\right],
 \nonumber\\
 m_{13,2}^2&=-\frac{z_1\alpha_{13,2}}{z^2}
 \left[\frac{|z_2\bk-z\bq|^2}{z_2\beta_{13,2}}+i0\right],&
 G_0^{(\epsilon)}(R)&=
 \mu^{2\epsilon}\int\frac{\dd^{2-2\epsilon}\bell}{(2\pi)^{2-2\epsilon}}
 \frac{e^{i\bell\cdot\boldsymbol R}}{\ell^2}.
 \label{eq:g13-spectator-definitions}
\end{align}
Momentum factors $p_i$ inside $n_{13,s}$ below mean
$-i\partial/\partial R_i$ and act only on the function immediately to their
right.  The first sector gives
\begin{align}
 \mathcal V_{13(1)}={}&
 \frac{e^2e_f^2g^2Q^2N_c}{(2\pi)^6}\sqrt{z_1}z_2^{3/2}
 \int\dd^2\bx_1\dd^2\bx_2\dd^2\bx_{1'}\dd^2\bx_{2'}
 e^{i\bq\cdot\bx_{2'2}}\mathcal C_{13}
 \frac{K_1(Q_\gamma|\bx_{1'2'}|)}{|\bx_{1'2'}|}
 \nonumber\\
 &\times\int_0^{z_2}\left(\frac{\nu^+}{zl^+}\right)^{\eta_Y}\dd z\,
 \sqrt{\alpha_{13,1}\beta_{13,1}}
 \frac{K_1(Q\sqrt{\alpha_{13,1}\beta_{13,1}}|\bx_{12}|)}{|\bx_{12}|}
 \int_{\bk}e^{-i\bk\cdot\bx_{12}}
 \nonumber\\
 &\times4\Reh\Biggl[x_{12,\epsilon}x_{1'2',\epsilon^*}
 \Biggl\{
 \frac{z_2\beta_{13,1}
 [z_1\alpha_{13,1}+z_2\beta_{13,1}]}{2z}
 \frac{\delta^{(2)}(\boldsymbol R)}{|z_2\bk-z\bq|^2}
 \nonumber\\
 &\quad+\left[z_1z_2z\,n_{13,1}^{\epsilon\epsilon^*}(-i\nabla_R)
 +\frac{z_2^2z\beta_{13,1}}{\alpha_{13,1}}
 n_{13,1}^{\epsilon^*\epsilon}(-i\nabla_R)\right]
 \frac{z_1\alpha_{13,1}e^{i(z_1/z)\bk\cdot\boldsymbol R}}
 {z^2|z_2\bk-z\bq|^2}
 \frac{K_0(m_{13,1}R)}{2\pi}
 \Biggr\}\Biggr].
 \label{eq:g131-result}
\end{align}
The second sector is
\begin{align}
 \mathcal V_{13(2)}={}&
 \frac{e^2e_f^2g^2Q^2N_c}{(2\pi)^6}\sqrt{z_1}z_2^{3/2}
 \int\dd^2\bx_1\dd^2\bx_2\dd^2\bx_{1'}\dd^2\bx_{2'}
 e^{i\bq\cdot\bx_{2'2}}\mathcal C_{13}
 \frac{K_1(Q_\gamma|\bx_{1'2'}|)}{|\bx_{1'2'}|}
 \nonumber\\
 &\times\int_0^{z_1}\left(\frac{\nu^+}{zl^+}\right)^{\eta_Y}\dd z\,
 \sqrt{\alpha_{13,2}\beta_{13,2}}
 \frac{K_1(Q\sqrt{\alpha_{13,2}\beta_{13,2}}|\bx_{12}|)}{|\bx_{12}|}
 \int_{\bk}e^{+i\bk\cdot\bx_{12}}
 \nonumber\\
 &\times4\Reh\Biggl[x_{12,\epsilon}x_{1'2',\epsilon^*}
 \Biggl\{
 \frac{z_1\alpha_{13,2}
 [z_1\alpha_{13,2}+z_2\beta_{13,2}]}{2z}
 \frac{e^{i(z_1/z)\bk\cdot\boldsymbol R}}{z^2}G_0^{(\epsilon)}(R)
 \nonumber\\
 &\quad+\left[z_1z_2z\,n_{13,2}^{\epsilon^*\epsilon}(-i\nabla_R)
 +\frac{z_1^2z\alpha_{13,2}}{\beta_{13,2}}
 n_{13,2}^{\epsilon\epsilon^*}(-i\nabla_R)\right]
 \frac{e^{i(z_1/z)\bk\cdot\boldsymbol R}}
 {z^2[|z_2\bk-z\bq|^2/(z_2\beta_{13,2})+i0]}\nonumber\\
 &\quad\times\left[\frac{K_0(m_{13,2}R)}{2\pi}-G_0^{(\epsilon)}(R)\right]
 \Biggr\}\Biggr].
 \label{eq:g132-result}
\end{align}
The full graph is $\mathcal V_{13}=\mathcal V_{13(1)}+\mathcal V_{13(2)}$.\footnote{Both
sectors have been checked against the original momentum-space Fourier
integral, including derivative contacts, principal-value and shell terms
with their original $i0$ signs. The check retains the angular measurement
and tests exchange terms that cancel in the inclusive moment.}

\paragraph{Graph $A_{14}$: quark--antiquark exchange before the shockwave.}
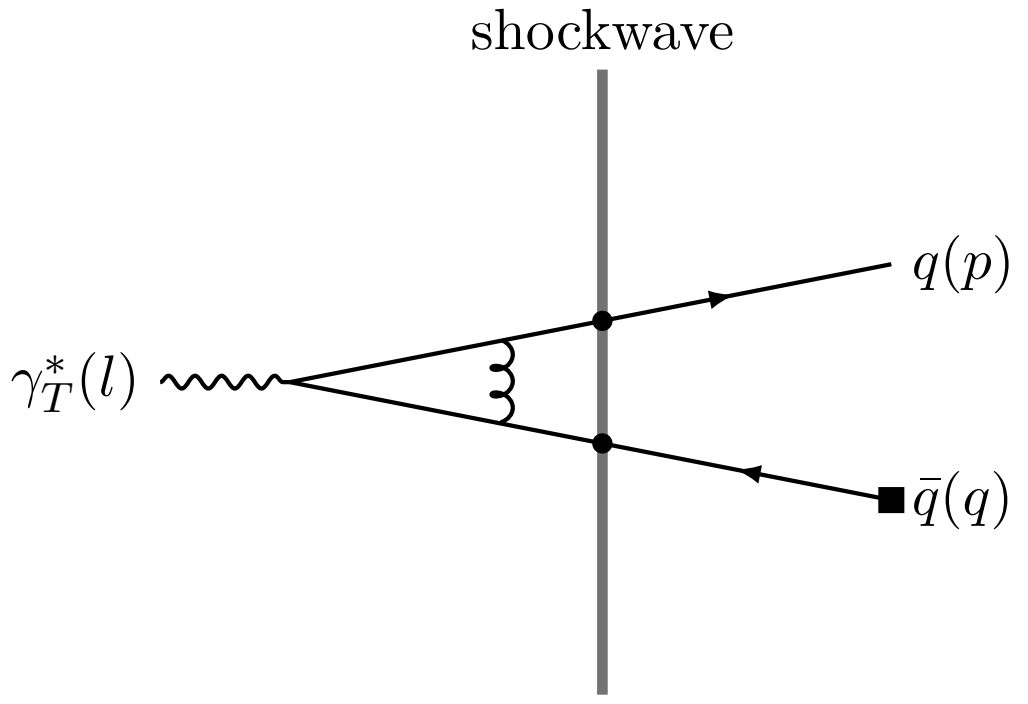
\virtualtopology{\draw[gluon] (1.8,{1.8/5.1})--(1.8,{-1.8/5.1});
\node at (2.65,-3.2) {$A_{14}$};}
This graph carries the Born matrices $\mathcal U_0$ and $\mathcal C_0$, but its
two longitudinal sectors have different photon and pair virtualities.
For sector $14(1)$, $0<z<z_1$, define
\begin{align}
 a_{14,1}&=z_1-z,& b_{14,1}&=z_2+z,&
 D_{\gamma,14(1)}&=\bk_1^2+a_{14,1}b_{14,1}Q^2,
 \nonumber\\
 D_{\gamma,0}&=\bk_2^2+Q_\gamma^2,&
 D_{qg,14(1)}&=
 \left|\bk_1-\frac{a_{14,1}}{z_1}\bk_2\right|^2
 +\frac{za_{14,1}}{z_2z_1^2}(\bk_2^2+Q_\gamma^2).
 \label{eq:g141-denominators}
\end{align}
The two helicity numerators are
\begingroup\small
\begin{align}
 \mathcal N_{14(1),+}={}&-
 \frac{32(l^+)^2\sqrt{z_1z_2}}{b_{14,1}z}
 \Biggl\{\frac{z_1z_2}{2}D_{\gamma,14(1)}
 +\frac{a_{14,1}b_{14,1}}2D_{\gamma,0}
 +z(z_2\bk_1+a_{14,1}\bk_2)_\epsilon
 (z_1\bk_1+b_{14,1}\bk_2)_{\epsilon^*}\Biggr\}k_{1\epsilon}
 \nonumber\\
 &+\frac{16(l^+)^2(z_1z_2)^{3/2}}
 {a_{14,1}zb_{14,1}}D_{\gamma,14(1)}
 (z_1\bk_1-a_{14,1}\bk_2)_\epsilon,
 \label{eq:g141-spin-plus}\\
 \mathcal N_{14(1),-}={}&
 \frac{32(l^+)^2\sqrt{z_1z_2}}{a_{14,1}z}
 \Biggl\{\frac{z_1z_2}{2}D_{\gamma,14(1)}
 +\frac{a_{14,1}b_{14,1}}2D_{\gamma,0}
 +z(z_2\bk_1+a_{14,1}\bk_2)_{\epsilon^*}
 (z_1\bk_1+b_{14,1}\bk_2)_\epsilon\Biggr\}k_{1\epsilon}.
 \label{eq:g141-spin-minus}
\end{align}
\endgroup
After both energy contours, the amplitude is
\begingroup\small
\begin{align}
 i\mathcal A_{14(1),+}={}&-
 \frac{ee_fg^2C_F}{4z_1l^+}
 \nonumber\\[-2pt]
 &\times
 \int_0^{z_1}\left(\frac{\nu^+}{zl^+}\right)^{\eta_Y}\frac{\dd z}{2\pi}\,a_{14,1}
 \int\dd^2\bx_1\dd^2\bx_2\,
 \mathcal U_0e^{-i\bp\cdot\bx_1-i\bq\cdot\bx_2}
 \int_{\bk_1,\bk_2}e^{i\bk_2\cdot\bx_{12}}
 \frac{\mathcal N_{14(1),+}}
 {D_{\gamma,14(1)}D_{\gamma,0}D_{qg,14(1)}},
 \label{eq:g141-final-amplitude-plus}\\
 i\mathcal A_{14(1),-}={}&-
 \frac{ee_fg^2C_F}{4z_1l^+}
 \nonumber\\[-2pt]
 &\times
 \int_0^{z_1}\left(\frac{\nu^+}{zl^+}\right)^{\eta_Y}\frac{\dd z}{2\pi}\,a_{14,1}
 \int\dd^2\bx_1\dd^2\bx_2\,
 \mathcal U_0e^{-i\bp\cdot\bx_1-i\bq\cdot\bx_2}
 \int_{\bk_1,\bk_2}e^{i\bk_2\cdot\bx_{12}}
 \frac{\mathcal N_{14(1),-}}
 {D_{\gamma,14(1)}D_{\gamma,0}D_{qg,14(1)}}.
 \label{eq:g141-final-amplitude-minus}
\end{align}
\endgroup

For sector $14(2)$, $0<z<z_2$, define
\begin{align}
 a_{14,2}&=z_2-z,& b_{14,2}&=z_1+z,&
 D_{\gamma,14(2)}&=\bk_1^2+b_{14,2}a_{14,2}Q^2,
 \nonumber\\
 D_{qg,14(2)}&=
 \left|\bk_1-\frac{a_{14,2}}{z_2}\bk_2\right|^2
 +\frac{za_{14,2}}{z_1z_2^2}(\bk_2^2+Q_\gamma^2).
 \label{eq:g142-denominators}
\end{align}
The numerator entries are
\begingroup\small
\begin{align}
 \mathcal N_{14(2),+}={}&-
 \frac{32(l^+)^2\sqrt{z_1z_2}}{a_{14,2}z}
 \Biggl\{\frac{z_1z_2}{2}D_{\gamma,14(2)}
 +\frac{b_{14,2}a_{14,2}}2D_{\gamma,0}
 +z(z_2\bk_1+b_{14,2}\bk_2)_\epsilon
 (z_1\bk_1+a_{14,2}\bk_2)_{\epsilon^*}\Biggr\}k_{1\epsilon},
 \label{eq:g142-spin-plus}\\
 \mathcal N_{14(2),-}={}&
 \frac{32(l^+)^2\sqrt{z_1z_2}}{b_{14,2}z}
 \Biggl\{\frac{z_1z_2}{2}D_{\gamma,14(2)}
 +\frac{b_{14,2}a_{14,2}}2D_{\gamma,0}
 +z(z_2\bk_1+b_{14,2}\bk_2)_{\epsilon^*}
 (z_1\bk_1+a_{14,2}\bk_2)_\epsilon\Biggr\}k_{1\epsilon}
 \nonumber\\
 &-\frac{16(l^+)^2(z_1z_2)^{3/2}}
 {b_{14,2}za_{14,2}}D_{\gamma,14(2)}
 (z_2\bk_1-a_{14,2}\bk_2)_\epsilon.
 \label{eq:g142-spin-minus}
\end{align}
\endgroup
The two amplitudes are
\begingroup\small
\begin{align}
 i\mathcal A_{14(2),+}={}&-
 \frac{ee_fg^2C_F}{4z_2l^+}
 \nonumber\\[-2pt]
 &\times
 \int_0^{z_2}\left(\frac{\nu^+}{zl^+}\right)^{\eta_Y}\frac{\dd z}{2\pi}\,a_{14,2}
 \int\dd^2\bx_1\dd^2\bx_2\,
 \mathcal U_0e^{-i\bp\cdot\bx_1-i\bq\cdot\bx_2}
 \int_{\bk_1,\bk_2}e^{i\bk_2\cdot\bx_{12}}
 \frac{\mathcal N_{14(2),+}}
 {D_{\gamma,14(2)}D_{\gamma,0}D_{qg,14(2)}},
 \label{eq:g142-final-amplitude-plus}\\
 i\mathcal A_{14(2),-}={}&-
 \frac{ee_fg^2C_F}{4z_2l^+}
 \nonumber\\[-2pt]
 &\times
 \int_0^{z_2}\left(\frac{\nu^+}{zl^+}\right)^{\eta_Y}\frac{\dd z}{2\pi}\,a_{14,2}
 \int\dd^2\bx_1\dd^2\bx_2\,
 \mathcal U_0e^{-i\bp\cdot\bx_1-i\bq\cdot\bx_2}
 \int_{\bk_1,\bk_2}e^{i\bk_2\cdot\bx_{12}}
 \frac{\mathcal N_{14(2),-}}
 {D_{\gamma,14(2)}D_{\gamma,0}D_{qg,14(2)}}.
 \label{eq:g142-final-amplitude-minus}
\end{align}
\endgroup

The helicity amplitudes above specify the physical four-dimensional
components. For the loop integration, the FDH numerator is evaluated
with spin dimension $D_s=4$ independent of momentum dimension
$D=4-2\epsilon$. Write
$\bk_1=\bk_{1,[2]}+\widehat{\bk}_1$, with the formal transverse trace
$\widehat\delta_{ii}=-2\epsilon$; $\bk_2$ and the external transverse vectors
have no hatted components.  Genuine-loop denominators contain the full
$(2-2\epsilon)$-dimensional squared momenta. The primitive
Born-interference numerator contains the evanescent term
\begin{equation*}
 \delta S_{14,s}=\widehat{\bk}_1^{\,2}\frac{z(z_s-z)}{z_s}
 (\bk_2\cdot\bx_{1 2'}),\qquad s=1,2,
\end{equation*}
where $s=1,2$ denotes the quark and antiquark sectors. After the helicity
sum and removal of shifted odd integrals, the interference kernels are
\begingroup\small
\begin{align}
 \mathcal K_{14(1)}={}&
 \frac{[z_1^2+z_2^2-2z_1z+z^2](\bk_2\cdot\bx_{1 2'})}
 {D_{\gamma,0}D_{qg,14(1)}}
 +\frac{\frac{a_{14,1}}{z_1}
 [1+z^2-2z_2a_{14,1}](\bk_1\cdot\bx_{1 2'})}
 {D_{\gamma,14(1)}D_{qg,14(1)}}
 \nonumber\\
 &-\frac{Q^2\frac{a_{14,1}}{z_1}
 [2z_1z_2z(\bk_1\cdot\bx_{1 2'})
 +z(z+z_2-z_1)^2(\bk_2\cdot\bx_{1 2'})]}
 {D_{\gamma,14(1)}D_{\gamma,0}D_{qg,14(1)}}
 \nonumber\\
 &+\frac{\frac{za_{14,1}}{z_1}
 \widehat{\bk}_1^{\,2}(\bk_2\cdot\bx_{1 2\prime})}
 {D_{\gamma,14(1)}D_{\gamma,0}D_{qg,14(1)}},
 \label{eq:g141-reduced-kernel}\\
 \mathcal K_{14(2)}={}&
 \frac{[z_2^2+z_1^2-2z_2z+z^2](\bk_2\cdot\bx_{1 2'})}
 {D_{\gamma,0}D_{qg,14(2)}}
 +\frac{\frac{a_{14,2}}{z_2}
 [1+z^2-2z_1a_{14,2}](\bk_1\cdot\bx_{1 2'})}
 {D_{\gamma,14(2)}D_{qg,14(2)}}
 \nonumber\\
 &-\frac{Q^2\frac{a_{14,2}}{z_2}
 [2z_1z_2z(\bk_1\cdot\bx_{1 2'})
 +z(z+z_1-z_2)^2(\bk_2\cdot\bx_{1 2'})]}
 {D_{\gamma,14(2)}D_{\gamma,0}D_{qg,14(2)}}
 \nonumber\\
 &+\frac{\frac{za_{14,2}}{z_2}
 \widehat{\bk}_1^{\,2}(\bk_2\cdot\bx_{1 2\prime})}
 {D_{\gamma,14(2)}D_{\gamma,0}D_{qg,14(2)}}.
 \label{eq:g142-reduced-kernel}
\end{align}
\endgroup
The two one-sided cross-section contributions are
\begingroup\small
\begin{align}
 \mathcal V_{14(1)}={}&-
 \frac{ie^2e_f^2g^2QN_cC_F}{(2\pi)^5}
 \nonumber\\[-2pt]
 &\times\sqrt{z_1}z_2^{3/2}
 \int\dd^2\bx_1\dd^2\bx_2\dd^2\bx_{2'}
 e^{i\bq\cdot\bx_{2'2}}\mathcal C_0
 \frac{K_1(Q_\gamma|\bx_{1 2'}|)}{|\bx_{1 2'}|}
 \int_0^{z_1}\left(\frac{\nu^+}{zl^+}\right)^{\eta_Y}\frac{\dd z}{z}
 \int_{\bk_1,\bk_2}e^{i\bk_2\cdot\bx_{12}}\mathcal K_{14(1)},
 \label{eq:g141-result}\\
 \mathcal V_{14(2)}={}&-
 \frac{ie^2e_f^2g^2QN_cC_F}{(2\pi)^5}
 \nonumber\\[-2pt]
 &\times\sqrt{z_1}z_2^{3/2}
 \int\dd^2\bx_1\dd^2\bx_2\dd^2\bx_{2'}
 e^{i\bq\cdot\bx_{2'2}}\mathcal C_0
 \frac{K_1(Q_\gamma|\bx_{1 2'}|)}{|\bx_{1 2'}|}
 \int_0^{z_2}\left(\frac{\nu^+}{zl^+}\right)^{\eta_Y}\frac{\dd z}{z}
 \int_{\bk_1,\bk_2}e^{i\bk_2\cdot\bx_{12}}\mathcal K_{14(2)}.
 \label{eq:g142-result}
\end{align}
\endgroup

For completeness, the remaining genuine loop in these two expressions is
already an explicit parameter integral.  For $s=1,2$, define
\begin{align}
 c_1&=\frac{a_{14,1}}{z_1},&
 \rho_1&=\frac{za_{14,1}}{z_2z_1^2},&
 M_1^2&=a_{14,1}b_{14,1}Q^2,\nonumber\\
 c_2&=\frac{a_{14,2}}{z_2},&
 \rho_2&=\frac{za_{14,2}}{z_1z_2^2},&
 M_2^2&=a_{14,2}b_{14,2}Q^2,\nonumber\\
 \Omega_{14(s)}(u,\bk_2)&=
 uM_s^2+(1-u)\rho_sQ_\gamma^2
 +(1-u)(\rho_s+uc_s^2)\bk_2^2 .
 \label{eq:g14-parameter-definitions}
\end{align}
Then
\begin{align}
 \mu^{2\epsilon}\int\frac{\dd^{2-2\epsilon}\bk_1}{(2\pi)^{2-2\epsilon}}
 \frac1{D_{qg,14(s)}}
 &=\frac{\mu^{2\epsilon}\Gamma(\epsilon)}{(4\pi)^{1-\epsilon}}
 [\rho_s(\bk_2^2+Q_\gamma^2)]^{-\epsilon},
 \nonumber\\
 \mu^{2\epsilon}\int\frac{\dd^{2-2\epsilon}\bk_1}{(2\pi)^{2-2\epsilon}}
 \frac1{D_{\gamma,14(s)}D_{qg,14(s)}}
 &=\frac{\mu^{2\epsilon}\Gamma(1+\epsilon)}{(4\pi)^{1-\epsilon}}
 \int_0^1\frac{\dd u}{\Omega_{14(s)}(u,\bk_2)^{1+\epsilon}},
 \nonumber\\
 \mu^{2\epsilon}\int\frac{\dd^{2-2\epsilon}\bk_1}{(2\pi)^{2-2\epsilon}}
 \frac{k_{1i}}{D_{\gamma,14(s)}D_{qg,14(s)}}
 &=\frac{\mu^{2\epsilon}\Gamma(1+\epsilon)}{(4\pi)^{1-\epsilon}}
 \int_0^1\dd u\,
 \frac{(1-u)c_sk_{2i}}{\Omega_{14(s)}(u,\bk_2)^{1+\epsilon}}.
 \label{eq:g14-loop-integrals}
\end{align}
The hatted terms in Eqs.~\eqref{eq:g141-reduced-kernel}
and~\eqref{eq:g142-reduced-kernel} require one further genuine-loop moment:
\begin{equation*}
 \mu^{2\epsilon}\int\frac{\dd^{2-2\epsilon}\bk_1}{(2\pi)^{2-2\epsilon}}
 \frac{\widehat{\bk}_1^{\,2}}
 {D_{\gamma,14(s)}D_{qg,14(s)}}
 =-\frac{\epsilon\mu^{2\epsilon}\Gamma(\epsilon)}{(4\pi)^{1-\epsilon}}
 \int_0^1\dd u\,\Omega_{14(s)}(u,\bk_2)^{-\epsilon}
 =-\frac1{4\pi}+O(\epsilon).
\end{equation*}
The Feynman-parameter shift is physical and leaves $\widehat{\bk}_1$ unchanged.
The factor $-\epsilon$ from the hatted trace multiplies the UV pole, so the
last term of each reduced kernel survives as the finite contribution
\begin{equation*}
 \delta\overline{\mathcal K}_{14(s)}
 =-\frac{za_{14,s}}{4\pi z_s}
 \frac{\bk_2\cdot\bx_{1 2'}}{\bk_2^2+Q_\gamma^2},
 \qquad s=1,2.
\end{equation*}
Here the overline denotes the $\bk_1$ integration of the evanescent term;
the common prefactors and Fourier transforms in
Eqs.~\eqref{eq:g141-result} and~\eqref{eq:g142-result} still apply.
After the $z$ integral, the corresponding one-sided Born ratios are
\begin{equation*}
 \frac{\delta\mathcal V_{14(s)}}{\mathcal V_{\Born}}
 =-\frac{\alpha_sC_F}{2\pi}\frac{z_s}{2(z_1^2+z_2^2)},
 \qquad
 \frac{\delta\mathcal V_{14}}{\mathcal V_{\Born}}
 =-\frac{\alpha_sC_F}{2\pi}\frac1{2(z_1^2+z_2^2)}.
\end{equation*}
These ratios refer to one virtual insertion; adding its conjugate gives
$2\Reh\,\delta\mathcal V_{14}$. This evanescent contribution is regular at
$z=0$ after the original $\dd z/z$ measure and introduces no rapidity pole.
It is included in both reduced kernels.

For every radial function left after Eq.~\eqref{eq:g14-loop-integrals}, the
last photon transform is the one-dimensional integral
\begin{equation}
 \int_{\bk_2}e^{i\bk_2\cdot\bx_{12}}
 (\bk_2\cdot\bx_{1 2'})f(\bk_2^2)
 =\frac{i\bx_{12}\cdot\bx_{1 2'}}{2\pi|\bx_{12}|}
 \int_0^\infty\dd k\,k^2J_1(k|\bx_{12}|)f(k^2).
 \label{eq:g14-final-photon-transform}
\end{equation}
The reduction completes the energy contours, spin traces and
spectator-momentum integrals in
$\mathcal V_{14}=\mathcal V_{14(1)}+\mathcal V_{14(2)}$.

The primitive spin chains have also been checked with integer-$D_s$
gamma matrices, a recursive Clifford trace and an independent
exterior-algebra implementation. The integer-dimensional reconstruction
uses $N(4)=2N(6)-N(8)$ and verifies $N(10)-2N(8)+N(6)=0$; the exact
rational tests have zero residual. These algebraic checks fix the
evanescent term before the inclusive comparison of
Sec.~\ref{sec:finite-validation}.

\paragraph{Measured virtual sum.}
For uniform notation let
\begin{equation}
 \mathcal V_{13}\equiv\mathcal V_{13(1)}+\mathcal V_{13(2)},
 \qquad
 \mathcal V_{14}\equiv\mathcal V_{14(1)}+\mathcal V_{14(2)}.
 \label{eq:virtual-sector-sums}
\end{equation}
Every graph is first assembled at fixed antiquark momentum.  Only then is the
energy-flow measurement applied:
\begin{align}
 \Sigma^T_{\bar q,V}(\theta)
 & =2\operatorname{Re}\sum_{n=5}^{14}
 \int\frac{\dd z_2}{z_2}\int\dd^2\bq\,
 \mathcal M_{\bar q}(\theta;z_2,\bq)\mathcal V_n(z_2,\bq)
 \nonumber\\
 & =2\operatorname{Re}\frac{x_BQ^2\sin\theta}{8\theta}
 \int_0^{\csc^2(\theta/2)}\dd v\,v^2\int_0^{2\pi}\dd\phi\,
 \sum_{n=5}^{14}\mathcal V_n\left(
 v\sin^2\frac\theta2,
 \frac{Qv}{2}\sin\theta\,\boldsymbol e_\phi\right).
 \label{eq:virtual-measured-sum}
\end{align}

The virtual contribution to the one-point energy flow also contains the
quark-tagged term.  It is obtained graph by graph by keeping $(z_1,\bp)$ fixed,
integrating the antiquark phase space, and exchanging
\begin{align}
 (z_1,\bp,\bx_1,\bx_{1'})
 &\longleftrightarrow
 (z_2,\bq,\bx_2,\bx_{2'}),\nonumber\\
 V(\bx_1)&\longleftrightarrow V^\dagger(\bx_2),\qquad
 V^\dagger(\bx_{1'})\longleftrightarrow V(\bx_{2'}),\qquad
 t^a_{q}\longleftrightarrow-\bigl(t^a\bigr)^{T}_{\bar q}
 \label{eq:virtual-quark-relabeling}
\end{align}
in the fermion-line Feynman rules before the target average is taken.  The
Wilson-line and generator order is inherited from the oriented quark or
antiquark line on which the shockwave acts.  With the superscript stating which
fermion momentum is kept fixed, the complete virtual energy flow is
\begin{align}
 \Sigma^T_{E,V}(\theta)
 ={}&2\operatorname{Re}\frac{x_BQ^2\sin\theta}{8\theta}
 \int_0^{\csc^2(\theta/2)}\dd v\,v^2\int_0^{2\pi}\dd\phi\,
 \sum_{n=5}^{14}\Bigg[
 \mathcal V_n^{(q\,\mathrm{fixed})}\left(
 v\sin^2\frac\theta2,
 \frac{Qv}{2}\sin\theta\,\boldsymbol e_\phi\right)
 \nonumber\\[-2pt]
 &\hspace{48mm}+
 \mathcal V_n^{(\bar q\,\mathrm{fixed})}\left(
 v\sin^2\frac\theta2,
 \frac{Qv}{2}\sin\theta\,\boldsymbol e_\phi\right)
 \Bigg].
 \label{eq:complete-virtual-measured-sum}
\end{align}
For a general target ensemble, Eq.~\eqref{eq:complete-virtual-measured-sum}
contains the quark- and antiquark-tagged contributions separately.

\subsubsection{Real-emission graphs and their measured contributions}
\label{sec:real-fixed-order}

There are four real-emission amplitudes.  In $A_1$ and $A_2$ the gluon is
emitted after the shockwave from the quark and antiquark, respectively; in
$A_3$ and $A_4$ it is emitted from the same two lines before the shockwave.
Squaring their sum gives four diagonal products and six independent
interferences.  Thus ten products have to be calculated; restoring the reverse
ordering of each interference gives the sixteen ordered terms that enter the
cross section.  We derive $A_2$ and $A_2A_2^\dagger$ in detail because this
pair contains the final-state collinear denominator of the detected antiquark.
For the other terms we display the graph, the final amplitude and the resulting
fixed-momentum kernel. The black square marks the antiquark
energy-flow insertion in each graph.
The independent products may be chosen as
\begin{equation}
 \{\Rij{1}{1},\Rij{2}{2},\Rij{3}{3},\Rij{4}{4}\},\qquad
 \{\Rij{2}{1},\Rij{3}{1},\Rij{4}{1},
   \Rij{3}{2},\Rij{4}{2},\Rij{4}{3}\},
 \label{eq:real-independent-products}
\end{equation}
with the reverse ordering of the second set supplied by complex conjugation.

For fixed $z_2$ it is convenient to write
\begin{align}
 \bar z_2&=1-z_2,&z_1&=\bar z_2-z,&0<z<\bar z_2,
 \nonumber\\
 Q_{2z}&=Q\sqrt{(\bar z_2-z)(z_2+z)},&
 Q_2&=Q\sqrt{z_2\bar z_2},
 \label{eq:real-longitudinal-definitions}
\end{align}
Every lower endpoint written as $z=0$ below is defined by the same analytic
continuation as Eq.~\eqref{eq:analytic-rapidity-regulator}.  Consequently an
exact cut identity carries the identical factor
$(\nu^+/(zl^+))^{\eta_Y}$ on its real and virtual sides.  Real and virtual terms with the same endpoint residue are combined
before expanding around $\eta_Y=0$.
The three-body transverse sizes are
\begin{align}
 X^2={}&z_2(\bar z_2-z)\bx_{21}^2+z_2z\bx_{23}^2
       +(\bar z_2-z)z\bx_{13}^2,
 \nonumber\\
 X'^2={}&z_2(\bar z_2-z)\bx_{2'1'}^2+z_2z\bx_{2'3}^2
       +(\bar z_2-z)z\bx_{1'3}^2,
 \nonumber\\
 \Delta^{(3)}_{ij}&\equiv
 \frac{\bx_{3i}\cdot\bx_{3j}}{\bx_{3i}^2\bx_{3j}^2}.
 \label{eq:real-coordinate-definitions}
\end{align}
Below, $\dd^8\bx$ denotes the four independent transverse-coordinate
integrals left after the unobserved $\bp$ and $\bk$ integrations. Their coordinate
choices are specified below. Before any particle is
integrated out, the three-body phase space and incident flux are
\begin{align}
 \dd\Phi_{\eta_Y}^{(3)}={}&2l^+\,
 \frac{\dd^2\bp\,\dd^2\bq\,\dd^2\bk\,
 \dd z_1\,\dd z_2\,\dd z}
 {(2\pi)^8(4l^+)^2z_1z_2z}
 \left(\frac{\nu^+}{zl^+}\right)^{\eta_Y}
 \delta(1-z_1-z_2-z),&
 \mathcal F&=2l^+.
 \label{eq:real-three-body-phase-space}
\end{align}
Equation~\eqref{eq:real-three-body-phase-space} is common to the quark,
antiquark and gluon tags. The corresponding detector insertion is applied
before integrating the unobserved momenta. For a fixed antiquark,
the ordered-product normalization is
\begin{equation}
 \Rij{i}{j}(z_2,\bq)=
 \int_0^{\bar z_2}\frac{\dd z}{z}\left(\frac{\nu^+}{zl^+}\right)^{\eta_Y}
 \int\dd^2\bp\,\dd^2\bk\,
 \frac{1}{16(l^+)^2(2\pi)^8z_1}\frac12
 \sum_{\lambda,h,\bar h,\lambda_g,a}
 \left\langle\tr\!\left[(i\mathcal M_i^a)
 (i\mathcal M_j^a)^\dagger\right]\right\rangle_Y .
 \label{eq:real-product-normalization}
\end{equation}

\paragraph{Detailed example: antiquark emission after the shockwave.}
The parent antiquark carries fraction $z_2+z$ through the target and splits
into the detected antiquark and the real gluon after the shockwave:
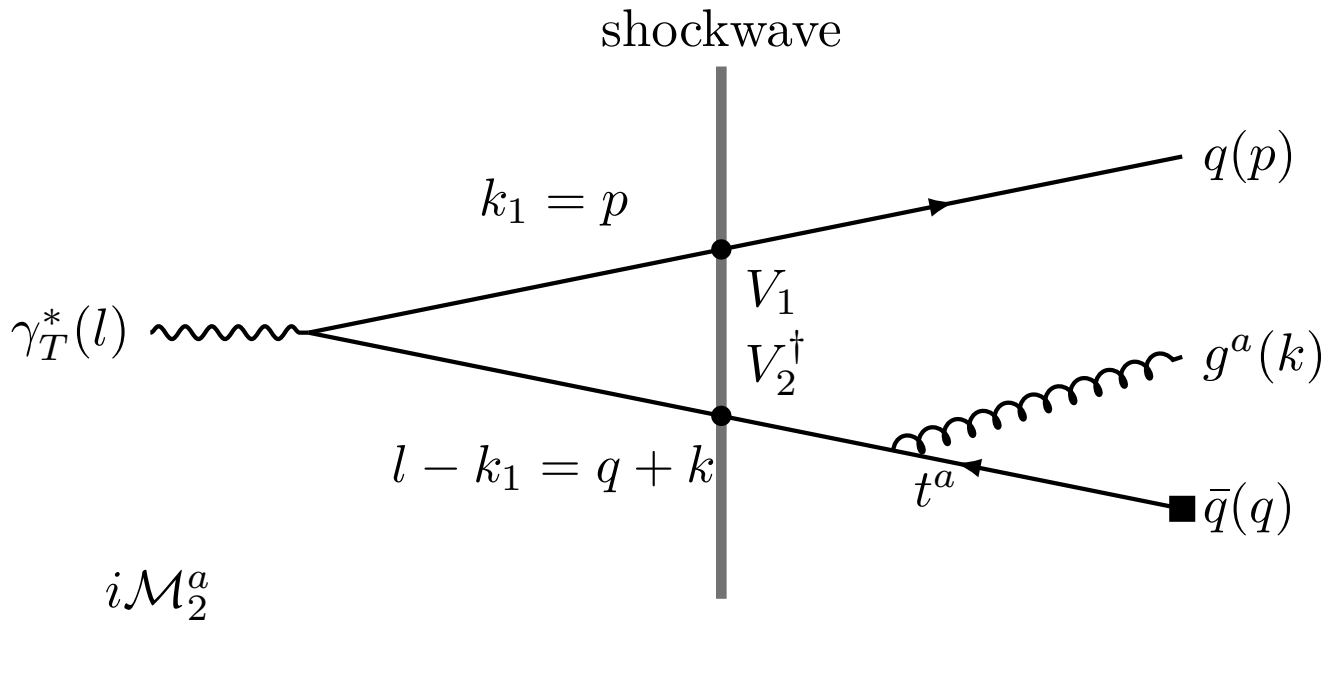
\realamplitudediagram{2}{
 \draw[fermion] (0,0)--(6.35,1.82);
 \draw[antifermion] (0,0)--(6.35,-1.82);
 \draw[gluon] (4.25,{-4.25*1.82/6.35})--(6.35,-.25);
 \node[shockvertex] at (3,{3*1.82/6.35}) {};
 \node[shockvertex] at (3,{-3*1.82/6.35}) {};
 \node[above=7pt] at (1.78,{1.78*1.82/6.35}) {$k_1=p$};
 \node[below=8pt] at (1.78,{-1.78*1.82/6.35}) {$l-k_1=q+k$};
 \node[below right] at (3.02,{3*1.82/6.35}) {$V_1$};
 \node[above right] at (3.02,{-3*1.82/6.35}) {$V_2^\dagger$};
 \node[below right] at (4.25,{-4.25*1.82/6.35}) {$t^a$};
 \node[right] at (6.35,1.82) {$q(p)$};
 \node[right] at (6.35,-.25) {$g^a(k)$};
 \node[energyinsertion] at (6.35,-1.82) {}; \node[right] at (6.35,-1.82) {$\bar q(q)$};
}
Direct insertion of the photon, shockwave and splitting vertices gives
\begin{align}
 i\mathcal M_2^a={}&+iee_fg\int_{k_1^-,\bk_1}
 \dd^2\bx_1\dd^2\bx_2\,
 \frac{16(l^+)^2N_2\,(V_1V_2^\dagger-1)t^a}
 {D(k_1)D(k_1-l)}
 e^{i\bk_1\cdot\bx_{12}}
 e^{-i\bp\cdot\bx_1-i(\bq+\bk)\cdot\bx_2},
 \qquad k_1^+=z_1l^+,
 \label{eq:real-a2-insertion}\\
 N_2={}&\frac{1}{16(l^+)^2}
 \frac{\bar u(p)\gamma^+\slashed k_1\slashed\epsilon_T(l)
 (\slashed k_1-\slashed l)\gamma^+
 (\slashed q+\slashed k)\slashed\epsilon_g^*(k)v(q)}
 {(q+k)^2+i0}.
 \label{eq:real-a2-numerator}
\end{align}
The two energy poles lie on opposite sides of the real axis,
\begin{equation}
 k_{1,a}^-=\frac{\bk_1^2}{2z_1l^+}-i0,
 \qquad
 k_{1,b}^-=l^--\frac{\bk_1^2}{2(z_2+z)l^+}+i0.
 \label{eq:real-a2-poles}
\end{equation}
Closing the contour in the upper half-plane yields
\begin{equation}
 \int\frac{\dd k_1^-}{2\pi}\frac1{D(k_1)D(k_1-l)}
 =\frac{i}{2l^+}\frac1{\bk_1^2+z_1(z_2+z)Q^2}.
 \label{eq:real-a2-residue}
\end{equation}
The amplitude after the energy integration is therefore
\begin{align}
 i\mathcal M_2^a={}&+iee_fg\int_{\bk_1}
 \dd^2\bx_1\dd^2\bx_2\,
 16(l^+)^2N_2\,(V_1V_2^\dagger-1)t^a
 \frac{i}{2l^+[\bk_1^2+z_1(z_2+z)Q^2]}
 \nonumber\\[-2pt]
 &\hspace{22mm}\times e^{i\bk_1\cdot\bx_{12}}
 e^{-i\bp\cdot\bx_1-i(\bq+\bk)\cdot\bx_2}.
 \label{eq:real-a2-post-contour}
\end{align}
For photon helicity $+$ the four nonzero helicity numerators are
\begin{align}
 N_2^{+;+,+}&=-z_1^{3/2}\sqrt{z_2}(z_2+z)
 \frac{(z_2\bk-z\bq)_\epsilon}{(z_2\bk-z\bq)^2}
 (\bk_1)_\epsilon,
 \nonumber\\
 N_2^{+;+,-}&=-(z_1z_2)^{3/2}
 \frac{(z_2\bk-z\bq)_{\epsilon^*}}{(z_2\bk-z\bq)^2}
 (\bk_1)_\epsilon,
 \nonumber\\
 N_2^{+;-,+}&=+z_2^{3/2}\sqrt{z_1}(z_2+z)
 \frac{(z_2\bk-z\bq)_\epsilon}{(z_2\bk-z\bq)^2}
 (\bk_1)_\epsilon,
 \nonumber\\
 N_2^{+;-,-}&=+\sqrt{z_1z_2}(z_2+z)^2
 \frac{(z_2\bk-z\bq)_{\epsilon^*}}{(z_2\bk-z\bq)^2}
 (\bk_1)_\epsilon.
 \label{eq:real-a2-helicities}
\end{align}
The numerator in Eq.~\eqref{eq:real-a2-numerator} is written with
$+(\slashed q+\slashed k)$.  The minus sign in the actual antiquark
propagator $S_F[-(q+k)]=-i(\slashed q+\slashed k)/[(q+k)^2+i0]$
is retained in the overall prefactor of
Eq.~\eqref{eq:real-a2-insertion}.  With the common photon phase
$e^{i\bk_1\cdot\bx_{12}}$, the antiquark-emission amplitude therefore has
the opposite overall sign to the quark-emission amplitude.

We next square Eq.~\eqref{eq:real-a2-post-contour}.  The unobserved-quark
integral sets $\bx_{1'}=\bx_1$, while the gluon integral gives
\begin{align}
 \int\dd^2\bp\,e^{i\bp\cdot\bx_{1'1}}
  &=(2\pi)^2\delta^{(2)}(\bx_{1'1}),
 \nonumber\\
 \int\dd^2\bk\,e^{i\bk\cdot\bx_{2'2}}
 \frac{(z_2\bk-z\bq)^i(z_2\bk-z\bq)^j}
 {(z_2\bk-z\bq)^4}
 &\longrightarrow
 e^{i(z/z_2)\bq\cdot\bx_{2'2}}
 \frac{x_{32'}^ix_{32}^j}{\bx_{32'}^2\bx_{32}^2}.
 \label{eq:real-r22-fourier}
\end{align}
The finite-$N_c$ color trace and the helicity sum are, respectively,
\begin{align}
 \frac1{N_c}\sum_a\tr[(V_1V_2^\dagger-1)t^at^a
 (V_{2'}V_1^\dagger-1)]
 &=C_F[S_{2'2}-S_{12}-S_{2'1}+1],
 \label{eq:real-r22-color}\\
 \mathcal S_{22}(z_2,z)
 &=\frac{(\bar z_2-z)(z_2+z)}{z_2}
 [z_2^2+(z_2+z)^2]
 [\,(\bar z_2-z)^2+(z_2+z)^2\,].
 \label{eq:real-r22-spin}
\end{align}
Combining the two photon transforms with Eqs.~\eqref{eq:real-r22-fourier}--
\eqref{eq:real-r22-spin} gives
\begin{align}
 \Rij{2}{2}(z_2,\bq)={}&
 \frac{e^2e_f^2g^2Q^2N_c}{(2\pi)^8}
 \int_0^{\bar z_2}\frac{\dd z}{z}\left(\frac{\nu^+}{zl^+}\right)^{\eta_Y}\,\mathcal S_{22}(z_2,z)
 \int\dd^8\bx\,
 C_F\langle S_{2'2}-S_{12}-S_{2'1}+1\rangle_Y
 \nonumber\\
 &\times K_1(|\bx_{21}|Q_{2z})K_1(|\bx_{2'1}|Q_{2z})
 \frac{\bx_{21}\cdot\bx_{2'1}}
 {|\bx_{21}|\,|\bx_{2'1}|}
 e^{i\frac{z_2+z}{z_2}\bq\cdot\bx_{2'2}}
 \Delta^{(3)}_{2'2},\qquad \bx_{1'}=\bx_1.
 \label{eq:real-r22-result}
\end{align}
Here and in the products below the four integrations in $\dd^8\bx$ are over
$\bx_1,\bx_2,\bx_{2'}$ and $\bx_3$, unless another constraint is displayed.

\paragraph{The remaining three amplitudes.}
For emission from the quark after the shockwave one finds
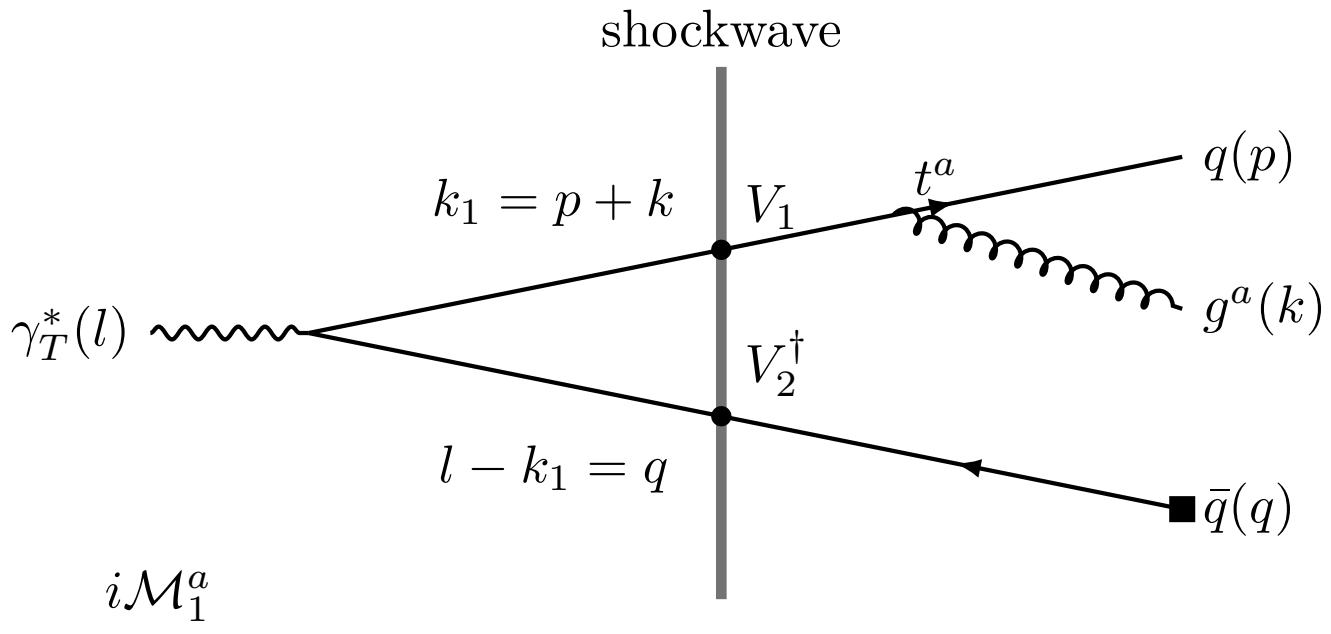
\realamplitudediagram{1}{
 \draw[fermion] (0,0)--(6.35,1.82);
 \draw[antifermion] (0,0)--(6.35,-1.82);
 \draw[gluon] (4.25,{4.25*1.82/6.35})--(6.35,.25);
 \node[shockvertex] at (3,{3*1.82/6.35}) {}; \node[shockvertex] at (3,{-3*1.82/6.35}) {};
 \node[above=7pt] at (1.78,{1.78*1.82/6.35}) {$k_1=p+k$};
 \node[below=8pt] at (1.78,{-1.78*1.82/6.35}) {$l-k_1=q$};
 \node[above right] at (3.02,{3*1.82/6.35}) {$V_1$};
 \node[above right] at (3.02,{-3*1.82/6.35}) {$V_2^\dagger$};
 \node[above right] at (4.25,{4.25*1.82/6.35}) {$t^a$};
 \node[right] at (6.35,1.82) {$q(p)$};
 \node[right] at (6.35,.25) {$g^a(k)$};
 \node[energyinsertion] at (6.35,-1.82) {}; \node[right] at (6.35,-1.82) {$\bar q(q)$};
}
Defining $D_{\gamma1}=\bk_1^2+(z_1+z)z_2Q^2$, the energy-integrated
amplitude is
\begin{align}
 i\mathcal M_1^a={}&-iee_fg\int_{\bk_1}\dd^2\bx_1\dd^2\bx_2\,
 16(l^+)^2N_1\,t^a(V_1V_2^\dagger-1)
 \frac{i}{2l^+D_{\gamma1}}
 e^{i\bk_1\cdot\bx_{12}}
 e^{-i(\bp+\bk)\cdot\bx_1-i\bq\cdot\bx_2},
 \nonumber\\[-2pt]
 & k_1^+=(z_1+z)l^+.
 \label{eq:real-a1-amplitude}
\end{align}
Here
\begin{equation}
 N_1=\frac{1}{16(l^+)^2}
 \frac{\bar u(p)\slashed\epsilon_g^*(k)(\slashed p+\slashed k)
 \gamma^+\slashed k_1\slashed\epsilon_T(l)
 (\slashed k_1-\slashed l)\gamma^+v(q)}{(p+k)^2+i0},
 \label{eq:real-a1-numerator}
\end{equation}
and the four nonzero helicity entries are
\begin{align}
 N_1^{+;+,+}&=-z_1^{3/2}\sqrt{z_2}(z_1+z)
 \frac{(z_1\bk-z\bp)_\epsilon}{(z_1\bk-z\bp)^2}(\bk_1)_\epsilon,
 \nonumber\\
 N_1^{+;+,-}&=-\sqrt{z_1z_2}(z_1+z)^2
 \frac{(z_1\bk-z\bp)_{\epsilon^*}}{(z_1\bk-z\bp)^2}(\bk_1)_\epsilon,
 \nonumber\\
 N_1^{+;-,+}&=+z_2^{3/2}\sqrt{z_1}(z_1+z)
 \frac{(z_1\bk-z\bp)_\epsilon}{(z_1\bk-z\bp)^2}(\bk_1)_\epsilon,
 \nonumber\\
 N_1^{+;-,-}&=+(z_1z_2)^{3/2}
 \frac{(z_1\bk-z\bp)_{\epsilon^*}}{(z_1\bk-z\bp)^2}(\bk_1)_\epsilon.
 \label{eq:real-a1-helicities}
\end{align}

When the gluon is emitted before the shockwave, all three final-state partons
cross the target.  The quark-emission graph is
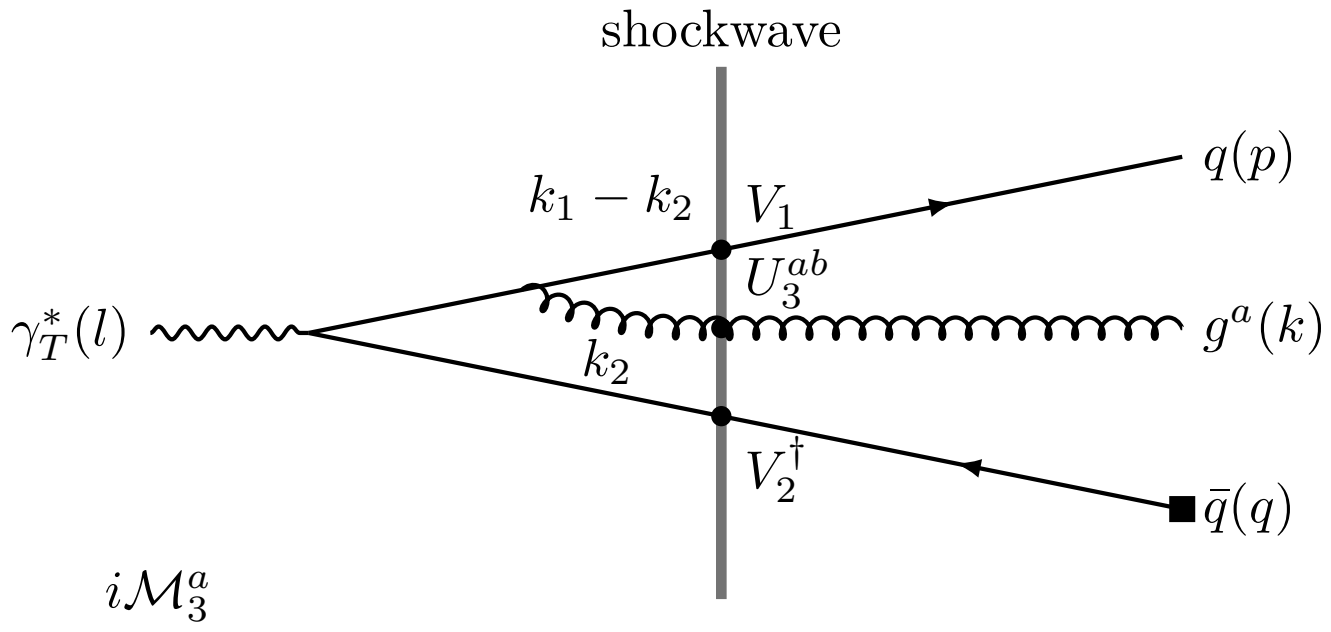
\realamplitudediagram{3}{
 \draw[fermion] (0,0)--(6.35,1.82);
 \draw[antifermion] (0,0)--(6.35,-1.82);
 \draw[gluon] (1.55,{1.55*1.82/6.35}) .. controls (2.05,.12) and (2.45,.05) .. (3,.05)--(6.35,.05);
 \node[shockvertex] at (3,{3*1.82/6.35}) {};
 \node[shockvertex] at (3,.05) {}; \node[shockvertex] at (3,{-3*1.82/6.35}) {};
 \node[above=6pt] at (2.20,{2.20*1.82/6.35}) {$k_1-k_2$};
 \node[below] at (2.17,.15) {$k_2$};
 \node[above right] at (3.02,{3*1.82/6.35}) {$V_1$};
 \node[above right] at (3.02,.05) {$U_3^{ab}$};
 \node[below right] at (3.02,{-3*1.82/6.35}) {$V_2^\dagger$};
 \node[right] at (6.35,1.82) {$q(p)$}; \node[right] at (6.35,.05) {$g^a(k)$};
 \node[energyinsertion] at (6.35,-1.82) {}; \node[right] at (6.35,-1.82) {$\bar q(q)$};
}
Define the two denominators left by the energy contours as
\begin{equation}
 D_{\gamma3}=\bk_1^2+(z_1+z)z_2Q^2,
 \qquad
 D_3=Q^2+\frac{\bk_1^2}{z_2}+\frac{\bk_2^2}{z}
 +\frac{(\bk_1-\bk_2)^2}{z_1}.
 \label{eq:real-a3-denominators}
\end{equation}
After both energy integrations the amplitude is
\begin{align}
 i\mathcal M_3^a={}&ee_fg\int_{\bk_1}\int_{\bk_2}
 \dd^2\bx_1\dd^2\bx_2\dd^2\bx_3\,
 \frac{16(l^+)^2N_3}{2z_1l^+D_{\gamma3}D_3}
 [V_1t^bV_2^\dagger U_3^{ab}-t^a]
 \nonumber\\[-2pt]
 &\times e^{i\bk_1\cdot\bx_{12}+i\bk_2\cdot\bx_{31}}
 e^{-i\bp\cdot\bx_1-i\bq\cdot\bx_2-i\bk\cdot\bx_3},
 \qquad k_1^+=(z_1+z)l^+,\qquad k_2^+=zl^+.
 \label{eq:real-a3-amplitude}
\end{align}
The numerator is
\begin{equation}
 N_3=\frac{\bar u(p)\gamma^+(\slashed k_1-\slashed k_2)
 \gamma^\mu\slashed k_1\slashed\epsilon_T(l)
 (\slashed k_1-\slashed l)\gamma^+v(q)
 d_{\mu\nu}(k_2)\epsilon_g^{*\nu}(k)}{16(l^+)^2},
 \label{eq:real-a3-numerator}
\end{equation}
and its nonzero helicity entries are
\begin{align}
 N_3^{+;+,+}&=z_1^{3/2}\sqrt{z_2}(z_1+z)L_\epsilon(\bk_1)_\epsilon,
 \nonumber\\
 N_3^{+;+,-}&=\sqrt{z_1z_2}(z_1+z)^2L_{\epsilon^*}(\bk_1)_\epsilon,
 \nonumber\\
 N_3^{+;-,+}&=-z_2^{3/2}\sqrt{z_1}(z_1+z)L_\epsilon(\bk_1)_\epsilon,
 \nonumber\\
 N_3^{+;-,-}&=-(z_1z_2)^{3/2}
 \left[L_{\epsilon^*}(\bk_1)_\epsilon+
 \frac{\bk_1^2+z_2(z_1+z)Q^2}{2z_2(z_1+z)}\right],
 \nonumber\\[-2pt]
 L_{\epsilon^{(*)}}&\equiv
 \frac{(\bk_2)_{\epsilon^{(*)}}}{z}
 -\frac{(\bk_1)_{\epsilon^{(*)}}}{z_1+z}.
 \label{eq:real-a3-helicities}
\end{align}

The corresponding antiquark-emission graph and amplitude are
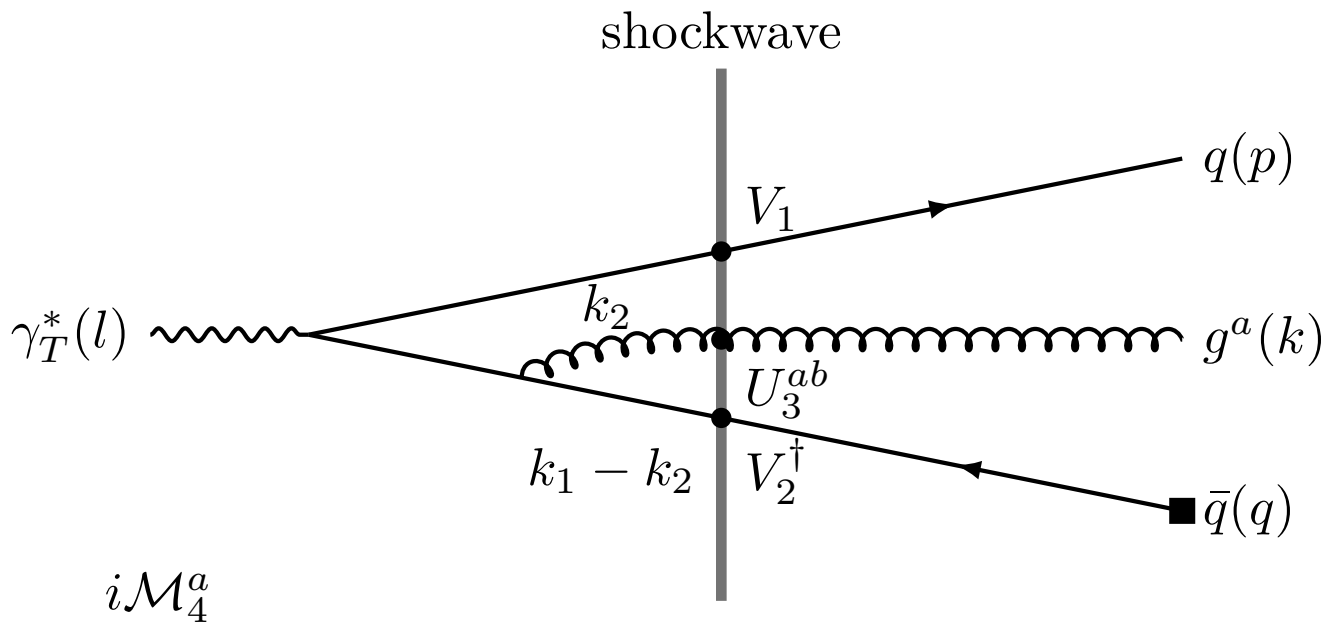
\realamplitudediagram{4}{
 \draw[fermion] (0,0)--(6.35,1.82);
 \draw[antifermion] (0,0)--(6.35,-1.82);
 \draw[gluon] (1.55,{-1.55*1.82/6.35}) .. controls (2.05,-.12) and (2.45,-.05) .. (3,-.05)--(6.35,-.05);
 \node[shockvertex] at (3,{3*1.82/6.35}) {};
 \node[shockvertex] at (3,-.05) {}; \node[shockvertex] at (3,{-3*1.82/6.35}) {};
 \node[below=6pt] at (2.20,{-2.20*1.82/6.35}) {$k_1-k_2$}; \node[above] at (2.17,-.15) {$k_2$};
 \node[above right] at (3.02,{3*1.82/6.35}) {$V_1$};
 \node[below right] at (3.02,-.05) {$U_3^{ab}$};
 \node[below right] at (3.02,{-3*1.82/6.35}) {$V_2^\dagger$};
 \node[right] at (6.35,1.82) {$q(p)$}; \node[right] at (6.35,-.05) {$g^a(k)$};
 \node[energyinsertion] at (6.35,-1.82) {}; \node[right] at (6.35,-1.82) {$\bar q(q)$};
}
The antiquark-side denominators are
\begin{equation}
 D_{\gamma4}=\bk_1^2+z_1(z_2+z)Q^2,
 \qquad
 D_4=Q^2+\frac{\bk_1^2}{z_1}+\frac{\bk_2^2}{z}
 +\frac{(\bk_1-\bk_2)^2}{z_2}.
 \label{eq:real-a4-denominators}
\end{equation}
The final amplitude is
\begin{align}
 i\mathcal M_4^a={}&ee_fg\int_{\bk_1}\int_{\bk_2}
 \dd^2\bx_1\dd^2\bx_2\dd^2\bx_3\,
 \frac{16(l^+)^2N_4}{2z_2l^+D_{\gamma4}D_4}
 [V_1t^bV_2^\dagger U_3^{ab}-t^a]
 \nonumber\\[-2pt]
 &\times e^{i\bk_1\cdot\bx_{21}+i\bk_2\cdot\bx_{32}}
 e^{-i\bp\cdot\bx_1-i\bq\cdot\bx_2-i\bk\cdot\bx_3},
 \qquad k_1^+=(z_2+z)l^+,\qquad k_2^+=zl^+.
 \label{eq:real-a4-amplitude}
\end{align}
With
$\bar L_{\epsilon^{(*)}}=(\bk_2)_{\epsilon^{(*)}}/z
-(\bk_1)_{\epsilon^{(*)}}/(z_2+z)$, the four helicity entries are
\begin{align}
 N_4^{+;+,+}&=z_1^{3/2}\sqrt{z_2}(z_2+z)\bar L_\epsilon(\bk_1)_\epsilon,
 \nonumber\\
 N_4^{+;+,-}&=(z_1z_2)^{3/2}
 \left[\bar L_{\epsilon^*}(\bk_1)_\epsilon+
 \frac{\bk_1^2+z_1(z_2+z)Q^2}{2z_1(z_2+z)}\right],
 \nonumber\\
 N_4^{+;-,+}&=-z_2^{3/2}\sqrt{z_1}(z_2+z)\bar L_\epsilon(\bk_1)_\epsilon,
 \nonumber\\
 N_4^{+;-,-}&=-\sqrt{z_1z_2}(z_2+z)^2
 \bar L_{\epsilon^*}(\bk_1)_\epsilon.
 \label{eq:real-a4-helicities}
\end{align}

\paragraph{Remaining products: color and coordinate definitions.}
The after--after interference obeys
\begin{align}
 \bx_{1'}&=\bx_1-\frac{z}{\bar z_2}\bx_{13},&
 \bx_{1'3}&=\frac{\bar z_2-z}{\bar z_2}\bx_{13},
 \label{eq:real-r21-coordinate-definitions}\\
 \mathcal C_{21}&=C_F\langle S_{12}S_{2'1'}-S_{12}-S_{2'1'}+1\rangle_Y
 -\frac1{2N_c}\langle S_{122'1'}-S_{12}S_{2'1'}\rangle_Y.
 \label{eq:real-r21-color}
\end{align}
Its result is
\begingroup\small
\begin{align}
 \Rij{2}{1}={}&\frac{e^2e_f^2g^2Q^2N_c}{(2\pi)^8}
 \int_0^{\bar z_2}\frac{\dd z}{z}\left(\frac{\nu^+}{zl^+}\right)^{\eta_Y}
 \frac{(\bar z_2-z)^2}{\bar z_2^2}
 \frac{\sqrt{z_2\bar z_2(\bar z_2-z)(z_2+z)}}{\bar z_2-z}
 \int\dd^8\bx\,\mathcal C_{21}
 \nonumber\\[-2pt]
 &\times K_1(|\bx_{21}|Q_{2z})K_1(|\bx_{2'1'}|Q_2)
 \nonumber\\[-2pt]
 &\times e^{i\bq\cdot(\bx_{2'2}+\frac z{z_2}\bx_{32})}
 4\Reh\!\left\{
 \frac{(\bx_{21})_\epsilon(\bx_{2'1'})_{\epsilon^*}}
 {|\bx_{21}|\,|\bx_{2'1'}|}
 \left[
 [z_2^2+(\bar z_2-z)^2]\bar z_2(z_2+z)
 \frac{(\bx_{32})_\epsilon(\bx_{1'3})_{\epsilon^*}}
 {\bx_{32}^2\bx_{1'3}^2}
 \right.\right.\nonumber\\[-2pt]
 &\left.\left.\hspace{39mm}
 +[\bar z_2^2+(z_2+z)^2]z_2(\bar z_2-z)
 \frac{(\bx_{32})_{\epsilon^*}(\bx_{1'3})_\epsilon}
 {\bx_{32}^2\bx_{1'3}^2}\right]\right\},
 \nonumber\\
 \Rij{1}{2}={}&\Rij{2}{1}^{\,*}.
 \label{eq:real-r21-result}
\end{align}
\endgroup

Emission from the unobserved quark after the shockwave is proportional to the
Born kernel.  Defining
\begin{align}
 \bell_q={}&\bk-\frac{z}{\bar z_2}(\bp+\bk),
 \nonumber\\
 \mathcal P_{qg}={}&\frac{g^2C_F}{2\pi}
 \int_0^{\bar z_2}\frac{\dd z}{z}\left(\frac{\nu^+}{zl^+}\right)^{\eta_Y}
 \frac{\bar z_2^2+(\bar z_2-z)^2}{\bar z_2^2}
 \mu^{2\epsilon}\int\frac{\dd^{2-2\epsilon}\bell_q}
 {(2\pi)^{2-2\epsilon}}\frac1{\ell_q^2},
 \label{eq:real-r11-definitions}
\end{align}
one obtains the exact regulated identity
\begin{equation}
 \Rij{1}{1}=\mathcal P_{qg}\,\mathcal V_\Born
 =-2\operatorname{Re}\mathcal V_9.
 \label{eq:real-r11-result}
\end{equation}

For the three products in which both amplitudes radiate before the shockwave,
all Wilson lines reduce to the common finite-$N_c$ operator
\begin{align}
 \mathcal C_{bb}={}&C_F\langle
 S_{2'2}-S_{13}S_{32}-S_{2'3}S_{31}+1\rangle_Y
 \nonumber\\[-2pt]
 &-\frac1{2N_c}\langle
 S_{13}S_{32}+S_{2'3}S_{31}-S_{12}-S_{2'1}\rangle_Y,
 \label{eq:real-before-before-color}\\
 \mathcal W_{bb}={}&
 \left[\frac{K_1(QX)K_1(QX')}{XX'}\right]_{\bx_{1'}=\bx_1}
 e^{i\bq\cdot\bx_{2'2}}\mathcal C_{bb}.
 \label{eq:real-before-before-weight}
\end{align}
To keep the cross sections readable, define
\begingroup\small
\begin{align}
 \mathcal H_{44}={}&4\Reh\Bigg\{
 \left([z_2^2+(\bar z_2-z)^2]
 \frac{(\bx_{32})_\epsilon(\bx_{32'})_{\epsilon^*}}{\bx_{32}^2\bx_{32'}^2}
 +\left[(z_2+z)^2+\frac{z_2^2(\bar z_2-z)^2}{(z_2+z)^2}\right]
 \frac{(\bx_{32})_{\epsilon^*}(\bx_{32'})_\epsilon}{\bx_{32}^2\bx_{32'}^2}\right)
 \nonumber\\[-2pt]
 &\hspace{4mm}\times
 (z_2\bx_{21}+z\bx_{31})_\epsilon
 (z_2\bx_{2'1}+z\bx_{31})_{\epsilon^*}\nonumber\\
 &-\frac{z_2^2z(\bar z_2-z)}{2(z_2+z)^2}
 \left[\frac{(\bx_{32})_{\epsilon^*}}{\bx_{32}^2}
 (z_2\bx_{21}+z\bx_{31})_\epsilon
 +\frac{(\bx_{32'})_\epsilon}{\bx_{32'}^2}
 (z_2\bx_{2'1}+z\bx_{31})_{\epsilon^*}\right]
 +\frac{z_2^2z^2}{4(z_2+z)^2}\Bigg\},
 \label{eq:real-h44}\\
 \mathcal H_{33}={}&4\Reh\Bigg\{
 \left([z_2^2+(\bar z_2-z)^2]
 \frac{(\bx_{31})_\epsilon(\bx_{31})_{\epsilon^*}}{\bx_{31}^4}
 +\left[\bar z_2^2+
 \frac{z_2^2(\bar z_2-z)^2}{\bar z_2^2}\right]
 \frac{(\bx_{31})_{\epsilon^*}(\bx_{31})_\epsilon}{\bx_{31}^4}\right)
 \nonumber\\[-2pt]
 &\hspace{4mm}\times
 [((\bar z_2-z)\bx_{12}+z\bx_{32})_\epsilon]
 [((\bar z_2-z)\bx_{12'}+z\bx_{32'})_{\epsilon^*}]
 -\frac{(\bar z_2-z)^2z_2z}{2\bar z_2^2}
 \left[\frac{(\bx_{31})_{\epsilon^*}}{\bx_{31}^2}
 ((\bar z_2-z)\bx_{12}+z\bx_{32})_\epsilon
 \right.\nonumber\\[-2pt]
 &\left.\hspace{43mm}
 +\frac{(\bx_{31})_\epsilon}{\bx_{31}^2}
 ((\bar z_2-z)\bx_{12'}+z\bx_{32'})_{\epsilon^*}\right]
 +\frac{(\bar z_2-z)^2z^2}{4\bar z_2^2}\Bigg\},
 \label{eq:real-h33}\\
 \mathcal H_{43}={}&4\Reh\Bigg\{
 \left([z_2^2+(\bar z_2-z)^2]
 \frac{(\bx_{32})_\epsilon(\bx_{31})_{\epsilon^*}}{\bx_{32}^2\bx_{31}^2}
 +\frac{z_2(\bar z_2-z)}{\bar z_2(z_2+z)}
 [\bar z_2^2+(z_2+z)^2]
 \frac{(\bx_{32})_{\epsilon^*}(\bx_{31})_\epsilon}{\bx_{32}^2\bx_{31}^2}\right)
 \nonumber\\[-2pt]
 &\hspace{4mm}\times(z_2\bx_{21}+z\bx_{31})_\epsilon
 ((\bar z_2-z)\bx_{12'}+z\bx_{32'})_{\epsilon^*}
 -\frac{z(\bar z_2-z)(z_2+z)}{2\bar z_2}
 \frac{(\bx_{32})_{\epsilon^*}}{\bx_{32}^2}
 (z_2\bx_{21}+z\bx_{31})_\epsilon
 \nonumber\\[-2pt]
 &\hspace{4mm}
 -\frac{z_2z\bar z_2}{2(z_2+z)}
 \frac{(\bx_{31})_\epsilon}{\bx_{31}^2}
 ((\bar z_2-z)\bx_{12'}+z\bx_{32'})_{\epsilon^*}\Bigg\}.
 \label{eq:real-h43}
\end{align}
\endgroup
The three fixed-momentum results are then
\begin{align}
 \Rij{4}{4}&=\frac{e^2e_f^2g^2Q^2N_c}{(2\pi)^8}
 \int_0^{\bar z_2}\frac{\dd z}{z}\left(\frac{\nu^+}{zl^+}\right)^{\eta_Y}\,z_2(\bar z_2-z)^2
 \int\dd^8\bx\,\mathcal W_{bb}\mathcal H_{44},
 \label{eq:real-r44-result}\\
 \Rij{3}{3}&=\frac{e^2e_f^2g^2Q^2N_c}{(2\pi)^8}
 \int_0^{\bar z_2}\frac{\dd z}{z}\left(\frac{\nu^+}{zl^+}\right)^{\eta_Y}\,z_2^3
 \int\dd^8\bx\,\mathcal W_{bb}\mathcal H_{33},
 \label{eq:real-r33-result}\\
 \Rij{4}{3}&=\frac{e^2e_f^2g^2Q^2N_c}{(2\pi)^8}
 \int_0^{\bar z_2}\frac{\dd z}{z}\left(\frac{\nu^+}{zl^+}\right)^{\eta_Y}\,z_2^2(\bar z_2-z)
 \int\dd^8\bx\,\mathcal W_{bb}\mathcal H_{43},
 \qquad \Rij{3}{4}=\Rij{4}{3}^{\,*}.
 \label{eq:real-r43-result}
\end{align}

The two interferences involving $A_2$ and a before-shockwave amplitude share
the color operator and photon weight
\begin{align}
 \mathcal C_{2b}={}&C_F\langle
 S_{2'3}S_{32}-S_{2'3}S_{31}-S_{12}+1\rangle_Y
 \nonumber\\[-2pt]
 &-\frac1{2N_c}\langle
 S_{2'2}-S_{2'1}-S_{2'3}S_{32}+S_{2'3}S_{31}\rangle_Y,
 \label{eq:real-mixed-color}\\
 \mathcal W_{2b}={}&
 \left[\frac{K_1(|\bx_{21}|Q_{2z})K_1(QX')}{X'}
 \right]_{\bx_{1'}=\bx_1}
 e^{i\bq\cdot(\bx_{2'2}+\frac z{z_2}\bx_{32})}\mathcal C_{2b}.
 \label{eq:real-mixed-weight}
\end{align}
Their numerator kernels are
\begingroup\scriptsize
\begin{align}
 \mathcal H_{24}={}&4\Reh\Bigg\{
 (z_2+z)[z_2^2+(\bar z_2-z)^2]
 \frac{(\bx_{21})_\epsilon(\bx_{32'})_{\epsilon^*}}
 {|\bx_{21}|\bx_{32'}^2}
 \frac{(\bx_{32})_\epsilon(z_2\bx_{2'1}+z\bx_{31})_{\epsilon^*}}
 {\bx_{32}^2}
 \nonumber\\[-2pt]
 &+\left[(z_2+z)^3+\frac{z_2^2(\bar z_2-z)^2}{z_2+z}\right]
 \frac{(\bx_{21})_\epsilon(\bx_{32'})_\epsilon}
 {|\bx_{21}|\bx_{32'}^2}
 \frac{(\bx_{32})_{\epsilon^*}(z_2\bx_{2'1}+z\bx_{31})_{\epsilon^*}}
 {\bx_{32}^2}
 -\frac{z_2^2(\bar z_2-z)z}{2(z_2+z)}
 \frac{(\bx_{21})_\epsilon}{|\bx_{21}|}
 \frac{(\bx_{32})_{\epsilon^*}}{\bx_{32}^2}\Bigg\},
 \label{eq:real-h24}\\
 \mathcal H_{23}={}&4\Reh\Bigg\{
 (z_2+z)[z_2^2+(\bar z_2-z)^2]
 \frac{(\bx_{21})_\epsilon(\bx_{31})_{\epsilon^*}}
 {|\bx_{21}|\bx_{31}^2}
 \frac{(\bx_{32})_\epsilon
 ((\bar z_2-z)\bx_{12'}+z\bx_{32'})_{\epsilon^*}}
 {\bx_{32}^2}
 \nonumber\\[-2pt]
 &+\frac{z_2(\bar z_2-z)}{\bar z_2}[\bar z_2^2+(z_2+z)^2]
 \frac{(\bx_{21})_\epsilon(\bx_{31})_\epsilon}
 {|\bx_{21}|\bx_{31}^2}
 \frac{(\bx_{32})_{\epsilon^*}
 ((\bar z_2-z)\bx_{12'}+z\bx_{32'})_{\epsilon^*}}
 {\bx_{32}^2}
 -\frac{(\bar z_2-z)z(z_2+z)^2}{2\bar z_2}
 \frac{(\bx_{21})_\epsilon}{|\bx_{21}|}
 \frac{(\bx_{32})_{\epsilon^*}}{\bx_{32}^2}\Bigg\}.
 \label{eq:real-h23}
\end{align}
\endgroup
The corresponding products are
\begin{align}
 \Rij{2}{4}&=-\frac{e^2e_f^2g^2Q^2N_c}{(2\pi)^8}
 \int_0^{\bar z_2}\frac{\dd z}{z}\left(\frac{\nu^+}{zl^+}\right)^{\eta_Y}
 (\bar z_2-z)\sqrt{(\bar z_2-z)(z_2+z)}
 \int\dd^8\bx\,\mathcal W_{2b}\mathcal H_{24},
 \nonumber\\
 \Rij{4}{2}&=\Rij{2}{4}^{\,*},
 \label{eq:real-r24-result}\\
 \Rij{2}{3}&=-\frac{e^2e_f^2g^2Q^2N_c}{(2\pi)^8}
 \int_0^{\bar z_2}\frac{\dd z}{z}\left(\frac{\nu^+}{zl^+}\right)^{\eta_Y}
 z_2\sqrt{(\bar z_2-z)(z_2+z)}
 \int\dd^8\bx\,\mathcal W_{2b}\mathcal H_{23},
 \nonumber\\
 \Rij{3}{2}&=\Rij{2}{3}^{\,*}.
 \label{eq:real-r23-result}
\end{align}

The last two interference pairs are inclusive cuts of virtual graphs already
displayed above.  Equality of their residues, phases, helicity numerators and
finite-$N_c$ traces gives the regulated, pointwise relations
\begin{align}
 \Rij{1}{3}&=-\mathcal V_5^*,&\Rij{3}{1}&=-\mathcal V_5,
 \label{eq:real-r13-result}\\
 \Rij{1}{4}&=-\mathcal V_7^*,&\Rij{4}{1}&=-\mathcal V_7.
 \label{eq:real-r14-result}
\end{align}
The explicit coordinate kernels on the right-hand sides are those in
Eqs.~\eqref{eq:g5-result} and \eqref{eq:g7-result}; repeating them would obscure
the cut identities.

\paragraph{Measured real sum.}
Restoring both orderings of every interference, the fixed-antiquark real kernel
is
\begin{align}
 \mathcal R_{\bar q,R}={}&
 \Rij{1}{1}+\Rij{2}{2}+\Rij{3}{3}+\Rij{4}{4}
 +\Rij{2}{1}+\Rij{1}{2}+\Rij{3}{1}+\Rij{1}{3}
 \nonumber\\[-2pt]
 &+\Rij{4}{1}+\Rij{1}{4}+\Rij{3}{2}+\Rij{2}{3}
 +\Rij{4}{2}+\Rij{2}{4}+\Rij{4}{3}+\Rij{3}{4}.
 \label{eq:real-complete-sum}
\end{align}
The exact inclusive relations
\begin{align}
 \Rij{1}{1}+2\operatorname{Re}\mathcal V_9&=0,
 &\Rij{1}{3}+\mathcal V_5^*&=0,
 &\Rij{3}{1}+\mathcal V_5&=0,
 \nonumber\\[-2pt]
 \Rij{1}{4}+\mathcal V_7^*&=0,
 &\Rij{4}{1}+\mathcal V_7&=0
 \label{eq:real-exact-inclusive-relations}
\end{align}
hold before any singular limit is taken and will be used in the
real--virtual analysis of the complete sum in Eq.~\eqref{eq:real-complete-sum}.

At fixed antiquark momentum the detector is applied only after the complete
three-body kernel has been assembled:
\begin{align}
 \Sigma^T_{\bar q,R}(\theta)
 &=\int\frac{\dd z_2}{z_2}\int\dd^2\bq\,
 \mathcal M_{\bar q}(\theta;z_2,\bq)
 \mathcal R_{\bar q,R}(z_2,\bq)
 \nonumber\\
 &=\frac{x_BQ^2\sin\theta}{8\theta}
 \int_0^{\csc^2(\theta/2)}\dd v\,v^2\int_0^{2\pi}\dd\phi\,
 \mathcal R_{\bar q,R}\left(
 v\sin^2\frac\theta2,
 \frac{Qv}{2}\sin\theta\,\boldsymbol e_\phi\right).
 \label{eq:real-measured-sum}
\end{align}
As in the Born and virtual terms, the first line leaves the $z_2$ range
implicit; after resolving the measurement delta function the change of
variables $z_2=v\sin^2(\theta/2)$ gives
$0<v<\csc^2(\theta/2)$. The same range follows separately for each of
the three tags from the original linear measurement.

\paragraph{Quark-tagged and gluon-tagged real terms.}
The other two detector tags are constructed directly from the same four
amplitudes, before either unobserved momentum is integrated.  Keeping the quark
momentum $(z_1,\bp)$ fixed gives
\begin{align}
 \mathcal R_{q,R}(z_1,\bp)={}&
 \sum_{i,j=1}^{4}
 \int_0^{1-z_1}\frac{\dd z}{z}\left(\frac{\nu^+}{zl^+}\right)^{\eta_Y}
 \int\dd^2\bq\,\dd^2\bk\,
 \frac{1}{16(l^+)^2(2\pi)^8(1-z_1-z)}
 \nonumber\\[-2pt]
 &\times\frac12
 \sum_{\lambda,h,\bar h,\lambda_g,a}
 \left.
 \left\langle\tr\!\left[(i\mathcal M_i^a)
 (i\mathcal M_j^a)^\dagger\right]\right\rangle_Y
 \right|_{z_2=1-z_1-z} .
 \label{eq:real-fixed-quark-kernel}
\end{align}
This construction retains the Wilson-line ordering of the quark-tagged
amplitudes before the target average. Keeping the gluon momentum $(z,\bk)$ fixed gives
\begin{align}
 \mathcal R_{g,R}(z,\bk)={}&
 \left(\frac{\nu^+}{zl^+}\right)^{\eta_Y}
 \sum_{i,j=1}^{4}
 \int_0^{1-z}\frac{\dd z_1}{z_1(1-z-z_1)}
 \int\dd^2\bp\,\dd^2\bq\,
 \frac{1}{16(l^+)^2(2\pi)^8}
 \nonumber\\[-2pt]
 &\times\frac12
 \sum_{\lambda,h,\bar h,\lambda_g,a}
 \left.
 \left\langle\tr\!\left[(i\mathcal M_i^a)
 (i\mathcal M_j^a)^\dagger\right]\right\rangle_Y
 \right|_{z_2=1-z-z_1} .
 \label{eq:real-fixed-gluon-kernel}
\end{align}
Equation~\eqref{eq:real-fixed-gluon-kernel} contains the two fermion--gluon
collinear regions, the leading small-angle region, and the finite noncollinear
remainder.

Resolving the detector delta function separately in the three tags gives the
complete real-emission energy flow
\begin{align}
 \Sigma^T_{E,R}(\theta)={}&
 \frac{x_BQ^2\sin\theta}{8\theta}
 \int_0^{\csc^2(\theta/2)}\dd v\,v^2\int_0^{2\pi}\dd\phi\,
 \Bigg[
 \mathcal R_{q,R}\left(
 v\sin^2\frac\theta2,
 \frac{Qv}{2}\sin\theta\,\boldsymbol e_\phi\right)
 \nonumber\\[-2pt]
 &\hspace{22mm}+
 \mathcal R_{\bar q,R}\left(
 v\sin^2\frac\theta2,
 \frac{Qv}{2}\sin\theta\,\boldsymbol e_\phi\right)
 +\mathcal R_{g,R}\left(
 v\sin^2\frac\theta2,
 \frac{Qv}{2}\sin\theta\,\boldsymbol e_\phi\right)
 \Bigg].
 \label{eq:complete-real-energy-flow}
\end{align}

\section{Singularity structure and small-\texorpdfstring{$x$}{x} factorization}
\label{sec:singularity-structure}

This section starts with the ultraviolet and rapidity limits because the two
limits have a common boundary.  At fixed nonzero gluon fraction, a transverse
loop momentum taken to infinity is ultraviolet.  In coordinate space the same
limit is a coincident emission point, $\bx_3\to\bx_i$.  At fixed nonzero
transverse momentum, $z\to0$ is a rapidity limit.  The simultaneous limit
$\bx_3\to\bx_i$ and $z\to0$ is therefore counted by both approximations unless
it is displayed separately. We also analyze soft and final-state collinear
singularities. In the soft limit the complete gluon momentum vanishes,
$z\to0$ and $\bk\to0$, and the soft poles cancel between real and virtual
contributions. In a final-state collinear limit the quark--gluon or
antiquark--gluon relative transverse momentum vanishes at fixed splitting
fraction; the two daughter energy weights add to the parent weight and
the collinear poles cancel.

\subsection{Coupled ultraviolet and rapidity limits}
\label{sec:uv-rapidity-coupled}

\subsubsection{Decomposition and order of the checks}

Let $T_U$ extract the leading coincident-coordinate or large-transverse-
momentum term of a regulated graph, and let $T_Y$ extract the coefficient of
$\dd z/z$ at $z=0$ while all transverse coordinates are held fixed.  For the
local logarithmic coefficients tabulated below, the two iterated residue
extractions give the same common term. We evaluate it from the displayed
graph kernel to confirm:
\begin{equation}
 T_{UY}\mathcal I_G\equiv T_UT_Y\mathcal I_G
 =T_YT_U\mathcal I_G .
 \label{eq:joint-uv-rap-projector}
\end{equation}
Every real or virtual contribution is decomposed as
\begin{align}
 \mathcal I_G={}&
 \left(\mathcal I_G-T_U\mathcal I_G-T_Y\mathcal I_G
 +T_{UY}\mathcal I_G\right)
 +\left(T_U\mathcal I_G-T_{UY}\mathcal I_G\right)
 \nonumber\\
 &+\left(T_Y\mathcal I_G-T_{UY}\mathcal I_G\right)
 +T_{UY}\mathcal I_G .
 \label{eq:joint-uv-rap-decomposition}
\end{align}
The four terms in Eq.~\eqref{eq:joint-uv-rap-decomposition} are the remainder
finite in both limits, the pure ultraviolet term, the pure rapidity term, and
the ultraviolet--rapidity overlap.

The calculation keeps the transverse and rapidity regulators distinct.
Transverse integrals are continued to $d_\perp=2-2\epsilon$ dimensions,
\begin{equation}
 J_{\rm UV}\equiv
 \left.\mu^{2\epsilon}\int
 \frac{\dd^{2-2\epsilon}\bell}{(2\pi)^{2-2\epsilon}}
 \frac1{\ell^2}\right|_{\rm UV}
 =\frac1{4\pi\epsilon_{\rm UV}}.
 \label{eq:joint-uv-master-integral}
\end{equation}
We use $\alpha_s=g^2/(4\pi)$ throughout this section.
Every longitudinal gluon integral contains the analytic factor displayed in
Eq.~\eqref{eq:analytic-rapidity-regulator}.  Thus a graph whose physical upper
endpoint is $z_{\max}$ obeys Eq.~\eqref{eq:analytic-rapidity-expansion}.  The
coefficient of $-1/\eta_Y$ is its rapidity pole, while
$F(0)\ln(z_{\max}l^+/\nu^+)$ and the endpoint-subtracted longitudinal integral
are retained in the finite NLO impact factor.

The common Born color operator for an antiquark-tagged kernel is
\begin{equation}
 \mathcal C_0=
 \left\langle S_{2'2}-S_{12}-S_{2'1}+1\right\rangle_Y,
 \label{eq:joint-born-color}
\end{equation}
and the Born kernel multiplying a transverse UV pole is written with its
physical normalization as
\begin{equation}
 \mathcal U_{\Born}(z_2,\bq)=
 \frac{g^2C_F}{2\pi}\,
 \mathcal V_{\Born}(z_2,\bq)J_{\rm UV}.
 \label{eq:joint-born-uv-kernel}
\end{equation}
For the rapidity residues we keep the complete hard factor. After setting $z_1=1-z_2$ in the coefficient of
$\dd z/z$, define
\begin{align}
 \mathcal W_T={}&
 \frac{e^2e_f^2Q^2N_c}{(2\pi)^5}
 z_1z_2^2(z_1^2+z_2^2)e^{i\bq\cdot\bx_{2'2}}
 \frac{\bx_{12}\cdot\bx_{1 2'}}
 {|\bx_{12}|\,|\bx_{1 2'}|}
 K_1(Q\sqrt{z_1z_2}|\bx_{12}|)
 K_1(Q\sqrt{z_1z_2}|\bx_{1 2'}|),
 \nonumber\\
 \dd\Gamma_Y={}&
 \frac{\alpha_s}{2\pi^2}\frac{\dd z}{z}
 \left(\frac{\nu^+}{zl^+}\right)^{\eta_Y}
 \dd^2\bx_1\dd^2\bx_2\dd^2\bx_{2'}\dd^2\bx_3\,
 \mathcal W_T,
 \label{eq:joint-rapidity-measure}\\
 \rho_i={}&\frac1{\bx_{i3}^2},\qquad
 D_{ij}=\frac{\bx_{3i}\cdot\bx_{3j}}
 {\bx_{3i}^2\bx_{3j}^2},\qquad
 K_{ij}=\rho_i+\rho_j-2D_{ij}
 =\frac{\bx_{ij}^2}{\bx_{i3}^2\bx_{j3}^2}.
 \label{eq:joint-transverse-kernels}
\end{align}
Whenever a rapidity coefficient contains $\mathcal C(\bx_3)/\bx_{3i}^2$, its
pure rapidity part is the explicit local subtraction
\begin{equation}
 \left[\frac{\mathcal C(\bx_3)}{\bx_{3i}^2}\right]_{U\text{-sub}}
 \equiv
 \frac{\mathcal C(\bx_3)
 -\mathcal C(\bx_i)\Theta(1-\mu^2\bx_{3i}^2)}
 {\bx_{3i}^2}.
 \label{eq:joint-coordinate-uv-subtraction}
\end{equation}
The term proportional to $\mathcal C(\bx_i)$ in
Eq.~\eqref{eq:joint-coordinate-uv-subtraction} is retained as
$T_{UY}\mathcal I_G$ and included in the overlap sum.

Equation~\eqref{eq:joint-coordinate-uv-subtraction} organizes the singular
ledger. A finite change of its local profile must be compensated between
the real remainder and the integrated virtual term.

The verification will be performed in the following order:
\begin{align}
 &\sum_G\left[T_{UY}\mathcal I_G\right]_{1/(\epsilon_{\rm UV}\eta_Y)}=0,
 \label{eq:joint-check-overlap}\\
 &\sum_G\left[\left(T_U-T_{UY}\right)\mathcal I_G\right]_{1/\epsilon_{\rm UV}}=0,
 \label{eq:joint-check-pure-uv}\\
 &\sum_G\left[T_Y\mathcal I_G\right]_{-1/\eta_Y}
 =H_{\rm JIMWLK}\mathcal V_{\Born}.
 \label{eq:joint-check-jimwlk}
\end{align}
In the last line, $H_{\rm JIMWLK}$ acts on the target operator in
the Born kernel $\mathcal V_\Born$ of Eq.~\eqref{eq:born-result}.

\subsubsection{Virtual example: graph \texorpdfstring{$A_5$}{A5}}
\label{sec:joint-virtual-example}

We begin with the exact one-sided kernel in Eq.~\eqref{eq:g5-result}.  The
spectator integral in Eq.~\eqref{eq:g5-spectator-delta} imposes
\begin{equation}
 \bx_{1'}=\bx_1+\frac{z}{z_1}\bx_{31}.
 \label{eq:joint-a5-spectator}
\end{equation}
For the ultraviolet limit introduce the physical separation
$\boldsymbol r=\bx_{31}$ and keep $z>0$ fixed.  Equation
\eqref{eq:joint-a5-spectator} and the photon size appearing in the exact graph
give
\begin{align}
 \bx_{1'}&\xrightarrow[\boldsymbol r\to0]{}\bx_1,
 &\bx_{1'2'}&\xrightarrow[\boldsymbol r\to0]{}\bx_{12'},
 \nonumber\\
 X_5^2&=z_2(z_1-z)\bx_{12}^2
 +z(z_1-z)\boldsymbol r^2+z_2z\bx_{23}^2
 \xrightarrow[\boldsymbol r\to0]{}
 z_1z_2\bx_{12}^2.
 \label{eq:joint-a5-uv-kinematics}
\end{align}
The finite-$N_c$ target operator in Eq.~\eqref{eq:g5-color} becomes
\begin{equation}
 \mathcal C_5\xrightarrow[\bx_3\to\bx_1]{}
 C_F\left\langle S_{2'2}-S_{12}-S_{2'1}+1\right\rangle_Y
 =C_F\mathcal C_0.
 \label{eq:joint-a5-uv-color}
\end{equation}
Keeping its full $z$ dependence, the numerator in Eq.~\eqref{eq:g5-result} reduces to
\begin{align}
 &\bx_{1'2'}\cdot\Bigg\{
 (z_1^2+z_2^2)[z_1^2+(z_1-z)^2]
 \left(\bx_{12}+\frac{z}{z_1}\boldsymbol r\right)
 -\frac{z_2z(z_1-z)^2}{z_1}\boldsymbol r\Bigg\}
 \nonumber\\
 &\hspace{12mm}\xrightarrow[\boldsymbol r\to0]{}
 (z_1^2+z_2^2)[z_1^2+(z_1-z)^2]
 \bx_{12'}\cdot\bx_{12}.
 \label{eq:joint-a5-uv-numerator}
\end{align}
All remaining factors reproduce the Born wave-function overlap.  The local
integral is
\begin{equation}
 \left.\mu^{-2\epsilon}
 \int\frac{\dd^{2-2\epsilon}\boldsymbol r}{r^2}
 \right|_{\rm UV}=(2\pi)^2J_{\rm UV}.
 \label{eq:joint-a5-local-integral}
\end{equation}
Consequently the complete fixed-$z$ ultraviolet term is
\begin{align}
 T_U\mathcal V_5
 ={}&\mathcal U_{\Born}
 \int_0^{z_1}\frac{\dd z}{z}
 \left(\frac{\nu^+}{zl^+}\right)^{\eta_Y}
 \left[1+\left(1-\frac{z}{z_1}\right)^2\right]
 \nonumber\\
 ={}&\left[-\frac{2}{\eta_Y}
 +2\ln\frac{z_1l^+}{\nu^+}-\frac32\right]
 \mathcal U_{\Born}+\mathcal O(\eta_Y).
 \label{eq:joint-a5-full-uv}
\end{align}

We next take the rapidity limit directly in Eq.~\eqref{eq:g5-result}, before
the coincident-coordinate limit.  The coordinate constraint, photon size and
numerator become
\begin{align}
 \bx_{1'}&\xrightarrow[z\to0]{}\bx_1,
 &X_5^2&\xrightarrow[z\to0]{}z_1z_2\bx_{12}^2,
 \nonumber\\
 &\bx_{1'2'}\cdot\Bigg\{
 (z_1^2+z_2^2)[z_1^2+(z_1-z)^2]
 \left(\bx_{12}+\frac{z}{z_1}\bx_{31}\right)
 -\frac{z_2z(z_1-z)^2}{z_1}\bx_{31}\Bigg\}
 \nonumber\\[-2pt]
 &\hspace{12mm}\xrightarrow[z\to0]{}
 2z_1^2(z_1^2+z_2^2)\bx_{12'}\cdot\bx_{12}.
 \label{eq:joint-a5-rapidity-kinematics}
\end{align}
The color operator that remains in this limit is, explicitly,
\begin{align}
 \mathcal C_5^{(0)}=\Bigg\langle&
 \frac{N_c}{2}S_{13}(S_{322'1}-S_{32})
 +C_F(1-S_{2'1})
 -\frac{S_{122'1}-S_{12}}{2N_c}
 \Bigg\rangle_Y.
 \label{eq:joint-a5-rapidity-color}
\end{align}
Substitution of Eq.~\eqref{eq:joint-a5-rapidity-kinematics} into the exact
normalization in Eq.~\eqref{eq:g5-result} gives the endpoint density
below, where $\rho_1=1/\bx_{13}^2$ is the kernel defined in
Eq.~\eqref{eq:joint-transverse-kernels}:
\begin{equation}
 T_Y\mathcal V_5
 =\dd\Gamma_Y\,2\mathcal C_5^{(0)}\rho_1.
 \label{eq:joint-a5-full-rapidity}
\end{equation}

The overlap follows by taking $\bx_3\to\bx_1$ in
Eq.~\eqref{eq:joint-a5-full-rapidity}.  Direct multiplication of the Wilson
lines gives
\begin{equation}
 \mathcal C_5^{(0)}\xrightarrow[\bx_3\to\bx_1]{}
 C_F\mathcal C_0,
 \qquad
 T_UT_Y\mathcal V_5
 =\dd\Gamma_Y\,2C_F\mathcal C_0\rho_1.
 \label{eq:joint-a5-overlap-density}
\end{equation}
Integrating the common $1/r^2$ term over the full physical longitudinal range
with the analytic regulator gives
\begin{align}
 T_{UY}\mathcal V_5
 ={}&2\mathcal U_{\Born}
 \int_0^{z_1}\frac{\dd z}{z}
 \left(\frac{\nu^+}{zl^+}\right)^{\eta_Y}
 \nonumber\\
 ={}&2\left[-\frac1{\eta_Y}
 +\ln\frac{z_1l^+}{\nu^+}\right]\mathcal U_{\Born}
 +\mathcal O(\eta_Y).
 \label{eq:joint-a5-overlap}
\end{align}
Subtracting Eq.~\eqref{eq:joint-a5-overlap} from
Eq.~\eqref{eq:joint-a5-full-uv} gives the pure ultraviolet pole
\begin{align}
 (T_U-T_{UY})\mathcal V_5
 ={}&\mathcal U_{\Born}
 \int_0^{z_1}\frac{\dd z}{z}
 \left(\frac{\nu^+}{zl^+}\right)^{\eta_Y}
 \left[\left(1-\frac{z}{z_1}\right)^2-1\right]
 \nonumber\\
 ={}&-\frac32\mathcal U_{\Born}+\mathcal O(\eta_Y).
 \label{eq:joint-a5-pure-uv}
\end{align}
The pure rapidity term is equally explicit.  Its pole and finite analytic
endpoint logarithm are both retained:
\begin{align}
 (T_Y-T_{UY})\mathcal V_5
 ={}&\frac{\alpha_s}{2\pi^2}
 \int_0^{z_1}\frac{\dd z}{z}
 \left(\frac{\nu^+}{zl^+}\right)^{\eta_Y}
 \int\dd^2\bx_1\dd^2\bx_2\dd^2\bx_{2'}\dd^2\bx_3\,
 \mathcal W_T\,
 2\left[\mathcal C_5^{(0)}\rho_1\right]_{U\text{-sub}}
 \nonumber\\
 ={}&\frac{\alpha_s}{2\pi^2}
 \left[-\frac1{\eta_Y}+\ln\frac{z_1l^+}{\nu^+}\right]
 \int\dd^2\bx_1\dd^2\bx_2\dd^2\bx_{2'}\dd^2\bx_3\,
 \mathcal W_T\,
 2\left[\mathcal C_5^{(0)}\rho_1\right]_{U\text{-sub}}
 +\mathcal O(\eta_Y).
 \label{eq:joint-a5-pure-rapidity}
\end{align}
Equations~\eqref{eq:joint-a5-overlap},
\eqref{eq:joint-a5-pure-uv}, and
\eqref{eq:joint-a5-pure-rapidity} are obtained independently from the exact
graph.  Their sum with the finite remainder reconstructs
Eq.~\eqref{eq:g5-result}.
The same antiquark detector insertion as in
Eq.~\eqref{eq:g5-measured-final} acts on each of the three pieces.

\subsubsection{Real example: \texorpdfstring{$\mathcal R_{3\times3}$}{R3x3}}
\label{sec:joint-real-example}

The starting point is the exact real cross section in
Eq.~\eqref{eq:real-r33-result}, including its three-body phase space.
Here $\bar z_2=1-z_2$ is defined in
Eq.~\eqref{eq:real-longitudinal-definitions}; $\mathcal C_{bb}$,
$\mathcal W_{bb}$ and $\mathcal H_{33}$ are defined in
Eqs.~\eqref{eq:real-before-before-color},
\eqref{eq:real-before-before-weight} and~\eqref{eq:real-h33}:

\begin{equation}
 \Rij{3}{3}=
 \frac{e^2e_f^2g^2Q^2N_c}{(2\pi)^8}
 \int_0^{\bar z_2}\frac{\dd z}{z}
 \left(\frac{\nu^+}{zl^+}\right)^{\eta_Y}z_2^3
 \int\dd^8\bx\,\mathcal W_{bb}\mathcal H_{33}.
 \label{eq:joint-r33-start}
\end{equation}
For the ultraviolet limit put $\boldsymbol r=\bx_{31}$ and keep $z>0$.
The Wilson-line operator, the photon sizes and the numerator then give
\begin{align}
 \mathcal C_{bb}&\xrightarrow[\bx_3\to\bx_1]{}C_F\mathcal C_0,
 &X&\longrightarrow\sqrt{z_2\bar z_2}\,|\bx_{12}|,
 &X'&\longrightarrow\sqrt{z_2\bar z_2}\,|\bx_{12'}|,
 \nonumber\\
 \mathcal H_{33}&\xrightarrow[\boldsymbol r\to0]{}
 \frac{[\bar z_2^2+(\bar z_2-z)^2]
 [\bar z_2^2+z_2^2]}{r^2}
 \bx_{12}\cdot\bx_{12'}.
 \label{eq:joint-r33-uv-limit}
\end{align}
Using Eq.~\eqref{eq:joint-a5-local-integral} and comparing every remaining
factor with the Born kernel gives
\begin{align}
 T_U\Rij{3}{3}
 ={}\mathcal U_{\Born}
 \int_0^{\bar z_2}\frac{\dd z}{z}
 \left(\frac{\nu^+}{zl^+}\right)^{\eta_Y}
 \left[1+\left(1-\frac{z}{\bar z_2}\right)^2\right]
 \nonumber\\
 ={}\left[-\frac2{\eta_Y}+2\ln\frac{\bar z_2l^+}{\nu^+}
 -\frac32\right]\mathcal U_{\Born}+\mathcal O(\eta_Y).
 \label{eq:joint-r33-full-uv}
\end{align}
The Born momentum fraction in
Eq.~\eqref{eq:joint-r33-full-uv} is $z_1^{\Born}=\bar z_2$.

For the rapidity limit, no coordinate is collapsed.  Directly in
Eqs.~\eqref{eq:real-coordinate-definitions},
\eqref{eq:real-before-before-weight}, and \eqref{eq:real-h33},
\begin{align}
 X&\xrightarrow[z\to0]{}\sqrt{z_1z_2}|\bx_{12}|,
 &X'&\xrightarrow[z\to0]{}\sqrt{z_1z_2}|\bx_{12'}|,
 \nonumber\\
 ((\bar z_2-z)\bx_{12}+z\bx_{32})&\xrightarrow[z\to0]{}
 z_1\bx_{12},
 &((\bar z_2-z)\bx_{12'}+z\bx_{32'})&\xrightarrow[z\to0]{}
 z_1\bx_{12'}.
 \label{eq:joint-r33-rapidity-kinematics}
\end{align}
The two terms containing one explicit power of $z$ and the contact term
containing $z^2$ vanish.  Using
$x_{31,\epsilon}x_{31,\epsilon^*}=\bx_{31}^2/2$ and
$2\operatorname{Re}_\epsilon(x_{12,\epsilon}
x_{12',\epsilon^*})=\bx_{12}\cdot\bx_{12'}$, the first line of
Eq.~\eqref{eq:real-h33} reduces to the Born transverse overlap times
$2\rho_1$.  Hence the exact $z=0$ coefficient is
\begin{equation}
 T_Y\Rij{3}{3}
 =\dd\Gamma_Y\,2\mathcal C_{bb}^{(0)}\rho_1,
 \label{eq:joint-r33-full-rapidity}
\end{equation}
where the physical finite-$N_c$ operator is
\begin{align}
 \mathcal C_{bb}^{(0)}={}&C_F\left\langle
 S_{2'2}-S_{13}S_{32}-S_{2'3}S_{31}+1\right\rangle_Y
 \nonumber\\[-2pt]
 &-\frac1{2N_c}\left\langle
 S_{13}S_{32}+S_{2'3}S_{31}-S_{12}-S_{2'1}\right\rangle_Y.
 \label{eq:joint-r33-rapidity-color}
\end{align}
At $\bx_3=\bx_1$, direct use of $V_1V_1^\dagger=1$ gives
\begin{equation}
 \mathcal C_{bb}^{(0)}\xrightarrow[\bx_3\to\bx_1]{}
 C_F\mathcal C_0.
 \label{eq:joint-r33-overlap-color}
\end{equation}
The overlap, pure ultraviolet term and pure rapidity term are therefore
\begin{align}
 T_{UY}\Rij{3}{3}
 &={}2\left[-\frac1{\eta_Y}+\ln\frac{\bar z_2l^+}{\nu^+}\right]
 \mathcal U_{\Born}+\mathcal O(\eta_Y),
 \label{eq:joint-r33-overlap}\\
 (T_U-T_{UY})\Rij{3}{3}
 &={}-\frac32\mathcal U_{\Born}+\mathcal O(\eta_Y),
 \label{eq:joint-r33-pure-uv}\\
 (T_Y-T_{UY})\Rij{3}{3}
 &={}\frac{\alpha_s}{2\pi^2}
 \left[-\frac1{\eta_Y}+\ln\frac{\bar z_2l^+}{\nu^+}\right]
 \int\dd^2\bx_1\dd^2\bx_2\dd^2\bx_{2'}\dd^2\bx_3\,
 \nonumber\\[-2pt]
 &\hspace{12mm}\times\mathcal W_T\,
 2\left[\mathcal C_{bb}^{(0)}\rho_1\right]_{U\text{-sub}}
 +\mathcal O(\eta_Y).
 \label{eq:joint-r33-pure-rapidity}
\end{align}
The mixed term in Eq.~\eqref{eq:joint-r33-overlap} was obtained twice: from
the $z\to0$ endpoint of Eq.~\eqref{eq:joint-r33-full-uv} and from the
$\bx_3\to\bx_1$ endpoint of Eq.~\eqref{eq:joint-r33-full-rapidity}.  Their
agreement explicitly verifies commutativity for this real graph.
The same detector operation acts on all three real pieces,
\begin{align}
 [\Sigma^T_{\bar q,3\times3}(\theta)]_X
 ={}&\frac{x_BQ^2\sin\theta}{8\theta}
 \int_0^{\csc^2(\theta/2)}\dd v\,v^2\int_0^{2\pi}\dd\phi\,
 [\Rij{3}{3}]_X\left(
 v\sin^2\frac\theta2,
 \frac{Qv}{2}\sin\theta\,\boldsymbol e_\phi\right),
 \nonumber\\[-2pt]
 &X=UY,\quad U\setminus Y,\quad Y\setminus U.
 \label{eq:joint-r33-measured-three-pieces}
\end{align}
The extra $\int\dd z/z\,\dd^2\bx_3$ real-emission phase space remains inside
every entry on the right-hand side.

\subsubsection{Graph-by-graph ledger}
\label{sec:joint-graph-ledger}

We now list the overlap, pure ultraviolet and pure rapidity contributions
of every antiquark-tagged graph. For clarity, the color operators entering
the $z=0$ residues are written in full:
\begin{align}
 \mathcal C_{21}^{(0)}={}&C_F\left\langle
 S_{12}S_{2'1}-S_{12}-S_{2'1}+1\right\rangle_Y
 -\frac1{2N_c}\left\langle
 S_{122'1}-S_{12}S_{2'1}\right\rangle_Y,
 \label{eq:joint-c21-color}\\
 \mathcal C_{2b}^{(0)}={}&
 \frac{N_c}{2}\left\langle S_{2'3}S_{32}-S_{2'3}S_{31}\right\rangle_Y
 +C_F\left\langle1-S_{12}\right\rangle_Y
 -\frac1{2N_c}\left\langle S_{2'2}-S_{2'1}\right\rangle_Y,
 \label{eq:joint-c2b-color}\\
 \mathcal C_6^{(0)}={}&
 \frac{N_c}{2}\left\langle S_{2'3}S_{32}-S_{13}S_{32}\right\rangle_Y
 +C_F\left\langle1-S_{2'1}\right\rangle_Y
 -\frac1{2N_c}\left\langle S_{2'2}-S_{12}\right\rangle_Y.
 \label{eq:joint-c6-color}
\end{align}
The operator $\mathcal C_5^{(0)}$ is given explicitly in
Eq.~\eqref{eq:joint-a5-rapidity-color}, and
$\mathcal C_{bb}^{(0)}$ is given in
Eq.~\eqref{eq:joint-r33-rapidity-color}.  A dagger on an entry means complex
conjugation followed by the dummy-coordinate interchange
$\bx_2\leftrightarrow\bx_{2'}$.  Thus
$D_{12}\leftrightarrow D_{2'1}$, while $D_{22'}$ is invariant under this
change of integration variables in the displayed cross section.

\paragraph{Virtual cross sections.}
The overlap terms obtained from the short-distance endpoint of each direct
$z=0$ residue, including their finite analytic endpoint logarithms, are
\begin{align}
 T_{UY}\mathcal V_5
 &=+2\left[-\frac1{\eta_Y}+\ln\frac{z_1l^+}{\nu^+}\right]
 \mathcal U_{\Born},&
 T_{UY}\mathcal V_6
 &=+2\left[-\frac1{\eta_Y}+\ln\frac{z_2l^+}{\nu^+}\right]
 \mathcal U_{\Born},
 \nonumber\\
 T_{UY}\mathcal V_9
 &=-\left[-\frac1{\eta_Y}+\ln\frac{z_1l^+}{\nu^+}\right]
 \mathcal U_{\Born},&
 T_{UY}\mathcal V_{10}
 &=-\left[-\frac1{\eta_Y}+\ln\frac{z_2l^+}{\nu^+}\right]
 \mathcal U_{\Born},
 \nonumber\\
 T_{UY}\mathcal V_{11}
 &=-2\left[-\frac1{\eta_Y}+\ln\frac{z_1l^+}{\nu^+}\right]
 \mathcal U_{\Born},&
 T_{UY}\mathcal V_{12}
 &=-2\left[-\frac1{\eta_Y}+\ln\frac{z_2l^+}{\nu^+}\right]
 \mathcal U_{\Born},
 \nonumber\\
 T_{UY}\mathcal V_{14(1)}
 &=+\left[-\frac1{\eta_Y}+\ln\frac{z_1l^+}{\nu^+}\right]
 \mathcal U_{\Born},&
 T_{UY}\mathcal V_{14(2)}
 &=+\left[-\frac1{\eta_Y}+\ln\frac{z_2l^+}{\nu^+}\right]
 \mathcal U_{\Born},
 \label{eq:joint-virtual-overlap-list}\\
 T_{UY}\mathcal V_7&=T_{UY}\mathcal V_8
 =T_{UY}\mathcal V_{13(1)}=T_{UY}\mathcal V_{13(2)}=0.
 \nonumber
\end{align}
Every nonzero entry in Eq.~\eqref{eq:joint-virtual-overlap-list} has an omitted
$\mathcal O(\eta_Y)$ remainder.  For example, the coefficient $+1$ in each
graph-14 sector follows by setting $z=0$ in its explicit fixed-$z$ UV
numerator
\begin{equation}
 \frac{z_1^2+z_2^2-2z_jz+z^2}{z_1^2+z_2^2}
 \xrightarrow[z\to0]{}1,
 \qquad j=1,2,
 \label{eq:joint-v14-overlap-origin}
\end{equation}
whereas $D_{12}$ is locally proportional to $1/|\bx_{31}|$ or
$1/|\bx_{32}|$ and produces no logarithmic transverse UV pole.

Subtracting Eq.~\eqref{eq:joint-virtual-overlap-list} from the complete
analytically regulated fixed-$z$ UV terms gives
\begin{align}
 (T_U-T_{UY})\mathcal V_5
 &=-\frac32\mathcal U_{\Born},&
 (T_U-T_{UY})\mathcal V_6
 &=-\frac32\mathcal U_{\Born},
 \nonumber\\
 (T_U-T_{UY})\mathcal V_9
 &=+\frac34\mathcal U_{\Born},&
 (T_U-T_{UY})\mathcal V_{10}
 &=+\frac34\mathcal U_{\Born},
 \nonumber\\
 (T_U-T_{UY})\mathcal V_{11}
 &=+\frac32\mathcal U_{\Born},&
 (T_U-T_{UY})\mathcal V_{12}
 &=+\frac32\mathcal U_{\Born},
 \nonumber\\
 (T_U-T_{UY})\mathcal V_{14(j)}
 &=-\frac32\frac{z_j^2}{z_1^2+z_2^2}\mathcal U_{\Born},
 \label{eq:joint-virtual-pure-uv-list}\\
 (T_U-T_{UY})\mathcal V_7
 &=(T_U-T_{UY})\mathcal V_8=0,
 \nonumber\\
 (T_U-T_{UY})\mathcal V_{13(1)}
 &=(T_U-T_{UY})\mathcal V_{13(2)}=0.
 \nonumber
\end{align}
In the graph-14 line of
Eq.~\eqref{eq:joint-virtual-pure-uv-list}, $j=1,2$.
All entries in Eq.~\eqref{eq:joint-virtual-pure-uv-list} are the complete
$\eta_Y^0$ coefficients of the transverse UV pole.  Their dependence on the
rapidity scale has canceled between $T_U$ and $T_{UY}$ graph by graph.

The pure rapidity coefficients, in units of the regulated measure
$\dd\Gamma_Y$, are listed graph by graph in
Table~\ref{tab:joint-virtual-rapidity-ledger}.  The endpoint column determines
both the pole and finite logarithm through
Eq.~\eqref{eq:analytic-rapidity-expansion}.
\begin{table}[H]
\centering
\small
\setlength{\tabcolsep}{4pt}
\begin{tabular}{@{}>{\raggedright\arraybackslash}p{19mm}
>{\raggedright\arraybackslash}p{31mm}
>{\raggedright\arraybackslash}p{15mm}
>{\raggedright\arraybackslash}p{86mm}@{}}
\toprule
virtual graph & exact kernel & $z_{\max}$ & pure-rapidity coefficient after
the local UV overlap is retained separately\\
\midrule
$\mathcal V_5$ & Eq.~\eqref{eq:g5-result}
& $z_1$ & $+2[\mathcal C_5^{(0)}\rho_1]_{U\text{-sub}}$\\
$\mathcal V_6$ & Eq.~\eqref{eq:g6-result}
& $z_2$ & $+2[\mathcal C_6^{(0)}\rho_2]_{U\text{-sub}}$\\
$\mathcal V_7$ & Eq.~\eqref{eq:g7-result}
& $z_1$ & $-2\mathcal C_5^{(0)}D_{12}$\\
$\mathcal V_8$ & Eq.~\eqref{eq:g8-result}
& $z_2$ & $-2\mathcal C_6^{(0)}D_{12}$\\
$\mathcal V_9$ & Eq.~\eqref{eq:g9-result}
& $z_1$ & $-[C_F\mathcal C_0\rho_1]_{U\text{-sub}}$\\
$\mathcal V_{10}$ & Eq.~\eqref{eq:g10-result}
& $z_2$ & $-[C_F\mathcal C_0\rho_2]_{U\text{-sub}}$\\
$\mathcal V_{11}$ & Eq.~\eqref{eq:g11-result}
& $z_1$ & $-2[C_F\mathcal C_0\rho_1]_{U\text{-sub}}$\\
$\mathcal V_{12}$ & Eq.~\eqref{eq:g12-result}
& $z_2$ & $-2[C_F\mathcal C_0\rho_2]_{U\text{-sub}}$\\
$\mathcal V_{13(1)}$ & Eq.~\eqref{eq:g131-result}
& $z_2$ & $+\mathcal C_{21}^{(0)}D_{12}$\\
$\mathcal V_{13(2)}$ & Eq.~\eqref{eq:g132-result}
& $z_1$ & $+\mathcal C_{21}^{(0)}D_{12}$\\
$\mathcal V_{14(1)}$ & Eq.~\eqref{eq:g141-result}
& $z_1$ & $+[C_F\mathcal C_0\rho_1]_{U\text{-sub}}
 +C_F\mathcal C_0D_{12}$\\
$\mathcal V_{14(2)}$ & Eq.~\eqref{eq:g142-result}
& $z_2$ & $+[C_F\mathcal C_0\rho_2]_{U\text{-sub}}
 +C_F\mathcal C_0D_{12}$\\
\bottomrule
\end{tabular}
\caption{Virtual source map and pure-rapidity contributions. The second column
identifies the fixed-order kernel, the third its longitudinal upper endpoint,
and the fourth its pure rapidity coefficient. The overlap and pure-UV
contributions for the same rows are Eqs.~\eqref{eq:joint-virtual-overlap-list}
and \eqref{eq:joint-virtual-pure-uv-list}.}
\label{tab:joint-virtual-rapidity-ledger}
\end{table}
Each conjugate-side virtual graph has the dagger of the displayed pure
rapidity coefficient and the same real longitudinal UV coefficient before the
overall real part is taken.

\paragraph{Real cross sections.}
Only four ordered real products have a logarithmic UV--rapidity overlap in the
antiquark-tagged kernel.  The exact cut relations in
Eqs.~\eqref{eq:real-r11-result} and \eqref{eq:real-r13-result} fix their signs
after their amplitudes and analytically regulated phase-space measures have
already been matched:
\begin{align}
 T_{UY}\Rij{1}{1}
 &=+2\left[-\frac1{\eta_Y}+\ln\frac{\bar z_2l^+}{\nu^+}\right]
 \mathcal U_{\Born},&
 T_{UY}\Rij{3}{3}
 &=+2\left[-\frac1{\eta_Y}+\ln\frac{\bar z_2l^+}{\nu^+}\right]
 \mathcal U_{\Born},
 \nonumber\\
 T_{UY}\Rij{3}{1}
 &=-2\left[-\frac1{\eta_Y}+\ln\frac{\bar z_2l^+}{\nu^+}\right]
 \mathcal U_{\Born},&
 T_{UY}\Rij{1}{3}
 &=-2\left[-\frac1{\eta_Y}+\ln\frac{\bar z_2l^+}{\nu^+}\right]
 \mathcal U_{\Born}.
 \label{eq:joint-real-overlap-list}
\end{align}
Every nonzero entry in Eq.~\eqref{eq:joint-real-overlap-list} has an omitted
$\mathcal O(\eta_Y)$ remainder.  Every other ordered real product has zero
$T_{UY}$ coefficient.  Their candidate coincident kernels contain at most one
inverse power of a transverse separation, or their Wilson-line subtraction
vanishes at coincidence.

The corresponding complete $\eta_Y^0$ pure UV terms are
\begin{align}
 (T_U-T_{UY})\Rij{1}{1}
 &=-\frac32\mathcal U_{\Born},&
 (T_U-T_{UY})\Rij{3}{3}
 &=-\frac32\mathcal U_{\Born},
 \nonumber\\
 (T_U-T_{UY})\Rij{3}{1}
 &=+\frac32\mathcal U_{\Born},&
 (T_U-T_{UY})\Rij{1}{3}
 &=+\frac32\mathcal U_{\Born}.
 \label{eq:joint-real-pure-uv-list}
\end{align}
The pure UV term of every other ordered real product is zero.

The direct $z=0$ evaluation of all sixteen products gives the following pure
rapidity ledger.  Each row of Table~\ref{tab:joint-real-rapidity-ledger} is in
units of $\dd\Gamma_Y$.
\begin{table}[H]
\centering
\small
\setlength{\tabcolsep}{4pt}
\begin{tabular}{@{}p{25mm}p{31mm}p{18mm}p{77mm}@{}}
\toprule
real product & exact kernel & $z_{\max}$ & pure-rapidity coefficient after the
local UV overlap is retained separately\\
\midrule
$\Rij{2}{2}$ & Eq.~\eqref{eq:real-r22-result}
& $\bar z_2$ & $+2C_F\mathcal C_0D_{22'}$\\
$\Rij{2}{1}$ & Eq.~\eqref{eq:real-r21-result}
& $\bar z_2$ & $-2\mathcal C_{21}^{(0)}D_{12}$\\
$\Rij{1}{2}$ & Eq.~\eqref{eq:real-r21-result}
& $\bar z_2$ & dagger of the preceding coefficient\\
$\Rij{1}{1}$ & Eq.~\eqref{eq:real-r11-result}
& $\bar z_2$ & $+2[C_F\mathcal C_0\rho_1]_{U\text{-sub}}$\\
$\Rij{4}{4}$ & Eq.~\eqref{eq:real-r44-result}
& $\bar z_2$ & $+2\mathcal C_{bb}^{(0)}D_{22'}$\\
$\Rij{3}{3}$ & Eq.~\eqref{eq:real-r33-result}
& $\bar z_2$ & $+2[\mathcal C_{bb}^{(0)}\rho_1]_{U\text{-sub}}$\\
$\Rij{4}{3}$ & Eq.~\eqref{eq:real-r43-result}
& $\bar z_2$ & $-2\mathcal C_{bb}^{(0)}D_{12}$\\
$\Rij{3}{4}$ & Eq.~\eqref{eq:real-r43-result}
& $\bar z_2$ & dagger of the preceding coefficient\\
$\Rij{2}{4}$ & Eq.~\eqref{eq:real-r24-result}
& $\bar z_2$ & $-2\mathcal C_{2b}^{(0)}D_{22'}$\\
$\Rij{4}{2}$ & Eq.~\eqref{eq:real-r24-result}
& $\bar z_2$ & dagger of the preceding coefficient\\
$\Rij{2}{3}$ & Eq.~\eqref{eq:real-r23-result}
& $\bar z_2$ & $+2\mathcal C_{2b}^{(0)}D_{12}$\\
$\Rij{3}{2}$ & Eq.~\eqref{eq:real-r23-result}
& $\bar z_2$ & dagger of the preceding coefficient\\
$\Rij{3}{1}$ & Eq.~\eqref{eq:real-r13-result}
& $\bar z_2$ & $-2[\mathcal C_5^{(0)}\rho_1]_{U\text{-sub}}$\\
$\Rij{1}{3}$ & Eq.~\eqref{eq:real-r13-result}
& $\bar z_2$ & dagger of the preceding coefficient\\
$\Rij{4}{1}$ & Eq.~\eqref{eq:real-r14-result}
& $\bar z_2$ & $+2\mathcal C_5^{(0)}D_{12}$\\
$\Rij{1}{4}$ & Eq.~\eqref{eq:real-r14-result}
& $\bar z_2$ & dagger of the preceding coefficient\\
\bottomrule
\end{tabular}
\caption{Real source map and pure-rapidity contributions. The second column
identifies the fixed-order kernel, the third its longitudinal upper endpoint,
and the fourth its pure rapidity coefficient. The overlap and pure-UV
contributions for the same rows are Eqs.~\eqref{eq:joint-real-overlap-list}
and \eqref{eq:joint-real-pure-uv-list}.}
\label{tab:joint-real-rapidity-ledger}
\end{table}
The last four rows follow from the exact regulated cut identities only after
those identities have been established from the explicit amplitudes in
Eqs.~\eqref{eq:real-r13-result}--\eqref{eq:real-r14-result}.

\begin{table}[H]
\centering
\small
\setlength{\tabcolsep}{4pt}
\begin{tabular}{@{}
 >{\raggedright\arraybackslash}p{31mm}
 >{\raggedright\arraybackslash}p{72mm}
 >{\raggedright\arraybackslash}p{51mm}@{}}
\toprule
check & entries to be compared & resulting equation\\
\midrule
virtual overlap
& $\mathcal V_5+\mathcal V_{11}$,
$\mathcal V_6+\mathcal V_{12}$, and
$\mathcal V_9+\mathcal V_{10}+\mathcal V_{14(1)}+\mathcal V_{14(2)}$
from Eq.~\eqref{eq:joint-virtual-overlap-list}
& Eqs.~\eqref{eq:joint-virtual-overlap-cancellation} and
\eqref{eq:joint-total-overlap-cancellation}\\
real overlap
& $\Rij{1}{1}+\Rij{3}{3}+\Rij{3}{1}+\Rij{1}{3}$
from Eq.~\eqref{eq:joint-real-overlap-list}
& Eqs.~\eqref{eq:joint-real-overlap-cancellation} and
\eqref{eq:joint-total-overlap-cancellation}\\
virtual pure UV
& the same three virtual groups, now from
Eq.~\eqref{eq:joint-virtual-pure-uv-list}
& Eqs.~\eqref{eq:joint-local-pure-uv-cancellations}--
\eqref{eq:joint-virtual-pure-uv-cancellation}\\
real pure UV
& the same four real products, now from
Eq.~\eqref{eq:joint-real-pure-uv-list}
& Eqs.~\eqref{eq:joint-real-pure-uv-cancellation} and
\eqref{eq:joint-total-pure-uv-cancellation}\\
rapidity cut pairs
& named rows of Tables~\ref{tab:joint-virtual-rapidity-ledger} and
\ref{tab:joint-real-rapidity-ledger}
& Eqs.~\eqref{eq:joint-pointwise-rapidity-cancellations},
\eqref{eq:joint-pointwise-finite-endpoint}, and
\eqref{eq:joint-cut-rapidity-cancellations}\\
remaining rapidity rows
& both rapidity ledgers after the preceding cut-pair cancellations
 & pole: Eqs.~\eqref{eq:joint-amplitude-side-rapidity-sum}--
 \eqref{eq:joint-direct-finite-nc-rapidity-sum}; finite:
 Eqs.~\eqref{eq:joint-pointwise-finite-endpoint} and
\eqref{eq:joint-full-graph-finite-analytic}; combined pole and logarithm:
Eq.~\eqref{eq:joint-pure-rapidity-pole-log-master}; checked against
Eq.~\eqref{eq:joint-jimwlk-born-action}\\
\bottomrule
\end{tabular}
\caption{Graph entries used in the overlap and pure-UV cancellations and the
JIMWLK comparison. The second column specifies the input rows and the
third gives the corresponding cancellation or evolution result.}
\label{tab:joint-cancellation-map}
\end{table}

\subsubsection{Cancellation of the overlap poles}
\label{sec:joint-uv-cancellations}

The virtual overlap contributions are those in
Eq.~\eqref{eq:joint-virtual-overlap-list}; their fixed-order sources are
listed in Table~\ref{tab:joint-virtual-rapidity-ledger}. They cancel in
the following three graph groups:
\begin{align}
 T_{UY}(\mathcal V_5+\mathcal V_{11})&=0,
 \nonumber\\
 T_{UY}(\mathcal V_6+\mathcal V_{12})&=0,
 \nonumber\\
 T_{UY}(\mathcal V_9+\mathcal V_{10}
 +\mathcal V_{14(1)}+\mathcal V_{14(2)})
 ={}&\Bigg[
 -\left(-\frac1{\eta_Y}+\ln\frac{z_1l^+}{\nu^+}\right)
 -\left(-\frac1{\eta_Y}+\ln\frac{z_2l^+}{\nu^+}\right)
 \nonumber\\[-2pt]
 &\hspace{18mm}
 +\left(-\frac1{\eta_Y}+\ln\frac{z_1l^+}{\nu^+}\right)
 +\left(-\frac1{\eta_Y}+\ln\frac{z_2l^+}{\nu^+}\right)
 \Bigg]
 \nonumber\\[-2pt]
 &\hspace{18mm}\times\mathcal U_{\Born}=0.
 \label{eq:joint-virtual-overlap-cancellation}
\end{align}
The conjugate-side sum vanishes independently.  In the real sector, the
$\Rij{1}{1}$, $\Rij{3}{3}$, $\Rij{3}{1}$, and $\Rij{1}{3}$ entries of
Eq.~\eqref{eq:joint-real-overlap-list}, linked to their exact kernels in
Table~\ref{tab:joint-real-rapidity-ledger}, give
\begin{align}
 T_{UY}(\Rij{1}{1}+\Rij{3}{3}+\Rij{3}{1}+\Rij{1}{3})
 ={}&(2+2-2-2)
 \left[-\frac1{\eta_Y}+\ln\frac{\bar z_2l^+}{\nu^+}\right]
 \mathcal U_{\Born}=0.
 \label{eq:joint-real-overlap-cancellation}
\end{align}
Combining the virtual and real sums gives
\begin{equation}
 \sum_{G\in\mathrm{real+virtual}}T_{UY}\mathcal I_G=0
 \label{eq:joint-total-overlap-cancellation}
\end{equation}
The coefficient of the mixed
 $1/(\epsilon_{\rm UV}\eta_Y)$ pole vanishes, and the associated finite
 endpoint logarithms vanish in the same graph groups.

\subsubsection{Cancellation of the pure ultraviolet poles}
\label{sec:joint-pure-uv-cancellations}

The virtual pure-UV contributions are given in
Eq.~\eqref{eq:joint-virtual-pure-uv-list}, and the real contributions in
Eq.~\eqref{eq:joint-real-pure-uv-list}.  The $\mathcal V_5$ and
$\mathcal V_{11}$ rows, and separately the $\mathcal V_6$ and
$\mathcal V_{12}$ rows, cancel locally in the parent longitudinal variable:
\begin{align}
 (T_U-T_{UY})(\mathcal V_5+\mathcal V_{11})&=0,
 &(T_U-T_{UY})(\mathcal V_6+\mathcal V_{12})&=0.
 \label{eq:joint-local-pure-uv-cancellations}
\end{align}
For the remaining one-sided virtual group, the $\mathcal V_{14(1)}$ and
$\mathcal V_{14(2)}$ rows of Eq.~\eqref{eq:joint-virtual-pure-uv-list} add to
\begin{align}
 &\sum_{j=1}^2(T_U-T_{UY})\mathcal V_{14(j)}
 \nonumber\\
 &\quad=-\frac32\mathcal U_{\Born},
 \label{eq:joint-v14-pure-uv-sum}
\end{align}
where $z_1^2+z_2^2$ in the denominators of the two rows has canceled against
$z_1^2+z_2^2$ in their numerator sum.  The $\mathcal V_9$ and
$\mathcal V_{10}$ rows of the same input equation give
\begin{align}
 &(T_U-T_{UY})(\mathcal V_9+\mathcal V_{10})
 \nonumber\\
 &\quad=+\frac32\mathcal U_{\Born}.
 \label{eq:joint-v910-pure-uv-sum}
\end{align}
Equations~\eqref{eq:joint-v14-pure-uv-sum} and
\eqref{eq:joint-v910-pure-uv-sum} cancel after the analytic rapidity pole and
its finite endpoint logarithm have been removed graph by graph.  Thus
\begin{equation}
 \sum_{n=5}^{14}(T_U-T_{UY})\mathcal V_n=0,
 \label{eq:joint-virtual-pure-uv-cancellation}
\end{equation}
with the conjugate sum again vanishing separately.

The four nonzero real entries in
Eq.~\eqref{eq:joint-real-pure-uv-list} have the same longitudinal coefficient
and give
\begin{equation}
 (T_U-T_{UY})(\Rij{1}{1}+\Rij{3}{3}
 +\Rij{3}{1}+\Rij{1}{3})=0.
 \label{eq:joint-real-pure-uv-cancellation}
\end{equation}
All other real entries are individually free of a logarithmic transverse UV
pole.  Combining Eqs.~\eqref{eq:joint-virtual-pure-uv-cancellation} and
\eqref{eq:joint-real-pure-uv-cancellation} proves
\begin{equation}
 \sum_{G\in\mathrm{real+virtual}}
 (T_U-T_{UY})\mathcal I_G=0.
 \label{eq:joint-total-pure-uv-cancellation}
\end{equation}
Together with Eq.~\eqref{eq:joint-total-overlap-cancellation}, this also
proves cancellation of the full $T_U$ contribution.  The proof did not take a
soft, collinear, or small-angle limit.

\subsubsection{Direct sum of the pure rapidity residues}
\label{sec:joint-direct-rapidity-sum}

For graph $G$, write the exact fixed-order integrand multiplying $\dd z/z$ as
$\mathcal I_G(z)$.  One formula contains the pole, its associated logarithm,
and the remaining finite integral:
\begin{align}
 &\int_0^{z_{\max,G}}\frac{\dd z}{z}
 \left(\frac{\nu^+}{zl^+}\right)^{\eta_Y}\mathcal I_G(z)
 \nonumber\\[-2pt]
 &\quad=
 \left[-\frac1{\eta_Y}
 +\ln\frac{z_{\max,G}l^+}{\nu^+}\right]\mathcal I_G(0)
 +\int_0^{z_{\max,G}}\frac{\dd z}{z}
 [\mathcal I_G(z)-\mathcal I_G(0)]
 +\mathcal O(\eta_Y).
 \label{eq:joint-full-graph-finite-analytic}
\end{align}
Thus the rapidity ledger supplies $\mathcal I_G(0)$ and $z_{\max,G}$.  We first
sum the small-$z$ residues.  We then insert the graph endpoints to obtain the
finite logarithms.  The last integral is finite at $z=0$ and belongs to the
finite NLO impact factor.

Because the overlap has already canceled, the sum of the pure rapidity column
equals the sum of the full $T_Y$ residues.  Several residues cancel before any
color algebra is simplified.  The first line below compares the
$\Rij{1}{1}$ row of Table~\ref{tab:joint-real-rapidity-ledger} with the
$\mathcal V_9$ row of Table~\ref{tab:joint-virtual-rapidity-ledger} and its
conjugate.  The next two lines compare the $\Rij{2}{1}$ and $\Rij{1}{2}$ rows
with the two $\mathcal V_{13}$ rows and their conjugates:
\begin{align}
 \left[\Rij{1}{1}+\mathcal V_9+\mathcal V_9^*\right]_{z=0}&=0,
 \nonumber\\
 \left[\Rij{2}{1}+\mathcal V_{13(1)}
 +\mathcal V_{13(2)}\right]_{z=0}&=0,
 \nonumber\\
 \left[\Rij{1}{2}+\mathcal V_{13(1)}^*
 +\mathcal V_{13(2)}^*\right]_{z=0}&=0,
 \label{eq:joint-pointwise-rapidity-cancellations}
\end{align}

The four exact cut pairs cancel before the remaining residue is summed:
\begin{align}
 \Rij{3}{1}+\mathcal V_5&=0,
 &\Rij{1}{3}+\mathcal V_5^*&=0,
 \nonumber\\
 \Rij{4}{1}+\mathcal V_7&=0,
 &\Rij{1}{4}+\mathcal V_7^*&=0.
 \label{eq:joint-cut-rapidity-cancellations}
\end{align}
Equations~\eqref{eq:joint-pointwise-rapidity-cancellations} and
\eqref{eq:joint-cut-rapidity-cancellations} follow from the displayed graph
coefficients.

We now insert the residue and upper endpoint of every row which has not
canceled in Eqs.~\eqref{eq:joint-pointwise-rapidity-cancellations} and
\eqref{eq:joint-cut-rapidity-cancellations}.  At the rapidity endpoint of a
real graph, $\bar z_2=z_1$.  The amplitude-side sum is therefore
\begin{align}
 &\Bigg[
 \frac12\Rij{2}{2}+\frac12\Rij{4}{4}+\frac12\Rij{3}{3}
 +\Rij{4}{3}+\Rij{2}{4}+\Rij{2}{3}+\Rij{2}{1}
 \nonumber\\[-2pt]
 &\hspace{16mm}
 +\mathcal V_6+\mathcal V_8+\mathcal V_{10}+\mathcal V_{11}
 +\mathcal V_{12}+\mathcal V_{13(1)}+\mathcal V_{13(2)}
 +\mathcal V_{14(1)}+\mathcal V_{14(2)}
 \Bigg]_{Y,\,\mathrm{pole+log}}
 \nonumber\\
 &={}
 \left[-\frac1{\eta_Y}+\ln\frac{z_1l^+}{\nu^+}\right]
 \Big[
 \underbrace{C_F\mathcal C_0D_{22'}}_{\frac12\Rij{2}{2}}
 +\underbrace{\mathcal C_{bb}^{(0)}D_{22'}}_{\frac12\Rij{4}{4}}
 +\underbrace{\mathcal C_{bb}^{(0)}\rho_1}_{\frac12\Rij{3}{3}}
 -\underbrace{2\mathcal C_{bb}^{(0)}D_{12}}_{\Rij{4}{3}}
 \nonumber\\[-2pt]
 &\hspace{18mm}
 -\underbrace{2\mathcal C_{2b}^{(0)}D_{22'}}_{\Rij{2}{4}}
 +\underbrace{2\mathcal C_{2b}^{(0)}D_{12}}_{\Rij{2}{3}}
 -\underbrace{2\mathcal C_{21}^{(0)}D_{12}}_{\Rij{2}{1}}
 -\underbrace{2C_F\mathcal C_0\rho_1}_{\mathcal V_{11}}
 \nonumber\\[-2pt]
 &\hspace{18mm}
 +\underbrace{\mathcal C_{21}^{(0)}D_{12}}_{\mathcal V_{13(2)}}
 +\underbrace{C_F\mathcal C_0(\rho_1+D_{12})}_{\mathcal V_{14(1)}}
 \Big]
 \nonumber\\
 &\quad+
 \left[-\frac1{\eta_Y}+\ln\frac{z_2l^+}{\nu^+}\right]
 \Big[
 \underbrace{2\mathcal C_6^{(0)}\rho_2}_{\mathcal V_6}
 -\underbrace{2\mathcal C_6^{(0)}D_{12}}_{\mathcal V_8}
 -\underbrace{C_F\mathcal C_0\rho_2}_{\mathcal V_{10}}
 -\underbrace{2C_F\mathcal C_0\rho_2}_{\mathcal V_{12}}
 \nonumber\\[-2pt]
 &\hspace{18mm}
 +\underbrace{\mathcal C_{21}^{(0)}D_{12}}_{\mathcal V_{13(1)}}
 +\underbrace{C_F\mathcal C_0(\rho_2+D_{12})}_{\mathcal V_{14(2)}}
 \Big].
 \label{eq:joint-amplitude-side-rapidity-sum}
\end{align}
The factors $1/2$ on the diagonal real products prevent double counting when
the conjugate-side sum is added.  Equation
\eqref{eq:joint-amplitude-side-rapidity-sum} is the direct substitution of the
graph results.

The independently evaluated conjugate-side rows give
\begin{align}
 &\Bigg[
 \frac12\Rij{2}{2}+\frac12\Rij{4}{4}+\frac12\Rij{3}{3}
 +\Rij{3}{4}+\Rij{4}{2}+\Rij{3}{2}+\Rij{1}{2}
 \nonumber\\[-2pt]
 &\hspace{16mm}
 +\mathcal V_6^*+\mathcal V_8^*+\mathcal V_{10}^*+\mathcal V_{11}^*
 +\mathcal V_{12}^*+\mathcal V_{13(1)}^*+\mathcal V_{13(2)}^*
 +\mathcal V_{14(1)}^*+\mathcal V_{14(2)}^*
 \Bigg]_{Y,\,\mathrm{pole+log}}
 \nonumber\\
 &={}
 \left[-\frac1{\eta_Y}+\ln\frac{z_1l^+}{\nu^+}\right]
 \Big[
 \underbrace{(C_F\mathcal C_0D_{22'})^\dagger}_{\frac12\Rij{2}{2}}
 +\underbrace{(\mathcal C_{bb}^{(0)}D_{22'})^\dagger}_{\frac12\Rij{4}{4}}
 +\underbrace{(\mathcal C_{bb}^{(0)}\rho_1)^\dagger}_{\frac12\Rij{3}{3}}
 -\underbrace{(2\mathcal C_{bb}^{(0)}D_{12})^\dagger}_{\Rij{3}{4}}
 \nonumber\\[-2pt]
 &\hspace{18mm}
 -\underbrace{(2\mathcal C_{2b}^{(0)}D_{22'})^\dagger}_{\Rij{4}{2}}
 +\underbrace{(2\mathcal C_{2b}^{(0)}D_{12})^\dagger}_{\Rij{3}{2}}
 -\underbrace{(2\mathcal C_{21}^{(0)}D_{12})^\dagger}_{\Rij{1}{2}}
 -\underbrace{(2C_F\mathcal C_0\rho_1)^\dagger}_{\mathcal V_{11}^*}
 \nonumber\\[-2pt]
 &\hspace{18mm}
 +\underbrace{(\mathcal C_{21}^{(0)}D_{12})^\dagger}_{\mathcal V_{13(2)}^*}
 +\underbrace{[C_F\mathcal C_0(\rho_1+D_{12})]^\dagger}_{\mathcal V_{14(1)}^*}
 \Big]
 \nonumber\\
 &\quad+
 \left[-\frac1{\eta_Y}+\ln\frac{z_2l^+}{\nu^+}\right]
 \Big[
 \underbrace{(2\mathcal C_6^{(0)}\rho_2)^\dagger}_{\mathcal V_6^*}
 -\underbrace{(2\mathcal C_6^{(0)}D_{12})^\dagger}_{\mathcal V_8^*}
 -\underbrace{(C_F\mathcal C_0\rho_2)^\dagger}_{\mathcal V_{10}^*}
 -\underbrace{(2C_F\mathcal C_0\rho_2)^\dagger}_{\mathcal V_{12}^*}
 \nonumber\\[-2pt]
 &\hspace{18mm}
 +\underbrace{(\mathcal C_{21}^{(0)}D_{12})^\dagger}_{\mathcal V_{13(1)}^*}
 +\underbrace{[C_F\mathcal C_0(\rho_2+D_{12})]^\dagger}_{\mathcal V_{14(2)}^*}
 \Big].
 \label{eq:joint-conjugate-side-rapidity-sum}
\end{align}

To separate the common pole from the finite endpoint terms, write
\begin{equation}
 -\frac1{\eta_Y}+\ln\frac{z_2l^+}{\nu^+}
 =-\frac1{\eta_Y}+\ln\frac{z_1l^+}{\nu^+}
 +\ln\frac{z_2}{z_1}.
 \label{eq:joint-endpoint-log-split}
\end{equation}
The coefficient multiplying the common pole and the same rapidity-scale
logarithm on the amplitude side then becomes, term by term,
\begin{align}
 &C_F\mathcal C_0D_{22'}
 +\mathcal C_{bb}^{(0)}D_{22'}
 +\mathcal C_{bb}^{(0)}\rho_1
 -2\mathcal C_{bb}^{(0)}D_{12}
 -2\mathcal C_{2b}^{(0)}D_{22'}
 +2\mathcal C_{2b}^{(0)}D_{12}
 \nonumber\\[-2pt]
 &\quad
 -2\mathcal C_{21}^{(0)}D_{12}
 -2C_F\mathcal C_0\rho_1
 +\mathcal C_{21}^{(0)}D_{12}
 +C_F\mathcal C_0(\rho_1+D_{12})
 \nonumber\\[-2pt]
 &\quad
 +2\mathcal C_6^{(0)}\rho_2
 -2\mathcal C_6^{(0)}D_{12}
 -C_F\mathcal C_0\rho_2
 -2C_F\mathcal C_0\rho_2
 +\mathcal C_{21}^{(0)}D_{12}
 +C_F\mathcal C_0(\rho_2+D_{12})
 \nonumber\\
 &={}
 C_F\mathcal C_0(D_{22'}-\rho_2-K_{12})
 +\mathcal C_{bb}^{(0)}(D_{22'}+\rho_1-2D_{12})
 +2\mathcal C_{2b}^{(0)}(D_{12}-D_{22'})
 +2\mathcal C_6^{(0)}(\rho_2-D_{12}),
 \label{eq:joint-common-rapidity-coefficient-amplitude}
\end{align}
where $K_{12}=\rho_1+\rho_2-2D_{12}$ has been inserted only in the last line.
The $\mathcal C_{21}^{(0)}D_{12}$ terms cancel in the common coefficient.

We now substitute the color operators in
Eqs.~\eqref{eq:joint-born-color}, \eqref{eq:joint-r33-rapidity-color},
\eqref{eq:joint-c2b-color} and~\eqref{eq:joint-c6-color} into the
amplitude- and conjugate-side sums.
For the conjugate-side sum, $S_{ij}^*=S_{ji}$ is followed by the
dummy-coordinate interchange $\bx_2\leftrightarrow\bx_{2'}$.
Using $2C_F=N_c-1/N_c$, every term proportional to $1/N_c$ has the
same color factor $S_{2'2}-S_{12}-S_{2'1}+1$. Their transverse coefficients
cancel explicitly:
\begin{align*}
 &(D_{22'}-\rho_1-\rho_2-\rho_{2'}+D_{12}+D_{2'1})
 +(D_{22'}+\rho_1-D_{12}-D_{2'1})\\[-2pt]
 &\quad +(D_{12}-D_{22'}+\rho_{2'}-D_{2'1})
 +(\rho_2-D_{12}+D_{2'1}-D_{22'})=0.
\end{align*}
The complete substitution and collection of the remaining terms give
\begin{align}
 &\Bigg\langle
 \left(N_c-\frac1{N_c}\right)(S_{2'2}-S_{12}-S_{2'1}+1)
 \nonumber\\[-2pt]
 &\qquad\times
 (D_{22'}-\rho_1-\rho_2-\rho_{2'}+D_{12}+D_{2'1})
 \nonumber\\
 &\quad+
 \left[N_c(S_{2'2}-S_{13}S_{32}-S_{2'3}S_{31}+1)
 -\frac{S_{2'2}-S_{12}-S_{2'1}+1}{N_c}\right]
 \nonumber\\[-2pt]
 &\qquad\times(D_{22'}+\rho_1-D_{12}-D_{2'1})
 \nonumber\\
 &\quad+
 \left[N_c(S_{2'3}S_{32}-S_{2'3}S_{31}+1-S_{12})
 -\frac{S_{2'2}-S_{12}-S_{2'1}+1}{N_c}\right]
 \nonumber\\[-2pt]
 &\qquad\times(D_{12}-D_{22'}+\rho_{2'}-D_{2'1})
 \nonumber\\
 &\quad+
 \left[N_c(S_{2'3}S_{32}-S_{13}S_{32}+1-S_{2'1})
 -\frac{S_{2'2}-S_{12}-S_{2'1}+1}{N_c}\right]
 \nonumber\\[-2pt]
 &\qquad\times(\rho_2-D_{12}+D_{2'1}-D_{22'})
 \Bigg\rangle_Y
 \nonumber\\
 &={}N_c\Big\langle
 (\rho_2+\rho_{2'}-2D_{22'})(S_{2'3}S_{32}-S_{2'2})
 \nonumber\\[-2pt]
 &\qquad
 -(\rho_1+\rho_2-2D_{12})(S_{13}S_{32}-S_{12})
 \nonumber\\[-2pt]
 &\qquad
 -(\rho_1+\rho_{2'}-2D_{2'1})(S_{2'3}S_{31}-S_{2'1})
 \Big\rangle_Y
 \nonumber\\
 &={}N_c\Big\langle
 K_{22'}(S_{2'3}S_{32}-S_{2'2})
 -K_{12}(S_{13}S_{32}-S_{12})
 -K_{2'1}(S_{2'3}S_{31}-S_{2'1})
 \Big\rangle_Y.
 \label{eq:joint-direct-finite-nc-rapidity-sum}
\end{align}
The last line uses $K_{ij}=\rho_i+\rho_j-2D_{ij}$.
Products of dipoles remain inside one common target average.

Consequently, the surviving pole and the rapidity-scale logarithm with exactly
the same coefficient are
\begin{align}
 &\left[-\frac1{\eta_Y}+\ln\frac{z_1l^+}{\nu^+}\right]N_c
 \Big\langle
 K_{22'}(S_{2'3}S_{32}-S_{2'2})
 -K_{12}(S_{13}S_{32}-S_{12})
 \nonumber\\[-2pt]
 &\hspace{44mm}
 -K_{2'1}(S_{2'3}S_{31}-S_{2'1})
 \Big\rangle_Y.
 \label{eq:joint-common-rapidity-pole-log}
\end{align}

The terms in the second square bracket of
Eq.~\eqref{eq:joint-amplitude-side-rapidity-sum} also multiply the last term
in Eq.~\eqref{eq:joint-endpoint-log-split}.  Their explicit sum, together with
the separately summed conjugate-side graphs, is
\begin{align}
 &\ln\frac{z_2}{z_1}\Bigg\{
 2\mathcal C_6^{(0)}\rho_2
 -2\mathcal C_6^{(0)}D_{12}
 -C_F\mathcal C_0\rho_2
 -2C_F\mathcal C_0\rho_2
 +\mathcal C_{21}^{(0)}D_{12}
 +C_F\mathcal C_0(\rho_2+D_{12})
 \nonumber\\[-2pt]
 &\hspace{29mm}
 +\Big[
 2\mathcal C_6^{(0)}\rho_2
 -2\mathcal C_6^{(0)}D_{12}
 -C_F\mathcal C_0\rho_2
 -2C_F\mathcal C_0\rho_2
 +\mathcal C_{21}^{(0)}D_{12}
 +C_F\mathcal C_0(\rho_2+D_{12})
 \Big]^\dagger\Bigg\}
 \nonumber\\
 &={}\ln\frac{z_2}{z_1}\Bigg\{
 2\mathcal C_6^{(0)}(\rho_2-D_{12})
 -2C_F\mathcal C_0\rho_2
 +C_F\mathcal C_0D_{12}
 +\mathcal C_{21}^{(0)}D_{12}
 \nonumber\\[-2pt]
 &\hspace{31mm}
 +\Big[
 2\mathcal C_6^{(0)}(\rho_2-D_{12})
 -2C_F\mathcal C_0\rho_2
 +C_F\mathcal C_0D_{12}
 +\mathcal C_{21}^{(0)}D_{12}
 \Big]^\dagger\Bigg\}.
 \label{eq:joint-pointwise-finite-endpoint}
\end{align}
The $\mathcal C_{21}^{(0)}D_{12}\ln(z_2/z_1)$ term is visible before any
color simplification: it comes from the $z_2$ endpoint of
$\mathcal V_{13(1)}$, whereas $\Rij{2}{1}$ and $\mathcal V_{13(2)}$ end at
$z_1$.  The final integral in
Eq.~\eqref{eq:joint-full-graph-finite-analytic} supplies the remaining
endpoint-subtracted finite part.

Restoring the hard factor and the transverse integrations, the sum of
Eqs.~\eqref{eq:joint-common-rapidity-pole-log} and
\eqref{eq:joint-pointwise-finite-endpoint} is
\begin{align}
 &\left.\frac{\dd\sigma^T_{\bar q,\mathrm{NLO}}}
 {\dd^2\bq\,\dd y_2}\right|_{\mathrm{pure}\,Y,
 \mathrm{pole+endpoint}}
 \nonumber\\
 &={}\frac{\alpha_s}{2\pi^2}
 \int\dd^2\bx_1\dd^2\bx_2\dd^2\bx_{2'}\dd^2\bx_3\,
 \mathcal W_T
 \Bigg\{
 \left[-\frac1{\eta_Y}+\ln\frac{z_1l^+}{\nu^+}\right]N_c
 \nonumber\\[-2pt]
 &\hspace{9mm}\times\Big\langle
 K_{22'}(S_{2'3}S_{32}-S_{2'2})
 -K_{12}(S_{13}S_{32}-S_{12})
 \nonumber\\[-2pt]
 &\hspace{42mm}
 -K_{2'1}(S_{2'3}S_{31}-S_{2'1})
 \Big\rangle_Y
 \nonumber\\
 &\hspace{9mm}
 +\ln\frac{z_2}{z_1}\Bigg(
 2\mathcal C_6^{(0)}(\rho_2-D_{12})
 -2C_F\mathcal C_0\rho_2
 +C_F\mathcal C_0D_{12}
 +\mathcal C_{21}^{(0)}D_{12}
 \nonumber\\[-2pt]
 &\hspace{47mm}
 +\Big[
 2\mathcal C_6^{(0)}(\rho_2-D_{12})
 -2C_F\mathcal C_0\rho_2
 +C_F\mathcal C_0D_{12}
 +\mathcal C_{21}^{(0)}D_{12}
 \Big]^\dagger\Bigg)
 \Bigg\}.
 \label{eq:joint-pure-rapidity-pole-log-master}
\end{align}
The first term in Eq.~\eqref{eq:joint-pure-rapidity-pole-log-master} contains
the surviving pure rapidity pole and its corresponding JIMWLK scale logarithm.
The second term is the finite endpoint logarithm.

\subsubsection{Independent JIMWLK verification}
\label{sec:joint-jimwlk-verification}

Only after the direct graph sum has produced
Eq.~\eqref{eq:joint-direct-finite-nc-rapidity-sum} do we compare it with the
Hamiltonian.  We use
\begin{align}
 H_{\rm JIMWLK}={}&\frac{\alpha_s}{2\pi^2}
 \int\dd^2\boldsymbol u\,\dd^2\boldsymbol v\,\dd^2\bx_3\,
 D_{uv}\Big[2J_L^a(\boldsymbol u)U_3^{ab}J_R^b(\boldsymbol v)
 \nonumber\\[-2pt]
 &\hspace{50mm}
 -J_L^a(\boldsymbol u)J_L^a(\boldsymbol v)
 -J_R^a(\boldsymbol u)J_R^a(\boldsymbol v)\Big],
 \label{eq:joint-jimwlk-hamiltonian}\\
 U_3^{ab}={}&2\tr(t^aV_3t^bV_3^\dagger),
 \qquad
 D_{uv}=\frac{\bx_{3u}\cdot\bx_{3v}}
 {\bx_{3u}^2\bx_{3v}^2}.
 \nonumber
\end{align}
The left and right rotations of a fundamental line are
\begin{align}
 J_L^a(\boldsymbol u)V_x&=t^aV_x\delta_{ux},&
 J_L^a(\boldsymbol u)V_x^\dagger&=-V_x^\dagger t^a\delta_{ux},
 \nonumber\\
 J_R^a(\boldsymbol u)V_x&=V_xt^a\delta_{ux},&
 J_R^a(\boldsymbol u)V_x^\dagger&=-t^aV_x^\dagger\delta_{ux}.
 \label{eq:joint-jimwlk-rotations}
\end{align}
Applied once to a dipole, these rules give
\begin{align}
 J_L^a(\boldsymbol u)S_{xy}
 &=\frac{\delta_{ux}-\delta_{uy}}{N_c}
 \tr(t^aV_xV_y^\dagger),
 \nonumber\\
 J_R^a(\boldsymbol u)S_{xy}
 &=\frac{\delta_{ux}-\delta_{uy}}{N_c}
 \tr(V_xt^aV_y^\dagger).
 \label{eq:joint-jimwlk-first-rotation}
\end{align}
The two virtual rotations yield
\begin{equation}
 -\int\dd^2\boldsymbol u\dd^2\boldsymbol v\dd^2\bx_3\,
 D_{uv}(J_L^aJ_L^a+J_R^aJ_R^a)S_{xy}
 =-2C_F\int\dd^2\bx_3\,K_{xy}S_{xy}.
 \label{eq:joint-jimwlk-virtual-action}
\end{equation}
For the real rotation, the finite-$N_c$ contraction is calculated directly:
\begin{equation}
 \frac1{N_c}U_3^{ab}\tr(t^aV_xt^bV_y^\dagger)
 =\frac{N_c}{2}S_{x3}S_{3y}-\frac1{2N_c}S_{xy}.
 \label{eq:joint-jimwlk-fierz}
\end{equation}
It gives
\begin{equation}
 2\int\dd^2\boldsymbol u\dd^2\boldsymbol v\dd^2\bx_3\,
 D_{uv}J_L^a(\boldsymbol u)U_3^{ab}J_R^b(\boldsymbol v)S_{xy}
 =\int\dd^2\bx_3\,K_{xy}
 \left[N_cS_{x3}S_{3y}-\frac1{N_c}S_{xy}\right].
 \label{eq:joint-jimwlk-real-action}
\end{equation}
Adding Eqs.~\eqref{eq:joint-jimwlk-virtual-action} and
\eqref{eq:joint-jimwlk-real-action}, and again using
$2C_F=N_c-1/N_c$, gives
\begin{equation}
 H_{\rm JIMWLK}S_{xy}
 =\frac{\alpha_sN_c}{2\pi^2}
 \int\dd^2\bx_3\,K_{xy}(S_{x3}S_{3y}-S_{xy}).
 \label{eq:joint-jimwlk-dipole-action}
\end{equation}
Linearity and $H_{\rm JIMWLK}1=0$ now give
\begin{align}
 H_{\rm JIMWLK}(S_{2'2}-S_{12}-S_{2'1}+1)
 ={}&\frac{\alpha_sN_c}{2\pi^2}\int\dd^2\bx_3\Big[
 K_{22'}(S_{2'3}S_{32}-S_{2'2})
 \nonumber\\[-2pt]
 &-K_{12}(S_{13}S_{32}-S_{12})
 -K_{2'1}(S_{2'3}S_{31}-S_{2'1})\Big].
 \label{eq:joint-jimwlk-born-action}
\end{align}
The three Wilson-line structures and their signs agree one by one with the
independent graph sum in
Eq.~\eqref{eq:joint-direct-finite-nc-rapidity-sum}, reproducing the
JIMWLK evolution of the Born operator. The large-$N_c$, mean-field
replacement $\langle S_{a3}S_{3b}\rangle_Y\to
\langle S_{a3}\rangle_Y\langle S_{3b}\rangle_Y$ immediately recovers the
BK equation~\cite{Balitsky1996,Kovchegov1999}, providing the standard
starting point for phenomenological applications~\cite{BeufHERA2020}.

Taking the $-1/\eta_Y$ term of
Eq.~\eqref{eq:joint-pure-rapidity-pole-log-master}, the antiquark-tagged bare
rapidity pole is
\begin{align}
 \left.\frac{\dd\sigma^T_{\bar q,\mathrm{NLO}}}
 {\dd^2\bq\,\dd y_2}\right|_{1/\eta_Y}
 ={}&-\frac1{\eta_Y}
 \int\dd^2\bx_1\dd^2\bx_2\dd^2\bx_{2'}\,
 \mathcal W_T
 \frac{\alpha_sN_c}{2\pi^2}
 \int\dd^2\bx_3
 \nonumber\\[-2pt]
 &\times\Big\langle
 K_{22'}(S_{2'3}S_{32}-S_{2'2})
 -K_{12}(S_{13}S_{32}-S_{12})
 \nonumber\\[-2pt]
 &\hspace{18mm}
 -K_{2'1}(S_{2'3}S_{31}-S_{2'1})
 \Big\rangle_Y.
 \label{eq:joint-antiquark-jimwlk-result}
\end{align}
It is removed by rapidity minimal subtraction of the full Born operator,
\begin{equation}
 \mathcal C_{0,\mathrm{bare}}
 =\mathcal C_0(\nu^+)
 +\frac1{\eta_Y}H_{\rm JIMWLK}\mathcal C_0(\nu^+).
 \label{eq:joint-rapidity-renormalization}
\end{equation}
The counterterm generated by
Eq.~\eqref{eq:joint-rapidity-renormalization} has the opposite sign to
Eq.~\eqref{eq:joint-antiquark-jimwlk-result}.  The finite impact factor contains
$+H_{\rm JIMWLK}\mathcal C_0\ln(l^+/\nu^+)$, so cancellation of the artificial
$\nu^+$ dependence fixes the evolution convention used here:
\begin{equation}
 \frac{\partial}{\partial\ln\nu^+}\mathcal C_0(\nu^+)
 =H_{\rm JIMWLK}\mathcal C_0(\nu^+).
 \label{eq:joint-rapidity-rg}
\end{equation}
The finite NLO impact factor retains
Eqs.~\eqref{eq:joint-pointwise-finite-endpoint} and
\eqref{eq:joint-full-graph-finite-analytic}.

\paragraph{The other two energy-flow tags.}
For the quark tag, the unobserved-antiquark integral sets
$\bx_{2'}=\bx_2$.  Repeating the line-oriented Feynman-rule insertions gives the Born color operator
\begin{equation}
 \left\langle S_{11'}-S_{12}-S_{21'}+1\right\rangle_Y.
 \label{eq:joint-quark-born-color}
\end{equation}
The graph ledger has the same longitudinal coefficients as
Eqs.~\eqref{eq:joint-virtual-overlap-list}--
\eqref{eq:joint-real-pure-uv-list}, with the emitting quark and antiquark
coordinates interchanged before the unobserved momentum is integrated.  Its
overlap and pure UV sums therefore vanish separately, while its direct
rapidity sum is
\begin{align}
 H_{\rm JIMWLK}(S_{11'}-S_{12}-S_{21'}+1)
 ={}&\frac{\alpha_sN_c}{2\pi^2}\int\dd^2\bx_3\Big[
 K_{11'}(S_{13}S_{31'}-S_{11'})
 \nonumber\\[-2pt]
 &-K_{12}(S_{13}S_{32}-S_{12})
 -K_{21'}(S_{23}S_{31'}-S_{21'})\Big].
 \label{eq:joint-quark-jimwlk-result}
\end{align}
The Wilson-line ordering in Eq.~\eqref{eq:joint-quark-jimwlk-result}
follows directly from the quark-tagged amplitudes.

For the gluon tag, the detector weight is applied to the final gluon before
the two fermion momenta are integrated.  In the rapidity limit at fixed
nonzero $\bk$,
\begin{equation}
 \vartheta_g=2\arctan\frac{Qz}{|\bk|}\xrightarrow[z\to0]{}0.
 \label{eq:joint-gluon-rapidity-angle}
\end{equation}
Consequently the detector delta function has no support at a fixed measured
angle $\theta>0$.  Keeping $\theta>0$ while sending $z\to0$ instead requires
$|\bk|=Qz\cot(\theta/2)$ and belongs to the simultaneous soft region, not the
fixed-transverse-momentum rapidity region.  The fixed-gluon
hard-relative-momentum behavior derived from the four exact amplitudes is
$\dd^2\boldsymbol\Lambda/\Lambda^4$, and hence it has no logarithmic UV pole.
Therefore all three singular entries of the measured gluon tag are zero at
$\theta>0$:
\begin{equation}
 T_{UY}\Sigma^T_{g,R}(\theta)=0,\qquad
 (T_U-T_{UY})\Sigma^T_{g,R}(\theta)=0,\qquad
 T_Y\Sigma^T_{g,R}(\theta)=0.
 \label{eq:joint-gluon-tagged-three-zeroes}
\end{equation}

Combining the three one-particle tags, the NLO energy flow has no UV or mixed
UV--rapidity counterterm.  Its remaining rapidity logarithm is exactly the
finite-$N_c$ JIMWLK evolution of the complete Born $q+\bar q$ energy-flow
operator.  The NEEC detector weight depends only on the tagged Born momentum
in this limit and commutes with the JIMWLK action on the target operator.

\subsection{Soft singularities}
\label{sec:soft-singularities}

The rapidity limit used above and the soft limit are different boundaries of
the real-emission phase space.  The former sends $z\to0$ at fixed nonzero
transverse momentum.  The latter sends the complete gluon momentum to zero,
so that $z\to0$ and $|\bk|\to0$ together.  In transverse coordinate space the
soft transverse limit is $R\equiv|\bx_3|\to\infty$.  We take this joint limit
with the analytic factor $(\nu^+/(zl^+))^{\eta_Y}$ left in every real and
virtual term. The calculation below
subtracts the simultaneous-soft endpoint and retains the
fixed-transverse-momentum rapidity residue that generates JIMWLK evolution.

At large $R$, the transverse kernels defined in
Eq.~\eqref{eq:joint-transverse-kernels} obey
\begin{align}
 \rho_i&=\frac1{R^2}+\mathcal O(R^{-3}),&
 D_{ij}&=\frac1{R^2}+\mathcal O(R^{-3}),
 \nonumber\\
 K_{ij}&=\frac{\bx_{ij}^2}{R^4}+\mathcal O(R^{-5}).
 \label{eq:soft-transverse-limits}
\end{align}
Thus a logarithmic soft integral can arise only from a $1/R^2$ term.  For a
fixed fermion tag the detector insertion approaches its Born value.  By
contrast, the energy carried by a tagged soft gluon is of order $z$; its
measurement therefore removes the leading $\dd z/z$ singularity.  The soft
pole of the full energy flow can consequently be checked first in either
fermion-tagged channel.

\subsubsection{Detailed example: \texorpdfstring{$\Rij{2}{2}$ and graph
$A_{10}$}{R2x2 and graph A10}}
\label{sec:soft-r22-v10-example}

Consider the real product in Eq.~\eqref{eq:real-r22-result}.  At $z=0$ its
spin factor, photon scale, phase and color operator reduce to
\begin{align}
 \mathcal S_{22}(z_2,0)
 &=2z_1z_2^2(z_1^2+z_2^2),&
 Q_{2z}&\longrightarrow Q\sqrt{z_1z_2},
 \nonumber\\
 e^{i(z_2+z)\bq\cdot\bx_{2'2}/z_2}
 &\longrightarrow e^{i\bq\cdot\bx_{2'2}},&
 \mathcal C_{22}
 &\equiv C_F\langle S_{2'2}-S_{12}-S_{2'1}+1\rangle_Y
 =C_F\mathcal C_0,
 \label{eq:soft-r22-reductions}
\end{align}
where $z_1=1-z_2$ at the endpoint.  The radiation factor in the same formula
is
\begin{equation}
 \Delta_{2'2}^{(3)}
 =\frac{\bx_{32'}\cdot\bx_{32}}
 {\bx_{32'}^2\bx_{32}^2}
 \xrightarrow[R\to\infty]{}\frac1{R^2}.
 \label{eq:soft-r22-radiation-factor}
\end{equation}
It is useful to include the antiquark detector insertion in the common soft
measure,
\begin{align}
 \dd\Gamma_{\bar q,S}\equiv{}&
 \mathcal M_{\bar q}(\theta;z_2,\bq)
 \frac{\alpha_s}{2\pi^2}\frac{\dd z}{z}
 \left(\frac{\nu^+}{zl^+}\right)^{\eta_Y}
 \dd^2\bx_1\dd^2\bx_2\dd^2\bx_{2'}\dd^2\bx_3
 \nonumber\\[-2pt]
 &\times\mathcal W_T\frac1{R^2}.
 \label{eq:soft-common-measure}
\end{align}
Equations~\eqref{eq:soft-r22-reductions} and
\eqref{eq:soft-r22-radiation-factor} then give the real soft residue directly,
\begin{equation}
 T_S\!\left[\mathcal M_{\bar q}\Rij{2}{2}\right]
 =2C_F\mathcal C_0\,\dd\Gamma_{\bar q,S}.
 \label{eq:soft-r22-residue}
\end{equation}

The corresponding virtual term is the antiquark external-leg graph.  In
Eq.~\eqref{eq:g10-result},
\begin{equation}
 \frac{z_2^2+(z_2-z)^2}{z_2^2}\xrightarrow[z\to0]{}2,
 \label{eq:soft-v10-splitting-limit}
\end{equation}
and the large-distance part of the scaleless transverse integral supplies the
same $1/R^2$ kernel as Eq.~\eqref{eq:soft-r22-radiation-factor}.  The amplitude
and its conjugate therefore yield
\begin{equation}
 T_S[\mathcal M_{\bar q}\mathcal V_{10}]
 =-C_F\mathcal C_0\,\dd\Gamma_{\bar q,S},\qquad
 T_S[\mathcal M_{\bar q}\mathcal V_{10}^*]
 =-C_F\mathcal C_0\,\dd\Gamma_{\bar q,S}.
 \label{eq:soft-v10-residues}
\end{equation}
The three terms contain the same analytic regulator before it is expanded.
Their endpoint residue consequently cancels locally,
\begin{equation}
 T_S\!\left[\mathcal M_{\bar q}
 (\Rij{2}{2}+\mathcal V_{10}+\mathcal V_{10}^*)\right]
 =C_F\mathcal C_0\,\dd\Gamma_{\bar q,S}(2-1-1)=0.
 \label{eq:soft-r22-v10-cancellation}
\end{equation}
The singular local approximation cancels. The physical upper endpoints
of the three full longitudinal integrals supply the finite endpoint terms
in the NLO hard part through
Eq.~\eqref{eq:analytic-rapidity-expansion}.

There is one further overlap that is obscured if the transverse integral is
made scaleless too early.  At leading soft order, every self-energy before the
shockwave contains the physical photon-wave-function denominator
\begin{equation}
 \frac1{\displaystyle
 \ell^2+\frac{z}{z_1z_2}(\bk^2+Q_\gamma^2)}.
 \label{eq:soft-matched-virtuality}
\end{equation}
Keeping the complete $\bk^2+Q_\gamma^2$ dependence in this denominator, the
two sides of the photon wave function give
\begin{align}
 T_S(\mathcal V_{11}+\mathcal V_{14(1)})
 &\propto(-2+1+1)
 \int\frac{\dd^{2-2\epsilon}\bell}{(2\pi)^{2-2\epsilon}}
 \frac1{\displaystyle
 \ell^2+\frac{z}{z_1z_2}(\bk^2+Q_\gamma^2)}=0,
 \nonumber\\
 T_S(\mathcal V_{12}+\mathcal V_{14(2)})
 &\propto(-2+1+1)
 \int\frac{\dd^{2-2\epsilon}\bell}{(2\pi)^{2-2\epsilon}}
 \frac1{\displaystyle
 \ell^2+\frac{z}{z_1z_2}(\bk^2+Q_\gamma^2)}=0.
 \label{eq:soft-pretarget-self-energy-cancellation}
\end{align}
Because the three denominators in each row are identical, this cancellation
also removes the $\ln z/z$ term produced by expanding
$[z(\bk^2+Q_\gamma^2)/(z_1z_2)]^{-\epsilon}$.  It remains valid with the
analytic rapidity regulator in place.

\subsubsection{Complete graph sum}

For an antiquark tag, five sets cancel as exact identities of regulated
integrands before the singular regions are separated:
\begin{align}
 \Rij{1}{1}+\mathcal V_9+\mathcal V_9^*&=0,
 \nonumber\\
 \Rij{1}{3}+\mathcal V_5^*&=0,&
 \Rij{3}{1}+\mathcal V_5&=0,
 \nonumber\\
 \Rij{1}{4}+\mathcal V_7^*&=0,&
 \Rij{4}{1}+\mathcal V_7&=0.
 \label{eq:soft-exact-cut-identities}
\end{align}
These relations account for emission and reabsorption on the unobserved quark
line.  After they are removed, every remaining logarithmic soft residue is
listed in Table~\ref{tab:soft-graph-ledger}.  Entries in one row multiply the
same endpoint wave function, Wilson-line operator, analytic factor and
$1/R^2$ kernel; the table records only their numerical coefficients.

\begin{table}[H]
\centering
\small
\begin{tabular}{@{}p{92mm}p{39mm}r@{}}
\toprule
graphs or ordered real products & soft coefficients & sum\\
\midrule
$\Rij{2}{2},\ \mathcal V_{10},\ \mathcal V_{10}^*$
 & $+2,-1,-1$ & $0$\\
$\Rij{2}{1},\ \mathcal V_{13(1)},\ \mathcal V_{13(2)}$
 & $-2,+1,+1$ & $0$\\
$\Rij{1}{2},\ \mathcal V_{13(1)}^*,\ \mathcal V_{13(2)}^*$
 & $-2,+1,+1$ & $0$\\
$\Rij{4}{4},\ \Rij{3}{3},\ \Rij{4}{3},\ \Rij{3}{4}$
 & $+2,+2,-2,-2$ & $0$\\
$\Rij{2}{4},\ \Rij{2}{3}$
 & $-2,+2$ & $0$\\
$\Rij{4}{2},\ \Rij{3}{2}$
 & $-2,+2$ & $0$\\
$\mathcal V_6,\ \mathcal V_8$ and their conjugates
 & $+2,-2$ & $0$\\
$\mathcal V_{12},\ \mathcal V_{14(2)}$ and their conjugates
 & $-2,+2$ & $0$\\
$\mathcal V_{11},\ \mathcal V_{14(1)}$ and their conjugates
 & $-2,+2$ & $0$\\
\bottomrule
\end{tabular}
\caption{Leading simultaneous-soft residues in the antiquark-tagged channel.
The exact cut identities in Eq.~\eqref{eq:soft-exact-cut-identities} contain
the remaining real products and virtual graphs.}
\label{tab:soft-graph-ledger}
\end{table}

Each row cancels with its Wilson-line operator kept explicit. The virtual
eikonal current is canceled by the real cut with the same attachments.  The result is
\begin{equation}
 T_S\Sigma^T_{\bar q,\mathrm{NLO}}(\theta)=0.
 \label{eq:soft-antiquark-total}
\end{equation}
Repeating the line-oriented insertions with the quark momentum kept fixed
gives the same coefficient cancellations with the quark and antiquark
coordinates interchanged.  This establishes
\begin{equation}
 T_S\Sigma^T_{q,\mathrm{NLO}}(\theta)=0.
 \label{eq:soft-quark-total}
\end{equation}
for a general target ensemble. Finally,
$T_S\Sigma^T_{g,R}=0$ because the soft-gluon energy weight removes its
leading endpoint pole.  The complete one-point energy flow therefore obeys
\begin{equation}
 \left[\Sigma^T_{q,\mathrm{NLO}}(\theta)
 +\Sigma^T_{\bar q,\mathrm{NLO}}(\theta)
 +\Sigma^T_{g,R}(\theta)\right]_{\rm soft}=0.
 \label{eq:soft-complete-kln}
\end{equation}
Equation~\eqref{eq:soft-complete-kln} is the KLN cancellation for the measured
cross section.  It cancels both the ordinary soft pole and its overlap with
the analytically regulated longitudinal endpoint, while leaving the pure
rapidity term in Eq.~\eqref{eq:joint-antiquark-jimwlk-result} untouched.

\subsection{Final-state collinear singularities}
\label{sec:final-state-collinear}

The collinear limit is taken at a fixed, nonzero daughter momentum fraction.
It must therefore be applied after the soft endpoint has been separated.  At
this stage the analytic rapidity factor is common to all cuts of a parent
splitting, and the transverse collinear pole is isolated dimensionally.  Among the real
products, only the two diagonal emissions after the shockwave contain such a
pole:
\begin{equation}
 \Rij{1}{1}:\quad z_1\bk-z\bp\to0,
 \qquad
 \Rij{2}{2}:\quad z_2\bk-z\bq\to0.
 \label{eq:two-final-state-collinear-faces}
\end{equation}
Their virtual partners are $A_9$ and $A_{10}$, respectively.  The complete
graph-by-graph classification is summarized in
Table~\ref{tab:collinear-graph-ledger}.

\begin{table}[H]
\centering
\small
\begin{tabular}{@{}
>{\raggedright\arraybackslash}p{42mm}
>{\raggedright\arraybackslash}p{58mm}
>{\raggedright\arraybackslash}p{48mm}@{}}
\toprule
graphs & fixed-fraction limit & logarithmic collinear term\\
\midrule
$\Rij{1}{1}$ and $A_9$ & $z_1\bk-z\bp\to0$ & quark-parent pole\\
$\Rij{2}{2}$ and $A_{10}$ & $z_2\bk-z\bq\to0$ & antiquark-parent pole\\
$\Rij{3}{3},\Rij{4}{4}$ & emission before the shockwave & none; photon-formation gap\\
$\Rij{i}{j}$, $i\ne j$ & either face in
Eq.~\eqref{eq:two-final-state-collinear-faces} & none; at most one inverse
power of the relative momentum\\
$A_5,A_6,A_7,A_8$ & fixed loop fraction & none after the displayed numerator
and cut combinations\\
$A_{11},A_{12}$ & fixed loop fraction & none; nonzero wave-function virtuality\\
$A_{13(1)},A_{13(2)}$ & both signed longitudinal sectors & the nominal
$1/\ell^2$ coefficients vanish\\
$A_{14(1)},A_{14(2)}$ & fixed loop fraction & none; nonzero wave-function
virtuality\\
\bottomrule
\end{tabular}
\caption{Final-state collinear limits of the real products and virtual
graphs.  ``None'' means no logarithmic transverse integral at a fixed
nonzero longitudinal fraction.}
\label{tab:collinear-graph-ledger}
\end{table}

The two parent faces are independent.  We derive the antiquark-parent face in
detail and then state the quark-parent result obtained directly from
$\Rij{1}{1}$.

\subsubsection{Antiquark parent}

Introduce the parent fraction and momentum, the antiquark daughter fraction,
and the relative transverse momentum
\begin{equation}
 Z=z_2+z,\qquad u=\frac{z_2}{Z},\qquad
 \boldsymbol P=\bq+\bk,\qquad
 \bell_{\bar q}=\bk-\frac{z}{z_2}\bq.
 \label{eq:coll-antiquark-parent-variables}
\end{equation}
The inverse transformation and its transverse Jacobian are
\begin{align}
 \bq&=u(\boldsymbol P-\bell_{\bar q}),&
 \bk&=(1-u)\boldsymbol P+u\bell_{\bar q},
 \nonumber\\
 \dd^2\bq\,\dd^2\bk
 &=u^2\dd^2\boldsymbol P\,\dd^2\bell_{\bar q}.
 \label{eq:coll-antiquark-inverse-jacobian}
\end{align}
Thus $\bell_{\bar q}\to0$ gives $q^\mu=uP^\mu$ and
$k^\mu=(1-u)P^\mu$.  Moreover,
\begin{equation}
 \frac{\dd z_2}{z_2}=\frac{\dd Z}{Z},
 \qquad
 \left|\frac{\dd z}{z}\right|_{z_2\ {\rm fixed}}
 =\frac{\dd u}{u(1-u)}.
 \label{eq:coll-antiquark-longitudinal-jacobian}
\end{equation}

The collinear reduction can be read directly from
Eq.~\eqref{eq:real-r22-result}.  Dividing its longitudinal numerator by the
Born numerator evaluated at $(Z,\boldsymbol P)$ gives
\begin{equation}
 \frac{z_1Z[z_2^2+Z^2][z_1^2+Z^2]/z_2}
 {z_1Z^2[z_1^2+Z^2]}=\frac{1+u^2}{u}.
 \label{eq:coll-r22-longitudinal-ratio}
\end{equation}
Together with Eq.~\eqref{eq:coll-antiquark-longitudinal-jacobian}, this
produces the convolution Jacobian $u^{-2}$ and the real splitting kernel
\begin{equation}
 C_F\frac{1+u^2}{1-u}.
 \label{eq:coll-real-pqq}
\end{equation}
The Fourier phase, photon wave function and Wilson-line factor simultaneously
reduce to their Born expressions at the parent momentum.  With $d_\perp=2-2\epsilon$, the collinear region is continued from
$\operatorname{Re}\epsilon<0$.  Its pole is $-1/\epsilon_{\rm coll}$,
where the subscript labels the region of the same regulator.  Thus
\begin{equation}
 \frac{g^2}{2\pi}\mu^{2\epsilon}
 \int\frac{\dd^{2-2\epsilon}\bell_{\bar q}}
 {(2\pi)^{2-2\epsilon}}\frac1{\ell_{\bar q}^2}
 =-\frac{\alpha_s}{2\pi}\frac1{\epsilon_{\rm coll}}
 +\text{finite},
 \label{eq:coll-pole-definition}
\end{equation}
the singular part of the real product becomes
\begin{align}
 [\Rij{2}{2}(z_2,\bq)]_{\rm coll}
 ={}&-\frac{\alpha_s}{2\pi}\frac1{\epsilon_{\rm coll}}
 \int_{z_2}^1\frac{\dd u}{u^2}\,
 C_F\frac{1+u^2}{1-u}
 \mathcal V_{\Born}\!\left(\frac{z_2}{u},\frac{\bq}{u}\right).
 \label{eq:coll-r22-factorization}
\end{align}
Here and below ``coll'' denotes the fixed-fraction, soft-subtracted
coefficient.  A common endpoint regulator is understood in the separate real
and virtual terms and cancels from their sum.

\paragraph{Virtual endpoint from graph $A_{10}$.}
The plus prescription follows directly from the virtual graph.
In Eq.~\eqref{eq:g10-result} the external antiquark $(z_2,\bq)$ is the
unsplit parent.  If the loop gluon carries fraction $z$, the internal
antiquark carries $z_2-z$.  Set
\begin{equation}
 u=\frac{z_2-z}{z_2},\qquad z=(1-u)z_2,
 \qquad \frac{\dd z}{z}=-\frac{\dd u}{1-u}.
 \label{eq:coll-v10-variable-change}
\end{equation}
Since $z:0\to z_2$ maps to $u:1\to0$, reversing the limits removes the
minus sign.  The numerator of the external-state correction becomes
\begin{equation}
 \frac{z_2^2+(z_2-z)^2}{z_2^2}=1+u^2.
 \label{eq:coll-v10-numerator}
\end{equation}
Thus the same unregularized splitting function found from the real graph has
now been obtained independently from the virtual loop.

More explicitly, the collinear pole of
Eq.~\eqref{eq:g10-final-amplitudes} is
\begin{equation}
 [i\mathcal A_{10}]_{\rm coll}
 =\frac{\alpha_sC_F}{4\pi}\frac1{\epsilon_{\rm coll}}
 \int_0^1\dd u\,\frac{1+u^2}{1-u}\,i\mathcal A_0.
 \label{eq:coll-v10-amplitude-factorization}
\end{equation}
The interference with the Born amplitude on the two sides of the cut
therefore gives
\begin{equation}
 2\operatorname{Re}\!\left[
 (i\mathcal A_{10})(i\mathcal A_0)^*\right]_{\rm coll}
 =\frac{\alpha_sC_F}{2\pi}\frac1{\epsilon_{\rm coll}}
 |i\mathcal A_0|^2\int_0^1\dd u\,\frac{1+u^2}{1-u}.
 \label{eq:coll-v10-interference}
\end{equation}
Using Eq.~\eqref{eq:coll-pole-definition}, this gives the virtual counterpart
of Eq.~\eqref{eq:coll-r22-factorization}:
\begin{equation}
 [2\operatorname{Re}\mathcal V_{10}(z_2,\bq)]_{\rm coll}
 =\frac{\alpha_s}{2\pi}\frac1{\epsilon_{\rm coll}}
 \mathcal V_{\Born}(z_2,\bq)
 \int_0^1\dd u\,C_F\frac{1+u^2}{1-u}.
 \label{eq:coll-v10-factorization}
\end{equation}
The target color operator, photon wave function and external momentum in this
term are exactly those of the Born graph.  Adding Eqs.~\eqref{eq:coll-r22-factorization}
and \eqref{eq:coll-v10-factorization} therefore gives
\begin{align}
 [\Rij{2}{2}+2\operatorname{Re}\mathcal V_{10}]_{\rm coll}
 ={}&-\frac{\alpha_s}{2\pi}\frac1{\epsilon_{\rm coll}}
 \left[
 \int_{z_2}^1\frac{\dd u}{u^2}\,
 C_F\frac{1+u^2}{1-u}
 \mathcal V_{\Born}\!\left(\frac{z_2}{u},\frac{\bq}{u}\right)
 \right.\nonumber\\[-2pt]
 &\left.\hspace{18mm}
 -\mathcal V_{\Born}(z_2,\bq)
 \int_0^1\dd u\,C_F\frac{1+u^2}{1-u}
 \right].
 \label{eq:coll-r22-v10-endpoint-subtraction}
\end{align}
This is precisely the plus prescription.  For any smooth function
$\varphi(u)$,
\begin{equation}
 \int_0^1\dd u\,[f(u)]_+\varphi(u)
 =\int_0^1\dd u\,f(u)[\varphi(u)-\varphi(1)].
 \label{eq:coll-plus-definition}
\end{equation}
In the present case
\begin{equation}
 \frac{1+u^2}{1-u}=\frac2{1-u}-(1+u),
 \qquad \int_0^1\dd u\,(1+u)=\frac32.
 \label{eq:coll-pqq-endpoint-integral}
\end{equation}
It follows that
\begin{equation}
 P_{qq}(u)=C_F\left[\frac{1+u^2}{1-u}\right]_+
 =C_F\left[\frac{2}{(1-u)_+}-1-u
 +\frac32\delta(1-u)\right].
 \label{eq:coll-pqq-plus}
\end{equation}
Hence Eq.~\eqref{eq:coll-r22-v10-endpoint-subtraction} can be written directly
in terms of the physical Born kernel,
\begin{equation}
 [\Rij{2}{2}+2\operatorname{Re}\mathcal V_{10}]_{\rm coll}
 =-\frac{\alpha_s}{2\pi}\frac1{\epsilon_{\rm coll}}
 \int_{z_2}^1\frac{\dd u}{u^2}\,P_{qq}(u)
 \mathcal V_{\Born}\!\left(\frac{z_2}{u},\frac{\bq}{u}\right).
 \label{eq:coll-r22-v10-factorization}
\end{equation}
The $3\delta(1-u)/2$ term is the endpoint contribution of $A_{10}$ when its
subtraction is expressed in the conventional $1/(1-u)_+$ basis.

For a parent with fraction $Z$ and transverse momentum
$\boldsymbol P$, its energy-flow insertion is explicitly
\begin{equation}
 \frac{E_P}{E_A}\delta(\theta^2-\vartheta_P^2)
 =x_B\left(Z+\frac{\boldsymbol P^2}{Q^2Z}\right)
 \delta\!\left[\theta^2-4\arctan^2\!\frac{QZ}{|\boldsymbol P|}\right].
 \label{eq:coll-parent-measurement}
\end{equation}
Linearity of the unrestricted detector gives, at exact collinearity,
\begin{align}
 \vartheta_{\bar q}(uP)&=\vartheta_g((1-u)P)
 =2\arctan\frac{QZ}{|\boldsymbol P|},
 \nonumber\\
 \mathcal M_{\bar q}(\theta;uZ,u\boldsymbol P)
 &=u\,x_B\left(Z+\frac{\boldsymbol P^2}{Q^2Z}\right)
 \delta\!\left[\theta^2-4\arctan^2\!\frac{QZ}{|\boldsymbol P|}\right],
 \nonumber\\
 \mathcal M_g(\theta;(1-u)Z,(1-u)\boldsymbol P)
 &=(1-u)x_B\left(Z+\frac{\boldsymbol P^2}{Q^2Z}\right)
 \delta\!\left[\theta^2-4\arctan^2\!\frac{QZ}{|\boldsymbol P|}\right].
 \label{eq:coll-measurement-sum-rule}
\end{align}
Summing the daughter energy tags reproduces the virtual parent insertion
for every parent energy.

For the antiquark energy tag, substituting
Eq.~\eqref{eq:coll-measurement-sum-rule} into the real term and using the
explicit parent insertion in the virtual term gives
\begin{align}
 [\Sigma^T_{\bar q,R+V}]_{\rm coll}
 ={}&-\frac{\alpha_s}{2\pi}\frac1{\epsilon_{\rm coll}}
 \int_0^1\frac{\dd Z}{Z}\int\dd^2\boldsymbol P\,
 \mathcal V_{\Born}(Z,\boldsymbol P)
 \nonumber\\[-2pt]
 &\times x_B\left(Z+\frac{\boldsymbol P^2}{Q^2Z}\right)
 \delta\!\left[\theta^2-4\arctan^2\!\frac{QZ}{|\boldsymbol P|}\right]
 \int_0^1\dd u\,uP_{qq}(u),
 \label{eq:coll-fixed-antiquark-pole}\\
 \int_0^1\dd u\,uP_{qq}(u)
 ={}&C_F\int_0^1\dd u\,
 \frac{1+u^2}{1-u}(u-1)
 =-\frac43C_F.
 \label{eq:coll-fixed-antiquark-moment}
\end{align}
For an identified hadron, this moment enters the fragmentation
counterterm~\cite{Bergabo2023,Altinoluk2025}. The energy-flow sum also
contains the gluon tag in the same collinear face. With $w=z/Z=1-u$, its kernel and energy moment are
\begin{align}
 P_{gq}(w)&=C_F\frac{1+(1-w)^2}{w},
 \label{eq:coll-pgq}\\
 [\Sigma^T_{g\mid\bar q}]_{\rm coll}
 ={}&-\frac{\alpha_s}{2\pi}\frac1{\epsilon_{\rm coll}}
 \int_0^1\frac{\dd Z}{Z}\int\dd^2\boldsymbol P\,
 \mathcal V_{\Born}(Z,\boldsymbol P)
 \nonumber\\[-2pt]
 &\times x_B\left(Z+\frac{\boldsymbol P^2}{Q^2Z}\right)
 \delta\!\left[\theta^2-4\arctan^2\!\frac{QZ}{|\boldsymbol P|}\right]
 \int_0^1\dd w\,wP_{gq}(w),
 \nonumber\\[-2pt]
 \int_0^1\dd w\,wP_{gq}(w)
 ={}&C_F\int_0^1\dd w\,[1+(1-w)^2]
 =+\frac43C_F.
 \label{eq:coll-gluon-antiquark-parent-moment}
\end{align}
The two moments cancel.  More directly, the cancellation is pointwise in the
splitting fraction already at the level of the common real kernel:
\begin{align}
 &[\Sigma^T_{\bar q,R+V}+\Sigma^T_{g\mid\bar q}]_{\rm coll}
 \nonumber\\
 &\quad=-\frac{\alpha_s}{2\pi}\frac1{\epsilon_{\rm coll}}
 \int_0^1\frac{\dd Z}{Z}\int\dd^2\boldsymbol P\,
 \mathcal V_{\Born}(Z,\boldsymbol P)
 \nonumber\\[-2pt]
 &\times x_B\left(Z+\frac{\boldsymbol P^2}{Q^2Z}\right)
 \delta\!\left[\theta^2-4\arctan^2\!\frac{QZ}{|\boldsymbol P|}\right]
 \int_0^1\dd u\,C_F\frac{1+u^2}{1-u}
 \big[u+(1-u)-1\big]=0.
 \label{eq:coll-antiquark-pointwise-cancellation}
\end{align}
The three entries in brackets are the antiquark energy, the gluon energy and
the virtual parent, respectively.  The quark is a spectator in this face; its
real and virtual detector insertions are identical and cancel separately.

\subsubsection{Quark parent and the inclusive energy sum}

For the quark-parent face, use the same parent variables with
\begin{equation}
 Z=z_1+z,\qquad u=\frac{z_1}{z_1+z},\qquad
 \boldsymbol P=\bp+\bk,\qquad
 \bell_q=\bk-\frac{z}{z_1}\bp.
 \label{eq:coll-quark-parent-variables}
\end{equation}
The color square of $\Rij{1}{1}$ is $C_F$ times the quark-tagged Born
operator, and its longitudinal numerator gives the same real splitting
factor,
\begin{equation}
 C_F\frac{1+u^2}{1-u}.
 \label{eq:coll-quark-parent-kernel}
\end{equation}
The collinear pole of Eq.~\eqref{eq:g9-final-amplitudes} has the form
of Eq.~\eqref{eq:coll-v10-amplitude-factorization}; $A_9$ supplies the
parent endpoint subtraction for this face. Adding the quark tag, the
gluon tag and the virtual endpoint gives the explicit cancellation
\begin{align}
 [\Sigma^T_{q,R+V}+\Sigma^T_{g\mid q}]_{\rm coll}
 ={}&-\frac{\alpha_s}{2\pi}\frac1{\epsilon_{\rm coll}}
 \int_0^1\frac{\dd Z}{Z}\int\dd^2\boldsymbol P\,
 \mathcal V_{\Born}(Z,\boldsymbol P)
 \nonumber\\[-2pt]
 &\times x_B\left(Z+\frac{\boldsymbol P^2}{Q^2Z}\right)
 \delta\!\left[\theta^2-4\arctan^2\!\frac{QZ}{|\boldsymbol P|}\right]
 \nonumber\\[-2pt]
 &\times\int_0^1\dd u\,C_F\frac{1+u^2}{1-u}
 [u+(1-u)-1]=0,
 \label{eq:coll-quark-pointwise-cancellation}
 \\
 \int_0^1\dd u\,uP_{qq}(u)&=-\frac43C_F,
 \qquad
 \int_0^1\dd w\,wP_{gq}(w)=+\frac43C_F.
 \label{eq:coll-quark-parent-moments}
\end{align}
Here $\mathcal V_{\Born}$ is the quark-tagged Born kernel and $w=1-u$.
The spectator antiquark cancels separately between the real cut and the
$A_9$ interference.

The gluon-tagged real cross section has thus supplied exactly two singular
faces,
\begin{equation}
 [\Sigma^T_{g,R}]_{\rm coll}
 =[\Sigma^T_{g\mid q}]_{\rm coll}
 +[\Sigma^T_{g\mid\bar q}]_{\rm coll}.
 \label{eq:coll-gluon-two-faces}
\end{equation}
The fixed-order master formula retains the finite noncollinear contribution.

Combining Eqs.~\eqref{eq:coll-antiquark-pointwise-cancellation} and
\eqref{eq:coll-quark-pointwise-cancellation} yields
\begin{equation}
 \left[\Sigma_q^T(\theta)+\Sigma_{\bar q}^T(\theta)
 +\Sigma_g^T(\theta)\right]_{\rm final\text{-}state\ coll}=0.
 \label{eq:coll-complete-energy-flow}
\end{equation}
The inclusive one-point energy measurement is therefore final-state collinear
safe at NLO, with the cancellation completed in the partonic energy
measurement. A subsequent small-angle expansion will match its
target-collinear sector onto the NEEC.

\subsection{NLO CGC factorization and summary}
\label{sec:nlo-cgc-factorization-summary}

These cancellations establish NLO CGC factorization. To display it
directly, the target average is reopened according to
$\langle O\rangle_{Y_f}=\int D\rho\,W_{Y_f}[\rho]O[\rho]$.  Every amplitude in
the equation below is then evaluated in one fixed shockwave field $\rho$.
The rapidity boundary $Y_f$ is the boundary associated with the factorization
scale $\nu_f^+$, and the analytic regulator in all longitudinal integrals is
evaluated with $\nu^+=\nu_f^+$.

Before rapidity renormalization, the remaining pole of the complete energy
flow is
\begin{equation}
 \left[
 \Sigma^T_{E,V}(\theta)+\Sigma^T_{E,R}(\theta)
 \right]_{1/\eta_Y}
 =-\frac1{\eta_Y}H_{\rm JIMWLK}
 \Sigma^T_{E,\Born}(\theta).
 \label{eq:summary-bare-rapidity-pole}
\end{equation}
Here $H_{\rm JIMWLK}$ acts only on the Wilson lines inside the Born
coordinate-space convolutions.  The quark and antiquark detector weights
commute with that action.  Equation~\eqref{eq:summary-bare-rapidity-pole}
contains the two independently derived finite-$N_c$ actions in
Eqs.~\eqref{eq:joint-antiquark-jimwlk-result} and
\eqref{eq:joint-quark-jimwlk-result}; the gluon tag has no fixed-transverse-
momentum rapidity residue at $\theta>0$ by
Eq.~\eqref{eq:joint-gluon-tagged-three-zeroes}.

At fixed $\rho$, the sum of the complete virtual and real terms has the
Laurent expansion
\begin{align}
&\int\frac{\dd\Phi^{(2)}}{2l^+}\,
 \mathcal M_E^{(2)}\,
 2\operatorname{Re}\frac12\sum_{\lambda,h,\bar h}
 \sum_{n=5}^{14}
 \tr\!\left[(i\mathcal A_{n,\lambda h\bar h}[\rho])
 (i\mathcal A_{0,\lambda h\bar h}[\rho])^\dagger\right]
 \nonumber\\[-2pt]
&\quad+
 \int\frac{\dd\Phi_{\eta_Y}^{(3)}}{2l^+}\,
 \mathcal M_E^{(3)}
 \frac12\sum_{\lambda,h,\bar h,\lambda_g,a}
 \sum_{i,j=1}^{4}
 \tr\!\left[(i\mathcal M_{i,\lambda h\bar h\lambda_g}^{a}[\rho])
 (i\mathcal M_{j,\lambda h\bar h\lambda_g}^{a}[\rho])^\dagger\right]
 \nonumber\\
&={}-\frac1{\eta_Y}H_{\rm JIMWLK}
 \int\frac{\dd\Phi^{(2)}}{2l^+}\,
 \mathcal M_E^{(2)}
 \frac12\sum_{\lambda,h,\bar h}
 \tr\!\left[(i\mathcal A_{0,\lambda h\bar h}[\rho])
 (i\mathcal A_{0,\lambda h\bar h}[\rho])^\dagger\right]
 \nonumber\\[-2pt]
&\quad+\mathcal I_{E,\mathrm{NLO}}^{T,\mathrm{fin}}
 (\theta;\rho,\nu_f^+)
 +\mathcal O(\eta_Y,\epsilon).
 \label{eq:summary-finite-impact-factor-decomposition}
\end{align}
The finite quantity
$\mathcal I_{E,\mathrm{NLO}}^{T,\mathrm{fin}}$ is the inclusive NLO
projectile impact factor.  It contains the full $q+\bar q$ virtual measurement,
the full $q+\bar q+g$ real measurement, and all finite noncollinear parts of
the four real amplitudes.  The two- and three-body measures in
Eq.~\eqref{eq:summary-finite-impact-factor-decomposition} are the physical
measures in Eqs.~\eqref{eq:born-phase-space} and
\eqref{eq:real-three-body-phase-space}, retaining the complete fixed-order
kinematics in the finite remainder.

The rapidity logarithm is retained in this finite impact factor.  Its
scale-dependent part is
\begin{align}
 \left.\mathcal I_{E,\mathrm{NLO}}^{T,\mathrm{fin}}
 (\theta;\rho,\nu_f^+)\right|_{\ln(l^+/\nu_f^+)}
 ={}&\ln\frac{l^+}{\nu_f^+}\,
 H_{\rm JIMWLK}
 \int\frac{\dd\Phi^{(2)}}{2l^+}\,
 \mathcal M_E^{(2)}
 \frac12\sum_{\lambda,h,\bar h}
 \tr\!\left[(i\mathcal A_{0,\lambda h\bar h}[\rho])
 (i\mathcal A_{0,\lambda h\bar h}[\rho])^\dagger\right].
 \label{eq:summary-finite-impact-factor-log}
\end{align}
The graph-dependent endpoint logarithms in
Eq.~\eqref{eq:joint-pointwise-finite-endpoint} and the endpoint-subtracted
integrals in Eq.~\eqref{eq:joint-full-graph-finite-analytic} remain in the same
finite impact factor.

We now display the absorption of the rapidity pole before writing the
factorized cross section.  The ensemble renormalization equivalent to
Eq.~\eqref{eq:joint-rapidity-renormalization} is
\begin{equation}
 W_{\mathrm{bare}}[\rho]
 =W_{Y_f}[\rho]+\frac1{\eta_Y}
 H_{\rm JIMWLK}^{\dagger}W_{Y_f}[\rho],
 \label{eq:summary-target-weight-renormalization}
\end{equation}
where functional integration by parts gives
$\int D\rho\,(H_{\rm JIMWLK}^{\dagger}W)O
=\int D\rho\,W(H_{\rm JIMWLK}O)$.  Therefore the universal target ensemble
times the Born impact factor absorbs the inclusive NLO rapidity pole
explicitly:
\begin{align}
&\int D\rho\,W_{\mathrm{bare}}[\rho]
 \int\frac{\dd\Phi^{(2)}}{2l^+}\,
 \mathcal M_E^{(2)}
 \frac12\sum_{\lambda,h,\bar h}
 \tr\!\left[(i\mathcal A_{0,\lambda h\bar h}[\rho])
 (i\mathcal A_{0,\lambda h\bar h}[\rho])^\dagger\right]
 +\left[\Sigma^T_{E,V}+\Sigma^T_{E,R}\right]_{1/\eta_Y}
 \nonumber\\
={}&\int D\rho\,W_{Y_f}[\rho]
 \int\frac{\dd\Phi^{(2)}}{2l^+}\,
 \mathcal M_E^{(2)}
 \frac12\sum_{\lambda,h,\bar h}
 \tr\!\left[(i\mathcal A_{0,\lambda h\bar h}[\rho])
 (i\mathcal A_{0,\lambda h\bar h}[\rho])^\dagger\right]
 \nonumber\\
&+\frac1{\eta_Y}\int D\rho\,W_{Y_f}[\rho]
 H_{\rm JIMWLK}
 \int\frac{\dd\Phi^{(2)}}{2l^+}\,
 \mathcal M_E^{(2)}
 \frac12\sum_{\lambda,h,\bar h}
 \tr\!\left[(i\mathcal A_{0,\lambda h\bar h}[\rho])
 (i\mathcal A_{0,\lambda h\bar h}[\rho])^\dagger\right]
 \nonumber\\
&-\frac1{\eta_Y}\int D\rho\,W_{Y_f}[\rho]
 H_{\rm JIMWLK}
 \int\frac{\dd\Phi^{(2)}}{2l^+}\,
 \mathcal M_E^{(2)}
 \frac12\sum_{\lambda,h,\bar h}
 \tr\!\left[(i\mathcal A_{0,\lambda h\bar h}[\rho])
 (i\mathcal A_{0,\lambda h\bar h}[\rho])^\dagger\right]
 \nonumber\\
={}&\int D\rho\,W_{Y_f}[\rho]
 \int\frac{\dd\Phi^{(2)}}{2l^+}\,
 \mathcal M_E^{(2)}
 \frac12\sum_{\lambda,h,\bar h}
 \tr\!\left[(i\mathcal A_{0,\lambda h\bar h}[\rho])
 (i\mathcal A_{0,\lambda h\bar h}[\rho])^\dagger\right].
 \label{eq:summary-explicit-rapidity-absorption}
\end{align}

After Eq.~\eqref{eq:summary-explicit-rapidity-absorption}, every quantity in the factorization formula is finite:
\begin{equation}
\boxed{\begin{aligned}
 \Sigma_E^T(\theta)={}&
 \int D\rho\,
 \underbrace{W_{Y_f}[\rho]}_{\substack{\text{universal target}\\
                                      \text{CGC ensemble}}}
 \Bigg\{
 \int\frac{\dd\Phi^{(2)}}{2l^+}\,
 \mathcal M_E^{(2)}
 \frac12\sum_{\lambda,h,\bar h}
 \tr\!\left[(i\mathcal A_{0,\lambda h\bar h}[\rho])
 (i\mathcal A_{0,\lambda h\bar h}[\rho])^\dagger\right]
 \\[-2pt]
 &\hspace{31mm}+\mathcal I_{E,\mathrm{NLO}}^{T,\mathrm{fin}}
 (\theta;\rho,\nu_f^+)\Bigg\}
 +\mathcal O(\alpha_s^2).
\end{aligned}}
\label{eq:summary-nlo-cgc-factorization}
\end{equation}
This is the CGC factorization: $W_{Y_f}[\rho]$ is independent of the
virtual-photon wave function, detector angle and final-state tag, while the
finite quantity inside braces is the process- and observable-dependent
projectile impact factor in the fixed field $\rho$.  After the $\rho$
integral, the exact dipoles, quadrupoles and products of traces remain inside
one finite-$N_c$ target average. In the large-$N_c$ mean-field limit,
the dipole evolution reduces to the BK equation~\cite{Kovchegov1999}.

The finiteness of
$\mathcal I_{E,\mathrm{NLO}}^{T,\mathrm{fin}}$ follows from four independent checks.  The mixed ultraviolet--rapidity residues vanish by
Eq.~\eqref{eq:joint-total-overlap-cancellation}; the pure ultraviolet poles
vanish by Eq.~\eqref{eq:joint-total-pure-uv-cancellation}; the soft poles
cancel by Eq.~\eqref{eq:soft-complete-kln}; and the final-state collinear poles
cancel by Eq.~\eqref{eq:coll-complete-energy-flow}.

The dependence on the rapidity factorization scale cancels between the two finite
factors.  With the convention of Eq.~\eqref{eq:joint-rapidity-rg},
\begin{align}
 \frac{\partial}{\partial\ln\nu_f^+}W_{Y_f}[\rho]
 & =H_{\rm JIMWLK}^{\dagger}W_{Y_f}[\rho],
 \nonumber\\
 \frac{\partial}{\partial\ln\nu_f^+}
 \mathcal I_{E,\mathrm{NLO}}^{T,\mathrm{fin}}
 &=-H_{\rm JIMWLK}
 \int\frac{\dd\Phi^{(2)}}{2l^+}\,
 \mathcal M_E^{(2)}
 \frac12\sum_{\lambda,h,\bar h}
 \tr\!\left[(i\mathcal A_{0,\lambda h\bar h}[\rho])
 (i\mathcal A_{0,\lambda h\bar h}[\rho])^\dagger\right].
 \label{eq:summary-two-factor-rapidity-rg}
\end{align}
Functional integration by parts then gives
\begin{equation}
 \frac{\dd}{\dd\ln\nu_f^+}\Sigma_E^T(\theta)
 =\mathcal O(\alpha_s^2).
 \label{eq:summary-rapidity-scale-independence}
\end{equation}
Equation~\eqref{eq:summary-nlo-cgc-factorization} therefore establishes NLO
CGC factorization for the transverse one-point energy flow at finite angle,
within the leading-power eikonal setup used in this paper. A subsequent
small-$\theta$ expansion and matching of the target-collinear region onto
the space-like NEEC evolution kernel are left for future work.

\section{Comparison with inclusive DIS}
\label{sec:finite-validation}

The full-angle weighted moment of our energy-flow cross section equals
the inclusive photon--target cross section. We use this identity to compare
its finite transverse-photon impact factors with Ref.~\cite{Beuf2017},
arXiv:1708.06557v1. The comparison concerns the incoming virtual and real
terms that survive in Eq.~\eqref{eq:finite-inclusive-reduction}, using the
reference's CDR, antenna and gluon-cutoff conventions.

We compare the analytic coefficients of the dipole and three-parton
Wilson-line operators before convolution with a target model. The
comparison retains the complete spin, color, flux and photon-average
normalization and holds at arbitrary transverse coordinates and
longitudinal fractions.

\subsection{A weighted angular moment equal to inclusive DIS}
\label{sec:finite-inclusive-projector}

Using the exact massless kinematics of
Eq.~\eqref{eq:particle-energy-angle}, choose
\begin{equation}
 w_{\rm inc}(\theta)=\frac{1-\cos\theta}{2x_B},\qquad
 w_{\rm inc}(\vartheta_i)\frac{E_i}{E_A}=z_i.
 \label{eq:finite-inclusive-projector}
\end{equation}
This weight converts each detected energy into its longitudinal momentum
fraction. Since $z_1+z_2=1$ in the virtual final state and
$z_1+z_2+z=1$ in the real final state, the summed detector weight is one
in either case. Thus
\begin{equation}
 \int_0^{\pi^2}\dd\theta^2\,
 \frac{1-\cos\theta}{2x_B}\Sigma_E^T(\theta)
 =\sigma_T^{\gamma^*A}.
 \label{eq:finite-angular-pairing}
\end{equation}
The moment covers the complete angular domain and includes all quark,
antiquark and gluon tags, with the exact angular kinematics.

The real--virtual cancellations used here follow from the exact cut
relations already established in Eqs.~\eqref{eq:real-exact-inclusive-relations}
and~\eqref{eq:soft-exact-cut-identities}. The paired contributions,
including their finite parts, cancel when the full real and virtual
kernels are added, before taking the angular moment. For the fixed
antiquark tag, the displayed pairs concern radiation from the unobserved
quark line; the quark-tagged relations follow by interchange.

Equation~\eqref{eq:finite-angular-pairing} gives unit summed final-state
weight and identifies the result with inclusive DIS. Completing the cut
sum over all final-state lines, with the same regulator and phase-space
normalization, implements the unitarity of QCD evolution after the
shockwave. After-target real emissions and their virtual partners cancel,
as do before--after interferences and the corresponding virtual graphs.
The surviving NLO graph contributions in the moment are
\begin{equation}
 2\operatorname{Re}\left[
 (\mathcal A_{11}+\mathcal A_{12}
  +\mathcal A_{14(1)}+\mathcal A_{14(2)})\mathcal A_0^\dagger\right]
 +|\mathcal A_3+\mathcal A_4|^2,
 \label{eq:finite-inclusive-reduction}
\end{equation}
with the phase-space integrals, spin and color sums, transverse-photon
average and common subtractions of the original calculation. These are
the incoming virtual and real contributions compared with inclusive DIS.
The following comparison identifies their finite coefficients with the
inclusive DIS result.

\subsection{Regularization and subtraction conventions}
\label{sec:finite-uv-matching}

The individual NLO graphs contain four types of singular limits: UV at
large transverse loop momentum (or coincident emission coordinates),
rapidity at $z\to0$ with nonzero transverse momentum held fixed, ordinary
soft when the gluon momentum vanishes in all components, and final-state
collinear when a daughter is parallel to its parent at fixed momentum
sharing. Section~\ref{sec:singularity-structure} separates these limits
and their overlaps before the finite terms are combined.

The soft poles cancel between matched real and virtual cuts; the
soft-gluon tag has a vanishing energy weight. The final-state collinear
poles cancel because the two daughter energy weights sum to the parent
weight, $u+(1-u)=1$. These are the cancellations in
Eqs.~\eqref{eq:soft-complete-kln} and~\eqref{eq:coll-complete-energy-flow}.
The cut sums described in Sec.~\ref{sec:finite-inclusive-projector} leave
the incoming contributions in Eq.~\eqref{eq:finite-inclusive-reduction}.
The local UV
limit and the rapidity endpoint of the remaining incoming terms are
handled explicitly as follows.

\paragraph{FDH and CDR.}
Both schemes use momentum dimension $D=4-2\epsilon$. They differ in the
spin dimension: FDH sets $D_s=4$ and keeps observed helicities in four
dimensions, whereas CDR sets $D_s=D$ for the spin algebra and uses
$D-2$ transverse photon polarizations. The angular amplitudes in
Sec.~\ref{sec:fixed-order} use FDH. For the inclusive comparison we
evaluate the primitive numerators in CDR, with the reference's
$1/(D-2)$ photon average. The $D$-dimensional Born factor is common to the virtual
term and its integrated real UV counterterm and is retained until their
poles cancel. The UV-finite nonlocal remainder can then be evaluated in
four dimensions.

The primitive quark splitting trace gives
$P_{qq}^{\rm CDR}(u)=1+u^2-\epsilon(1-u)^2$.
The $D_s$ dependence of the initial vertex is reconstructed independently
from its raw Dirac chain. The resulting finite changes from FDH, in units
$\alpha_sC_F/\pi$ multiplying the Born cross section, are
\begin{equation}
 \Delta C_{\rm virtual}=\frac12,\qquad
 \Delta C_{{\rm UV},q}=\Delta C_{{\rm UV},\bar q}=-\frac14,
 \qquad \Delta C_{\rm dipole}=0.
 \label{eq:reference-CDR-cancellation}
\end{equation}
The common $O(\epsilon)$ Born factor likewise multiplies a zero sum of
UV residues. These cancellations are checked before comparing finite
coefficients; they are the CDR cancellation described in Sec.~IV.C of
Ref.~\cite{Beuf2017}.

\paragraph{UV subtraction in the reference convention.}
For emission from the quark, the reference antenna has the
$D$-dimensional coordinate kernel
\begin{equation}
 \frac{\bx_{31}}{(\bx_{31}^2)^{1-\epsilon}}\cdot
 \left[\frac{\bx_{31}}{(\bx_{31}^2)^{1-\epsilon}}
       -\frac{\bx_{32}}{(\bx_{32}^2)^{1-\epsilon}}\right]
 \xrightarrow[\epsilon\to0]{}
 \frac1{\bx_{31}^2}
 -\frac{\bx_{31}\cdot\bx_{32}}{\bx_{31}^2\bx_{32}^2}.
 \label{eq:reference-antenna-profile}
\end{equation}
The antiquark antenna follows by $1\leftrightarrow2$.
Including the dimensional Fourier factors, the transverse integral
multiplying the Born term is
\begin{equation}
 4\pi J_{\rm ant}(\epsilon,|\bx_{12}|)
 =\frac{\Gamma(1-\epsilon)}{\epsilon}
   (\pi\mu^2\bx_{12}^2)^\epsilon
 =\frac1\epsilon+\gamma_E+\ln(\pi\mu^2\bx_{12}^2)+O(\epsilon).
 \label{eq:reference-antenna-master}
\end{equation}
The antenna counterterm is subtracted from
$|\mathcal A_3+\mathcal A_4|^2$, the second term in
Eq.~\eqref{eq:finite-inclusive-reduction}. Its transverse integral in
Eq.~\eqref{eq:reference-antenna-master} is added to the virtual
interference, the first term in Eq.~\eqref{eq:finite-inclusive-reduction}.
This rearrangement preserves the total cross section; the integrated
counterterm cancels the UV poles of the virtual contributions. The
$-\epsilon(1-u)^2$ numerator is retained against the UV pole, giving the
real terms in Eq.~\eqref{eq:reference-CDR-cancellation}.

The antenna is fixed by the emitter and spectator coordinates. Its
integral over the whole gluon transverse plane is evaluated analytically
in $D-2=2-2\epsilon$ dimensions and expanded through the finite term.

\paragraph{Rapidity cutoff and target factorization.}
We also use the reference's minimal longitudinal cutoff prescription,
$z>\lambda\equiv k^+_{\min}/l^+$, with $z_c=0$. Let $F_e(z)$ be $z$
times the antenna-subtracted real density in sector $e$, and
$z_{\max,e}$ its upper gluon fraction. The actual cutoff integral obeys
\begin{align}
 \int_\lambda^{z_{\max,e}}\frac{\dd z}{z}F_e(z)
 ={}&\int_\lambda^{z_{\max,e}}\frac{\dd z}{z}[F_e(z)-F_e(0)]
 +F_e(0)\ln\frac{z_{\max,e}}\lambda,\nonumber\\
 \lim_{\lambda\to0}\left[
 \int_\lambda^{z_{\max,e}}\frac{\dd z}{z}F_e(z)
 -F_e(0)\ln\frac{z_f}\lambda\right]
 ={}&\int_0^{z_{\max,e}}\frac{\dd z}{z}[F_e(z)-F_e(0)]
 +F_e(0)\ln\frac{z_{\max,e}}{z_f}.
 \label{eq:reference-cutoff-factorization}
\end{align}
The endpoint coefficient combines with the virtual residues into the
JIMWLK action on the Born operator. The subtracted interval from
$\lambda$ to the common factorization boundary $z_f$ is assigned to the
target, while the finite upper-endpoint logarithm remains in the hard
kernel. Both expressions retain the same lower cutoff, target
factorization boundary and finite upper-endpoint logarithm.

\subsection{Comparison with the inclusive finite kernels}
\label{sec:inclusive-finite-check}

The first term in Eq.~\eqref{eq:finite-inclusive-reduction} contains
$\mathcal A_{11}$, $\mathcal A_{12}$, $\mathcal A_{14(1)}$ and
$\mathcal A_{14(2)}$
interfering with $\mathcal A_0$. Their cross-section contributions are
given in Eqs.~\eqref{eq:g11-result}, \eqref{eq:g12-result},
\eqref{eq:g141-result} and~\eqref{eq:g142-result}. After taking the
inclusive moment and adding the integrated antenna counterterm from the
second term, their finite sum determines $C_T(z_1)$ in
\begin{equation}
 \dd\sigma_{T,\mathrm{dip}}
 =\dd\sigma_{T,\Born}
 \left[1+\frac{\alpha_sC_F}{\pi}C_T(z_1)\right],\qquad
 C_T(z_1)=\frac12\ln^2\frac{z_1}{z_2}+\frac52-\frac{\pi^2}{6}.
 \label{eq:finite-inclusive-dipole-coefficient}
\end{equation}
Integrating the finite longitudinal kernels in the two emitter sectors
gives
\begin{equation*}
 C_T(z_1)=\frac52-\frac{\pi^2}{3}
 -\operatorname{Li}_2\!\left(-\frac{z_1}{z_2}\right)
 -\operatorname{Li}_2\!\left(-\frac{z_2}{z_1}\right).
\end{equation*}
The identity
$\operatorname{Li}_2(-x)+\operatorname{Li}_2(-1/x)
=-\pi^2/6-\tfrac12\ln^2x$, for $x>0$, yields
Eq.~\eqref{eq:finite-inclusive-dipole-coefficient}, exactly the finite
coefficient in Eq.~(147) of arXiv:1708.06557v1~\cite{Beuf2017}.
The cutoff logarithms and UV poles cancel before this integration, with
the scheme-dependent finite terms canceling as in
Eq.~\eqref{eq:reference-CDR-cancellation}.

For the real contribution, the unit weight in
Eq.~\eqref{eq:finite-angular-pairing} allows the unobserved momentum
integrals to identify the coordinates on the two sides of the cut.
The sum $\Rij{3}{3}+\Rij{4}{4}+2\operatorname{Re}\Rij{4}{3}$ in
Eqs.~\eqref{eq:real-r44-result}--\eqref{eq:real-r43-result} then contains
the target factor
\begin{equation*}
 N_c^2(1-S_{13}S_{32})-(1-S_{12}).
\end{equation*}
The product of dipoles remains inside one common target average.
Subtracting the two antenna terms and regrouping the helicity products
gives a quark-emission contribution and its quark--antiquark interchange.
For one quark flavor, the first contribution is
\begin{align}
 \frac{\dd\sigma_T|_{q\to g}}
 {\dd z_1\,\dd z\prod_{i=1}^3\dd^2\bx_i}
 ={}&\frac{N_cC_F\alpha_{\rm em}\alpha_se_f^2Q^2}{2\pi^4}
 z_2\Bigg\{
 [(z_1+z)^2+z_2^2]\frac{(z_1+z)^2+z_1^2}{z(z_1+z)}
 \nonumber\\[-2pt]
 &\times\left(\frac1{\bx_{31}^2}
 -\frac{\bx_{31}\cdot\bx_{32}}{\bx_{31}^2\bx_{32}^2}\right)
 \Bigg[K_1(QX)^2
 \operatorname{Re}\!\left(1-\frac{N_c^2S_{13}S_{32}-S_{12}}{N_c^2-1}\right)
 \nonumber\\[-2pt]
 &\hspace{13mm}
 -K_1\!\left(Q\sqrt{z_2(z_1+z)}\,|\bx_{12}|\right)^2
 \operatorname{Re}(1-S_{12})\Bigg]
 \nonumber\\
 &+\frac{z}{z_1+z}\Bigg[
 [(z_1+z)^2+z_2^2]
 \frac{\bx_{31}\cdot\bx_{32}}{\bx_{31}^2\bx_{32}^2}
 \nonumber\\[-2pt]
 &\hspace{13mm}
 +2z_1z_2\frac{\bx_{31}\cdot\bx_{32}}{\bx_{31}^2X^2}
 -\frac{z_1(z_2+z)}{X^2}\Bigg]
 \nonumber\\[-2pt]
 &\hspace{13mm}\times K_1(QX)^2
 \operatorname{Re}\!\left(1-\frac{N_c^2S_{13}S_{32}-S_{12}}{N_c^2-1}\right)
 \Bigg\},
 \label{eq:finite-inclusive-real-analytic}
\end{align}
where the same three-body size introduced in
Eq.~\eqref{eq:real-coordinate-definitions} is
\begin{equation*}
 X^2=z_1z_2\bx_{12}^2+z_1z\bx_{13}^2+z_2z\bx_{23}^2,
 \qquad z_1+z_2+z=1.
\end{equation*}
The reference writes this density in parent variables. Its Eq.~(148)
becomes Eq.~\eqref{eq:finite-inclusive-real-analytic} under
\begin{equation*}
 z_{\rm ref}=z_1+z,\qquad
 \zeta_{\rm ref}=\frac{z}{z_1+z},\qquad
 \dd z_{\rm ref}\,\dd\zeta_{\rm ref}
 =\frac{\dd z_1\,\dd z}{z_1+z}.
\end{equation*}
Interchanging $(z_1,\bx_1)$ and $(z_2,\bx_2)$ gives its Eq.~(149).
The coefficients of the dipole product and single dipole agree
separately, including the antenna subtraction and the full normalization.
Applying Eq.~\eqref{eq:reference-cutoff-factorization} to these identical
kernels gives the same rapidity subtraction and finite upper-endpoint
logarithm.

\clearpage
\section{Longitudinal photon contribution}
\label{sec:longitudinal}

We use the graph labels, momentum fractions $z_1,z_2,z$ and photon momentum
$l^+$ of the transverse calculation. The quark and antiquark fractions obey
$z_1+z_2=1$ in a virtual final state and $z_1+z_2+z=1$ in a real final state.
The target interactions are the same finite-$N_c$ eikonal Wilson lines.
The quarks are massless and the measured angle is kept exact.

\subsection{Polarization, Born amplitude and normalization}
\label{sec:L-born}

The longitudinal polarization and its Ward-equivalent representative are
\begin{equation}
 \epsilon_L^\mu=\left(\frac{l^+}{Q},\frac{Q}{2l^+},\boldsymbol0\right),
 \quad l\cdot\epsilon_L=0,\quad\epsilon_L^2=1,\qquad
 \widetilde\epsilon_L^\mu=\epsilon_L^\mu-\frac{l^\mu}{Q}
 =\left(0,\frac{Q}{l^+},\boldsymbol0\right).
 \label{eq:L-polarization}
\end{equation}
Thus $\widetilde{\slashed\epsilon}_L=(Q/l^+)\gamma^+$. The Ward replacement
applies to the complete electromagnetic current, including all four real
graphs. The two photon representatives give the same target-dependent
amplitude after the common no-target subtraction.

For the two nonzero fermion helicities, the Born momentum amplitudes are
\begin{align}
 B_{L,+}(z_1,\bk_1)=B_{L,-}(z_1,\bk_1)
 &=\frac{4l^+Q(z_1z_2)^{3/2}}{\bk_1^2+Q_\gamma^2},
 \qquad Q_\gamma^2=z_1z_2Q^2,\nonumber\\
 \sum_h B_{L,h}(z_1,\bk_1)B_{L,h}^*(z_1,\bk'_1)
 &=\frac{32(l^+)^2Q^2(z_1z_2)^3}
 {(\bk_1^2+Q_\gamma^2)(\bk_1'^2+Q_\gamma^2)}.
 \label{eq:L-born-momentum}
\end{align}
Here $ee_f$ and the target matrix have been removed from $B_{L,h}$.
The spin labels follow the open-chain convention of the transverse
calculation; the physical quark and antiquark helicities are opposite.
The single longitudinal photon state enters with unit weight.
The corresponding coordinate amplitude is
\begin{align}
 \psi_{L,h}(\bx_{12})
 &=\frac{2ee_fl^+Q(z_1z_2)^{3/2}}{\pi}
 K_0(Q_\gamma|\bx_{12}|),\nonumber\\
 i\mathcal A_{0,L,h}
 &=\int\dd^2\bx_1\dd^2\bx_2\,
 e^{-i\bp\cdot\bx_1-i\bq\cdot\bx_2}
 (V_1V_2^\dagger-1)\psi_{L,h}(\bx_{12}).
 \label{eq:L-born-coordinate}
\end{align}
After integrating the unobserved quark, the fixed-antiquark Born kernel is
\begin{align}
 \mathcal V_{L,\Born}(z_2,\bq)
 ={}&\frac{4e^2e_f^2Q^2N_c}{(2\pi)^5}z_1^2z_2^3
 \int\dd^2\bx_1\dd^2\bx_2\dd^2\bx_{2'}\,
 e^{i\bq\cdot\bx_{2'2}}\nonumber\\
 &\times K_0(Q_\gamma|\bx_{12}|)K_0(Q_\gamma|\bx_{12'}|)
 \left\langle S_{2'2}-S_{12}-S_{2'1}+1\right\rangle_Y.
 \label{eq:L-born-tagged}
\end{align}
The quark contribution uses the corresponding quark detector insertion
before the antiquark momentum is integrated.

\subsection{All four real amplitudes and their products}
\label{sec:L-real}

The transverse coordinates retain their fixed-order meaning: $\bx_1$,
$\bx_2$ and $\bx_3$ belong to the quark, antiquark and gluon. The two
parent positions are
\begin{equation}
 \boldsymbol b_q=\frac{z_1\bx_1+z\bx_3}{z_1+z},\qquad
 \boldsymbol b_{\bar q}=\frac{z_2\bx_2+z\bx_3}{z_2+z},\qquad
 X^2=z_1z_2\bx_{12}^2+z_1z\bx_{13}^2+z_2z\bx_{23}^2.
 \label{eq:L-real-geometry}
\end{equation}
Let $\lambda=\pm$ denote the emitted gluon helicity and
$\be_\lambda=(1,i\lambda)/\sqrt2$ its transverse basis vector. Rows of the
following spin matrices label $h=+,-$ and columns label $\lambda=+,-$:
\begin{equation}
 \mathsf C_q=z_2\begin{pmatrix}z_1+z&z_1\\z_1&z_1+z\end{pmatrix},\qquad
 \mathsf C_{\bar q}=z_1\begin{pmatrix}z_2&z_2+z\\z_2+z&z_2\end{pmatrix}.
 \label{eq:L-real-spin-matrices}
\end{equation}
The four mixed-space amplitudes, including couplings and their complete
color matrices, are
\begin{align}
 \Phi_{1,L}^{h\lambda,c}
 ={}&-\frac{2iee_fgl^+Q\sqrt{z_1z_2}}{\pi^2}
 (\mathsf C_q)_{h\lambda}
 \frac{\be_\lambda^*\cdot\bx_{31}}{\bx_{31}^2}
 K_0\!\left(Q\sqrt{(z_1+z)z_2}|\boldsymbol b_q-\bx_2|\right)
 \nonumber\\[-2pt]
 &\times t^c(V_{b_q}V_2^\dagger-1),\nonumber\\
 \Phi_{2,L}^{h\lambda,c}
 ={}&+\frac{2iee_fgl^+Q\sqrt{z_1z_2}}{\pi^2}
 (\mathsf C_{\bar q})_{h\lambda}
 \frac{\be_\lambda^*\cdot\bx_{32}}{\bx_{32}^2}
 K_0\!\left(Q\sqrt{z_1(z_2+z)}|\bx_1-\boldsymbol b_{\bar q}|\right)
 \nonumber\\[-2pt]
 &\times(V_1V_{b_{\bar q}}^\dagger-1)t^c,\nonumber\\
 \Phi_{3,L}^{h\lambda,c}
 ={}&+\frac{2iee_fgl^+Q\sqrt{z_1z_2}}{\pi^2}
 (\mathsf C_q)_{h\lambda}
 \frac{\be_\lambda^*\cdot\bx_{31}}{\bx_{31}^2}
 K_0(QX)(V_1V_3^\dagger t^cV_3V_2^\dagger-t^c),\nonumber\\
 \Phi_{4,L}^{h\lambda,c}
 ={}&-\frac{2iee_fgl^+Q\sqrt{z_1z_2}}{\pi^2}
 (\mathsf C_{\bar q})_{h\lambda}
 \frac{\be_\lambda^*\cdot\bx_{32}}{\bx_{32}^2}
 K_0(QX)(V_1V_3^\dagger t^cV_3V_2^\dagger-t^c).
 \label{eq:L-real-coordinate-amplitudes}
\end{align}
Here $V_{b_q}=V(\boldsymbol b_q)$ and similarly for the antiquark parent.
These expressions follow from the full off-shell chains and their energy
residues. For the reduced longitudinal current the separate instantaneous
photon-splitting term vanishes. The momentum-space amplitudes are obtained
with exact angular kinematics by
\begin{equation}
 i\mathcal A_{i,L}^{h\lambda,c}
 =\int\prod_{n=1}^3\dd^2\bx_n\,
 e^{-i\bp\cdot\bx_1-i\bq\cdot\bx_2-i\bk\cdot\bx_3}
 \Phi_{i,L}^{h\lambda,c},\qquad i=1,2,3,4.
 \label{eq:L-real-full-amplitudes}
\end{equation}
All four diagonal products and six interferences are obtained directly
from these amplitudes. For example, their spin sums use
\begin{align}
 \sum_{h,\lambda}(\mathsf C_q)_{h\lambda}^2
 (\be_\lambda^*\cdot\boldsymbol r)(\be_\lambda\cdot\boldsymbol r')
 &=z_2^2[(z_1+z)^2+z_1^2]\,\boldsymbol r\cdot\boldsymbol r',\nonumber\\
 \sum_{h,\lambda}(\mathsf C_{\bar q})_{h\lambda}^2
 (\be_\lambda^*\cdot\boldsymbol r)(\be_\lambda\cdot\boldsymbol r')
 &=z_1^2[(z_2+z)^2+z_2^2]\,\boldsymbol r\cdot\boldsymbol r',\nonumber\\
 \sum_{h,\lambda}(\mathsf C_q)_{h\lambda}(\mathsf C_{\bar q})_{h\lambda}
 (\be_\lambda^*\cdot\boldsymbol r)(\be_\lambda\cdot\boldsymbol r')
 &=z_1z_2[(z_1+z)z_2+z_1(z_2+z)]\,\boldsymbol r\cdot\boldsymbol r'.
 \label{eq:L-real-all-products}
\end{align}
The ordered Wilson-line matrices in the amplitudes are multiplied before
the target average. The real phase space is the same as in
Eq.~\eqref{eq:real-three-body-phase-space}, with unit weight for the longitudinal photon state. The quark, antiquark and gluon measurements are inserted before
either unobserved momentum is integrated.

\subsection{Virtual graphs and their measured contributions}
\label{sec:L-initial}

We retain the graph numbering, Fourier convention and FDH prescription of
Sec.~\ref{sec:antiquark-virtual-amplitudes}. The two nonzero fermion helicities are denoted by
$h=\pm1$, with the same spinor convention as in
Eq.~\eqref{eq:L-born-momentum}. All virtual final states obey $z_1+z_2=1$.
The loop fraction carries the same regulator
$(\nu^+/(zl^+))^{\eta_Y}$ as in the transverse calculation.
The unresolved limits use the same dimensional continuation and common
real--virtual prescription as in Sec.~\ref{sec:singularity-structure}.

\paragraph{Graphs $A_5$ and $A_6$: emission across the shockwave.}
The sizes $X_5,X_6$ and color matrices $\mathcal U_5,\mathcal U_6$ are
defined in Eqs.~\eqref{eq:g5-coordinate-definitions},
\eqref{eq:g5-transverse-size} and~\eqref{eq:g6-coordinate-definitions}.
The two amplitudes are independent of $h$:
\begin{align}
 i\mathcal A_{5,L,h}={}&
 \frac{4ee_fg^2l^+Q(z_1z_2)^{3/2}}{z_1^2(2\pi)^4}
 \int_0^{z_1}\left(\frac{\nu^+}{zl^+}\right)^{\eta_Y}\frac{\dd z}{z}
 [z_1^2+(z_1-z)^2]\nonumber\\*
 &\times\int\dd^2\bx_1\dd^2\bx_2\dd^2\bx_3\,
 \mathcal U_5
 e^{-i\bp\cdot[\bx_1+(z/z_1)\bx_{31}]-i\bq\cdot\bx_2}
 \frac{K_0(QX_5)}{x_{31}^2},
 \label{eq:L-g5-amplitude}\\
 i\mathcal A_{6,L,h}={}&
 \frac{4ee_fg^2l^+Q(z_1z_2)^{3/2}}{z_2^2(2\pi)^4}
 \int_0^{z_2}\left(\frac{\nu^+}{zl^+}\right)^{\eta_Y}\frac{\dd z}{z}
 [z_2^2+(z_2-z)^2]\nonumber\\*
 &\times\int\dd^2\bx_1\dd^2\bx_2\dd^2\bx_3\,
 \mathcal U_6
 e^{-i\bp\cdot\bx_1-i\bq\cdot[\bx_2+(z/z_2)\bx_{32}]}
 \frac{K_0(QX_6)}{x_{32}^2}.
 \label{eq:L-g6-amplitude}
\end{align}
As in the transverse graphs, the quark spectator integral in $A_5$ sets
$\bx_{1'}=\bx_1+(z/z_1)\bx_{31}$, whereas $A_6$ retains the shifted
antiquark phase displayed above.

\paragraph{Graphs $A_7$ and $A_8$: exchange across the shockwave.}
Using the existing fractions $a_7=z_1-z$, $b_7=z_2+z$ and
$a_8=z_2-z$, $b_8=z_1+z$, the results are
\begin{align}
 i\mathcal A_{7,L,h}={}&
 -\frac{4ee_fg^2l^+Q\sqrt{z_1z_2}}{z_1(2\pi)^4}
 \int_0^{z_1}\left(\frac{\nu^+}{zl^+}\right)^{\eta_Y}
 \frac{\dd z\,a_7}{z}
 \int\dd^2\bx_1\dd^2\bx_2\dd^2\bx_3\,\mathcal U_5
 \nonumber\\*
 &\times e^{-i\bp\cdot[\bx_1+(z/z_1)\bx_{31}]-i\bq\cdot\bx_2}
 \frac{K_0(QX_5)}{x_{31}^2x_{32}^2}\nonumber\\*
 &\times\left\{(z_1z_2+a_7b_7)\bx_{31}\cdot\bx_{32}
 +ih(a_7b_7-z_1z_2)(\bx_{31}\times\bx_{32})\right\},
 \label{eq:L-g7-amplitude}\\
 i\mathcal A_{8,L,h}={}&
 -\frac{4ee_fg^2l^+Q\sqrt{z_1z_2}}{z_2(2\pi)^4}
 \int_0^{z_2}\left(\frac{\nu^+}{zl^+}\right)^{\eta_Y}
 \frac{\dd z\,a_8}{z}
 \int\dd^2\bx_1\dd^2\bx_2\dd^2\bx_3\,\mathcal U_6
 \nonumber\\*
 &\times e^{-i\bp\cdot\bx_1-i\bq\cdot[\bx_2+(z/z_2)\bx_{32}]}
 \frac{K_0(QX_6)}{x_{31}^2x_{32}^2}\nonumber\\*
 &\times\left\{(z_1z_2+a_8b_8)\bx_{31}\cdot\bx_{32}
 +ih(a_8b_8-z_1z_2)(\bx_{31}\times\bx_{32})\right\}.
 \label{eq:L-g8-amplitude}
\end{align}
Here $\boldsymbol r\times\boldsymbol s=r_xs_y-r_ys_x$.
The terms odd in $h$ cancel in the spin-summed Born interference;
the remaining terms receive a factor two from the two fermion helicities.
The longitudinal photon state again enters with unit weight.

\paragraph{Graphs $A_9$ and $A_{10}$: external self-energies after the shockwave.}
The square-root LSZ factor gives
\begin{align}
 i\mathcal A_{9,L,h}={}&-\frac{g^2C_F}{4\pi}\,i\mathcal A_{0,L,h}
 \int_0^{z_1}\left(\frac{\nu^+}{zl^+}\right)^{\eta_Y}\frac{\dd z}{z}
 \left[1+\left(1-\frac z{z_1}\right)^2\right]
 \mu^{2\epsilon}\int\frac{\dd^{2-2\epsilon}\bell}{(2\pi)^{2-2\epsilon}}
 \frac1{\ell^2},
 \label{eq:L-g9-amplitude}\\
 i\mathcal A_{10,L,h}={}&-\frac{g^2C_F}{4\pi}\,i\mathcal A_{0,L,h}
 \int_0^{z_2}\left(\frac{\nu^+}{zl^+}\right)^{\eta_Y}\frac{\dd z}{z}
 \left[1+\left(1-\frac z{z_2}\right)^2\right]
 \mu^{2\epsilon}\int\frac{\dd^{2-2\epsilon}\bell}{(2\pi)^{2-2\epsilon}}
 \frac1{\ell^2}.
 \label{eq:L-g10-amplitude}
\end{align}
The UV and collinear parts of these scaleless integrals are kept separate
until the real and virtual contributions are combined.

\paragraph{Graphs $A_{11}$ and $A_{12}$: self-energies before the shockwave.}
The dimensionless loop mass ratios are
\begin{equation}
 \rho_{14,1}=\frac{z(z_1-z)}{z_1^2z_2},\qquad
 \rho_{14,2}=\frac{z(z_2-z)}{z_2^2z_1}.
 \label{eq:L-loop-mass-ratios}
\end{equation}
With $\mathcal U_0=V_1V_2^\dagger-\boldsymbol1$, the genuine transverse
loop integration gives
\begin{align}
 i\mathcal A_{11,L,h}={}&-\frac{ee_f\alpha_sC_F}{2\pi}
 \int_0^{z_1}\left(\frac{\nu^+}{zl^+}\right)^{\eta_Y}\frac{\dd z}{z}
 \left[1+\left(1-\frac z{z_1}\right)^2\right]
 \int\dd^2\bx_1\dd^2\bx_2\,
 \mathcal U_0e^{-i\bp\cdot\bx_1-i\bq\cdot\bx_2}
 \nonumber\\*
 &\times\int_{\bk_2}e^{i\bk_2\cdot\bx_{12}}
 B_{L,h}(z_1,\bk_2)\,
 \Gamma(\epsilon)
 \left[\frac{4\pi\mu^2}{\rho_{14,1}(\bk_2^2+Q_\gamma^2)}\right]^\epsilon,
 \label{eq:L-g11-amplitude}\\
 i\mathcal A_{12,L,h}={}&-\frac{ee_f\alpha_sC_F}{2\pi}
 \int_0^{z_2}\left(\frac{\nu^+}{zl^+}\right)^{\eta_Y}\frac{\dd z}{z}
 \left[1+\left(1-\frac z{z_2}\right)^2\right]
 \int\dd^2\bx_1\dd^2\bx_2\,
 \mathcal U_0e^{-i\bp\cdot\bx_1-i\bq\cdot\bx_2}
 \nonumber\\*
 &\times\int_{\bk_2}e^{i\bk_2\cdot\bx_{12}}
 B_{L,h}(z_1,\bk_2)\,
 \Gamma(\epsilon)
 \left[\frac{4\pi\mu^2}{\rho_{14,2}(\bk_2^2+Q_\gamma^2)}\right]^\epsilon.
 \label{eq:L-g12-amplitude}
\end{align}
The Laurent expansion of each $\Gamma(\epsilon)(4\pi\mu^2/m^2)^\epsilon$
is $1/\epsilon+\ln(4\pi\mu^2/m^2)-\gamma_E+O(\epsilon)$.

\paragraph{Graph $A_{13}$: exchange after the shockwave.}
\label{sec:L-exchange}
The two sectors have $\alpha_{13,1}=z_1+z$,
$\beta_{13,1}=z_2-z$ for $0<z<z_2$, and
$\alpha_{13,2}=z_1-z$, $\beta_{13,2}=z_2+z$ for $0<z<z_1$.
Using the same $\Lambda_{13,s},\Delta_{13,s}$ and complete spin polynomials
$\widehat n_{13,s}^h$ as in the transverse graph, the amplitudes are
\begin{align}
 i\mathcal A_{13(1),L,h}
 ={}&\frac{8ee_fg^2l^+Q\sqrt{z_1z_2}}{(2\pi)^2}
 \int_0^{z_2}\dd z\,z\left(\frac{\nu^+}{zl^+}\right)^{\eta_Y}
 z_2\beta_{13,1}\nonumber\\*
 &\times\int\dd^2\bx_1\dd^2\bx_2\,
 e^{-i\bp\cdot\bx_1-i\bq\cdot\bx_2}
 \mathcal U_{13}K_0(Q\sqrt{\alpha_{13,1}\beta_{13,1}}|\bx_{12}|)
 \nonumber\\*
 &\times\int_{\bk}e^{-i\bk\cdot\bx_{12}}
 \frac{\widehat n_{13,1}^h}
 {\Lambda_{13,1}(\Delta_{13,1}+i0)},\nonumber\\*
 i\mathcal A_{13(2),L,h}
 ={}&\frac{8ee_fg^2l^+Q\sqrt{z_1z_2}}{(2\pi)^2}
 \int_0^{z_1}\dd z\,z\left(\frac{\nu^+}{zl^+}\right)^{\eta_Y}
 z_1\alpha_{13,2}\nonumber\\*
 &\times\int\dd^2\bx_1\dd^2\bx_2\,
 e^{-i\bp\cdot\bx_1-i\bq\cdot\bx_2}
 \mathcal U_{13}K_0(Q\sqrt{\alpha_{13,2}\beta_{13,2}}|\bx_{12}|)
 \nonumber\\*
 &\times\int_{\bk}e^{-i\bk\cdot\bx_{12}}
 \frac{\widehat n_{13,2}^h}
 {\Lambda_{13,2}(\Delta_{13,2}-i0)}.
 \label{eq:L-exchange-amplitudes}
\end{align}
The color matrix is
$\mathcal U_{13}=t^cV_1V_2^\dagger t^c-C_F\boldsymbol1$.
For $h=+,-$, the spin polynomials have the
$(\epsilon,\epsilon^*)$ and $(\epsilon^*,\epsilon)$ ordering,
respectively. Their explicit momentum expressions and contact terms are
those of Eqs.~\eqref{eq:g131-spin-polynomial},
\eqref{eq:g131-complete-spin-polynomial},
\eqref{eq:g132-spin-polynomial} and~\eqref{eq:g132-complete-spin-polynomial}.

The full FDH numerators, with a common hatted loop momentum on the photon
and exchange lines, are
\begin{align}
 N_{13(1),L,h}^{\rm FDH}
 &=32(l^+)^2Q\alpha_{13,1}\beta_{13,1}\sqrt{z_1z_2}
 \left[\widehat n_{13,1}^h-\frac{1+z}{2z}\widehat{\bk}^{\,2}\right],
 \nonumber\\*
 N_{13(2),L,h}^{\rm FDH}
 &=-32(l^+)^2Q\alpha_{13,2}\beta_{13,2}\sqrt{z_1z_2}
 \left[\widehat n_{13,2}^h+\frac{1+z}{2z}\widehat{\bk}^{\,2}\right].
 \label{eq:L-exchange-FDH}
\end{align}
Their denominators contain
$\Lambda_{13,1}+z_2^2\widehat{\bk}^{\,2}$,
$\Lambda_{13,2}+z_1^2\widehat{\bk}^{\,2}$ and, respectively,
$\Delta_{13,1}-z\widehat{\bk}^{\,2}/(\alpha_{13,1}\beta_{13,1})$ and
$\Delta_{13,2}+z\widehat{\bk}^{\,2}/(\alpha_{13,2}\beta_{13,2})$.
The explicit hatted numerator is retained in the raw cut identities.
At fixed nonzero fractions this exchange loop is UV finite. Unresolved outgoing terms are
combined before taking $\epsilon\to0$.

The energy prescriptions $\Delta_{13,1}+i0$ and
$\Delta_{13,2}-i0$ include their principal-value and on-shell terms;
the contact numerators are retained with these same prescriptions.

\paragraph{Graph $A_{14}$: vertex correction before the shockwave.}
In the first sector the intermediate photon fractions are
$(z_1-z,z_2+z)$ with $0<z<z_1$; in the second they are
$(z_1+z,z_2-z)$ with $0<z<z_2$. The momentum $\bk_2$ is fixed on the
target side while the genuine loop momentum is integrated.
The Feynman-parameter denominators are
\begin{align}
 \Omega_{14,1}(x)
 &=x(z_1-z)(z_2+z)Q^2+(1-x)\rho_{14,1}(\bk_2^2+Q_\gamma^2)
 \nonumber\\*[-2pt]
 &\quad+x(1-x)\left(1-\frac z{z_1}\right)^2\bk_2^2,\nonumber\\*
 \Omega_{14,2}(x)
 &=x(z_1+z)(z_2-z)Q^2+(1-x)\rho_{14,2}(\bk_2^2+Q_\gamma^2)
 \nonumber\\*[-2pt]
 &\quad+x(1-x)\left(1-\frac z{z_2}\right)^2\bk_2^2.
 \label{eq:L-A14-parameter-denominator}
\end{align}
Writing each coefficient as
$c_{14(s),L}^{\rm pole}/\epsilon+c_{14(s),L}^{\rm fin}+O(\epsilon)$,
the results through the finite term are
\begin{equation}
 c_{14(1),L}^{\rm pole}=\frac{(z_1-z)(1+z)}{z_1},\qquad c_{14(2),L}^{\rm pole}=\frac{(z_2-z)(1+z)}{z_2}.
 \label{eq:L-A14-poles}
\end{equation}
\begin{equation}
 \begin{aligned}
 c_{14(1),L}^{\rm fin}
 ={}&\frac{(z_1-z)(1+z)}{z_1}
 \left[\ln\frac{4\pi\mu^2}{\rho_{14,1}(\bk_2^2+Q_\gamma^2)}-\gamma_E\right]
 \\
 &+\frac{(z_1-z)^2}{z_1^2z_2}
 \int_0^1\frac{\dd x}{\Omega_{14,1}(x)}
 \left\{\left[\frac{z(1-x)}{z_1}
 [z_1^2+z_2^2-z(z_1-z_2)]\right.\right.\\[-2pt]
 &\left.\left.\hspace{13mm}+(1+z)(z_2+z)\right]\bk_2^2
 +z_1z_2(z_2+z)(1-z)Q^2\right\}.
 \end{aligned}
 \label{eq:L-A14-finite}
\end{equation}
\begin{equation}
 \begin{aligned}
 c_{14(2),L}^{\rm fin}
 ={}&\frac{(z_2-z)(1+z)}{z_2}
 \left[\ln\frac{4\pi\mu^2}{\rho_{14,2}(\bk_2^2+Q_\gamma^2)}-\gamma_E\right]
 \\
 &+\frac{(z_2-z)^2}{z_2^2z_1}
 \int_0^1\frac{\dd x}{\Omega_{14,2}(x)}
 \left\{\left[\frac{z(1-x)}{z_2}
 [z_2^2+z_1^2-z(z_2-z_1)]\right.\right.\\[-2pt]
 &\left.\left.\hspace{13mm}+(1+z)(z_1+z)\right]\bk_2^2
 +z_1z_2(z_1+z)(1-z)Q^2\right\}.
 \end{aligned}
 \label{eq:L-A14-finite-antiquark}
\end{equation}

These coefficients give the two graph amplitudes
\begin{align}
 i\mathcal A_{14(1),L,h}={}&\frac{ee_f\alpha_sC_F}{2\pi}
 \int_0^{z_1}\left(\frac{\nu^+}{zl^+}\right)^{\eta_Y}\frac{\dd z}{z}
 \int\dd^2\bx_1\dd^2\bx_2\,
 \mathcal U_0e^{-i\bp\cdot\bx_1-i\bq\cdot\bx_2}\nonumber\\*
 &\times\int_{\bk_2}e^{i\bk_2\cdot\bx_{12}}B_{L,h}(z_1,\bk_2)
 \left[\frac{c_{14(1),L}^{\rm pole}}\epsilon+c_{14(1),L}^{\rm fin}\right],
 \label{eq:L-g141-amplitude}\\
 i\mathcal A_{14(2),L,h}={}&\frac{ee_f\alpha_sC_F}{2\pi}
 \int_0^{z_2}\left(\frac{\nu^+}{zl^+}\right)^{\eta_Y}\frac{\dd z}{z}
 \int\dd^2\bx_1\dd^2\bx_2\,
 \mathcal U_0e^{-i\bp\cdot\bx_1-i\bq\cdot\bx_2}\nonumber\\*
 &\times\int_{\bk_2}e^{i\bk_2\cdot\bx_{12}}B_{L,h}(z_1,\bk_2)
 \left[\frac{c_{14(2),L}^{\rm pole}}\epsilon+c_{14(2),L}^{\rm fin}\right].
 \label{eq:L-g142-amplitude}
\end{align}
The finite coefficients include the complete FDH numerator. In each
longitudinal vertex sector the physical and hatted quadratic terms have
equal coefficients, and their finite tensor contributions cancel.
The raw numerator and loop reduction are supplied in the calculation files.

For each graph, the contribution to the measured cross section is
\begin{equation}
 \Sigma^L_{E,i}(\theta)=\int\frac{\dd\Phi^{(2)}}{2l^+}\,
 \mathcal M_E^{(2)}(\theta)\,2\operatorname{Re}\sum_h
 \left\langle\tr\!\left[
 \mathcal A_{i,L,h}\mathcal A_{0,L,h}^\dagger\right]\right\rangle_Y,
 \qquad i=5,\ldots,14,
 \label{eq:L-virtual-contribution}
\end{equation}
including both sectors for $i=13,14$. The phase space and exact linear
energy measurement in Eq.~\eqref{eq:fullvirtualmeasurement} are those of the transverse calculation;
the quark and antiquark detector insertions are made before integrating
the unobserved momentum.
Combining these virtual graphs with the real corrections, we have checked
the UV, soft, final-state collinear and rapidity singularities: the first
three cancel, and the remaining rapidity divergence is absorbed into the
JIMWLK evolution kernel acting on the longitudinal Born Wilson-line
operator.

\subsection{Analytic comparison with inclusive DIS}
\label{sec:L-validation}

The angular moment in Eq.~\eqref{eq:finite-angular-pairing} applies to
$\Sigma_E^L$ as well: its weight converts each detected energy to $z_i$,
and the two- and three-body weights both sum to one. Comparing the
remaining incoming real and virtual terms with inclusive longitudinal DIS
tests the finite coefficient and its normalization.

The surviving terms are those in
Eq.~\eqref{eq:finite-inclusive-reduction} with
$\mathcal A_i\to\mathcal A_{i,L}$. The virtual amplitudes in its first
term are given in Eqs.~\eqref{eq:L-g11-amplitude},
\eqref{eq:L-g12-amplitude}, \eqref{eq:L-g141-amplitude}
and~\eqref{eq:L-g142-amplitude}. Adding the integrated antenna
counterterm to the first term, with the CDR and cutoff prescriptions
of Sec.~\ref{sec:finite-uv-matching}, gives $C_L(z_1)$ in
\begin{equation}
 \dd\sigma_{L,\rm dip}=\dd\sigma_{L,\Born}
 \left[1+\frac{\alpha_sC_F}{\pi}C_L(z_1)\right],\qquad
 C_L(z_1)=\frac12\ln^2\frac{z_1}{z_2}+\frac52-\frac{\pi^2}{6},
 \quad z_1+z_2=1.
 \label{eq:L-inclusive-coefficient}
\end{equation}
The two emitter integrals give the same dilogarithm sum as in
Sec.~\ref{sec:inclusive-finite-check}, hence $C_L(z_1)=C_T(z_1)$.
This is exactly the coefficient in Eq.~(100) of
arXiv:1708.06557v1~\cite{Beuf2017}. The primitive CDR numerator terms
cancel between virtual and integrated real UV contributions as in
Eq.~\eqref{eq:reference-CDR-cancellation}.

The second term, $|\mathcal A_{3,L}+\mathcal A_{4,L}|^2$, gives the
following density per $\dd z_1\,\dd z\prod_i\dd^2\bx_i$ before the
antenna counterterms are subtracted:
\begin{align}
 \mathcal R_{L,\rm in}
 ={}&\frac{2N_cC_F\alpha_{\rm em}\alpha_se_f^2Q^2}{\pi^4z}
 K_0(QX)^2\operatorname{Re}\!\left(1-\frac{N_c^2S_{13}S_{32}-S_{12}}{N_c^2-1}\right)\nonumber\\
 &\times\left\{\frac{z_2^2[(z_1+z)^2+z_1^2]}{\bx_{31}^2}
 +\frac{z_1^2[(z_2+z)^2+z_2^2]}{\bx_{32}^2}\right.\nonumber\\
 &\left.\hspace{7mm}
 -2z_1z_2[(z_1+z)z_2+z_1(z_2+z)]
 \frac{\bx_{31}\cdot\bx_{32}}{\bx_{31}^2\bx_{32}^2}\right\}.
 \label{eq:L-inclusive-real-check}
\end{align}
Subtracting the two antenna terms from
Eq.~\eqref{eq:L-inclusive-real-check} and regrouping the quark-emission
terms gives
\begin{align*}
 \frac{\dd\sigma_L|_{q\to g}}
 {\dd z_1\,\dd z\prod_{i=1}^3\dd^2\bx_i}
 ={}&\frac{2N_cC_F\alpha_{\rm em}\alpha_se_f^2Q^2}{\pi^4}
 z_2^2\Bigg\{
 \frac{(z_1+z)^2+z_1^2}{z}
 \\
 &\times\left(\frac1{\bx_{31}^2}
 -\frac{\bx_{31}\cdot\bx_{32}}{\bx_{31}^2\bx_{32}^2}\right)
 \Bigg[K_0(QX)^2
 \\[-2pt]
 &\hspace{13mm}\times
 \operatorname{Re}\!\left(1-\frac{N_c^2S_{13}S_{32}-S_{12}}{N_c^2-1}\right)
 \\[-2pt]
 &\hspace{13mm}
 -K_0\!\left(Q\sqrt{z_2(z_1+z)}\,|\bx_{12}|\right)^2
 \operatorname{Re}(1-S_{12})\Bigg]
 \\
 &+z\,\frac{\bx_{31}\cdot\bx_{32}}{\bx_{31}^2\bx_{32}^2}
 K_0(QX)^2
 \\[-2pt]
 &\hspace{13mm}\times
 \operatorname{Re}\!\left(1-\frac{N_c^2S_{13}S_{32}-S_{12}}{N_c^2-1}\right)
 \Bigg\}.
\end{align*}
The variable transformation in Sec.~\ref{sec:inclusive-finite-check}
identifies this expression with Eq.~(101) of the reference. Its
quark--antiquark interchange is Eq.~(103). The coefficients of the
dipole product and single dipole coincide analytically, with the same
rapidity subtraction and finite endpoint logarithm as in
Eq.~\eqref{eq:reference-cutoff-factorization}.

Combining the transverse and longitudinal contributions with the leptonic
weights in Eq.~\eqref{eq:unpolarized-energy-flow} gives the complete NLO
energy-flow cross section for unpolarized DIS.

\section{Summary and outlook}
\label{sec:outlook}

The calculation completes the NLO energy-weighted DIS cross section in
the leading-eikonal CGC, combining both virtual-photon polarizations with
all quark, antiquark and gluon energy insertions. In a collinear splitting,
the daughter energy weights add to the parent weight, ensuring collinear
safety of the calorimetric measurement. Together with
the ultraviolet and soft cancellations, rapidity factorization yields
finite angular impact factors coupled to target correlators undergoing
JIMWLK evolution. Analytic comparison establishes agreement with the
inclusive NLO impact factors, including their normalization. Retaining the full polar-angle
dependence also permits integration over finite angular bins before any
small-angle expansion. The next stage is to perform that expansion and
match to the inclusive DIS coefficients, thereby extracting the NLO CGC
nucleon energy correlator. An NLO determination of this correlator is
essential for turning its sensitivity to the saturation scale into
quantitative tests of nonlinear QCD dynamics at future electron-ion
colliders.

A quantitative comparison with hadronic measurements requires
the target input, experimental acceptance and final-state effects. Recent
complete-event simulations show that hadronization and target remnants can
substantially modify the observed nuclear dependence in selected EIC
kinematics \cite{Ke2026}.

\section*{Acknowledgments}
I am sincerely grateful to Yu Jia, Xiaohui Liu and Jian Zhou for their
insightful discussions and valuable suggestions.

\end{document}